\documentclass[prl,superscriptaddress,twocolumn]{revtex4-1} 

\usepackage{color}
\usepackage{mathtools}
\usepackage{tabularx}
\usepackage{ifpdf}
\usepackage{graphicx}
\usepackage[hyperindex=true]{hyperref}

\definecolor{linkcol}{rgb}{0.2,0.2,0.6}

\hypersetup{bookmarksopen=true,
pdfmenubar=true, 
pdfhighlight=/O, 
colorlinks=true, 
pdfpagemode=None, 
pdffitwindow=true, 
linkcolor=linkcol, 
citecolor=linkcol, 
urlcolor=linkcol 
}

\def\SCBO{SrCu$_2$(BO$_3$)$_2$}

\newcolumntype{L}[1]{>{\raggedright\arraybackslash}p{#1}}
\newcolumntype{C}[1]{>{\centering\arraybackslash}p{#1}}
\newcolumntype{R}[1]{>{\raggedleft\arraybackslash}p{#1}}

\begin{document}

\title{Field-induced bound-state condensation and spin-nematic phase in 
SrCu$_2$(BO$_3$)$_2$ revealed by neutron scattering up to 25.9 T}

\author{Ellen Fogh}
\thanks{These authors contributed equally to this work.}
\affiliation{Laboratory for Quantum Magnetism, Institute of Physics, 
Ecole Polytechnique F\'{e}d\'{e}rale de Lausanne (EPFL), CH-1015 
Lausanne, Switzerland}
\author{Mithilesh Nayak}
\thanks{These authors contributed equally to this work.}
\affiliation{Institute of Physics, 
Ecole Polytechnique F\'{e}d\'{e}rale de Lausanne (EPFL), CH-1015 
Lausanne, Switzerland}
\author{Oleksandr Prokhnenko}
\affiliation{Helmholtz-Zentrum Berlin f\"{u}r Materialien und Energie, 
D-14109 Berlin, Germany}
\author{Maciej Bartkowiak}
\affiliation{Helmholtz-Zentrum Berlin f\"{u}r Materialien und Energie, 
D-14109 Berlin, Germany}
\affiliation{ISIS Neutron and Muon Source, Rutherford Appleton Laboratory, Harwell OX11 0QX, UK}
\author{Koji Munakata}
\affiliation{Neutron Science and Technology Center, Comprehensive Research 
Organization for Science and Society (CROSS), Tokai, Ibaraki 319-1106, Japan}
\author{Jian-Rui Soh}
\affiliation{Laboratory for Quantum Magnetism, Institute of Physics, 
Ecole Polytechnique F\'{e}d\'{e}rale de Lausanne (EPFL), CH-1015 
Lausanne, Switzerland}
\author{Alexandra A. Turrini}
\affiliation{Laboratory for Neutron Scattering and Imaging, Paul 
Scherrer Institute, CH-5232 Villigen-PSI, Switzerland}
\affiliation{Laboratory for Quantum Magnetism, Institute of Physics, 
Ecole Polytechnique F\'{e}d\'{e}rale de Lausanne (EPFL), CH-1015 
Lausanne, Switzerland}
\author{Mohamed E. Zayed}
\affiliation{Department of Physics, Carnegie Mellon University in Qatar, 
Education City, PO Box 24866, Doha, Qatar}
\author{Ekaterina Pomjakushina}
\affiliation{Laboratory for Multiscale Materials Experiments, Paul Scherrer 
Institute, CH-5232 Villigen PSI, Switzerland}
\author{Hiroshi Kageyama}
\affiliation{Graduate School of Engineering, Kyoto University, Nishikyo-ku, Kyoto 615-8510, Japan}
\author{Hiroyuki Nojiri}
\affiliation{Institute for Materials Research, Tohoku University, 
Sendai 980-8577, Japan}
\author{Kazuhisa Kakurai}
\affiliation{Neutron Science and Technology Center, Comprehensive 
Research Organization for Science and Society (CROSS), Tokai, Ibaraki 
319-1106, Japan}
\author{Bruce Normand}
\affiliation{Laboratory for Quantum Magnetism, Institute of Physics, 
Ecole Polytechnique F\'{e}d\'{e}rale de Lausanne (EPFL), CH-1015 
Lausanne, Switzerland}
\affiliation{Laboratory for Theoretical and Computational Physics, 
Paul Scherrer Institute, CH-5232 Villigen-PSI, Switzerland}
\author{Fr\'{e}d\'{e}ric Mila}
\affiliation{Institute of Physics, 
Ecole Polytechnique F\'{e}d\'{e}rale de Lausanne (EPFL), CH-1015 
Lausanne, Switzerland}
\author{Henrik M. R\o nnow}
\affiliation{Laboratory for Quantum Magnetism, Institute of Physics, 
Ecole Polytechnique F\'{e}d\'{e}rale de Lausanne (EPFL), CH-1015 
Lausanne, Switzerland}

\maketitle

{\bf Bose-Einstein condensation (BEC) underpins exotic forms of order ranging 
from superconductivity to superfluid $^4$He. In quantum magnetic materials, 
ordered phases induced by an applied magnetic field can be described as the 
BEC of magnon excitations. With sufficiently strong magnetic frustration, 
exemplified by the system \SCBO, no clear magnon BEC is observed and the 
complex spectrum of multi-magnon bound states may allow a different type of 
condensation, but the high fields required to probe this physics have remained 
a barrier to detailed investigation. Here we exploit the first purpose-built 
high-field neutron scattering facility to measure the spin excitations of 
\SCBO~up to 25.9~T and use cylinder matrix-product-states (MPS) calculations 
to reproduce the experimental spectra with high accuracy. Multiple 
unconventional features point to a condensation of $S = 2$ bound states 
into a spin-nematic phase, including the gradients of the one-magnon branches, 
the presence of many novel composite two- and three-triplon excitations and 
the persistence of a one-magnon spin gap. This gap reflects a direct analogy 
with superconductivity, suggesting that the spin-nematic phase in \SCBO~is 
best understood as a condensate of bosonic Cooper pairs. Our results underline 
the wealth of unconventional states yet to be found in frustrated quantum 
magnetic materials under extreme conditions.} 

Condensation of a macroscopic number of particles into a single state is a 
purely quantum mechanical phenomenon exhibited by a wide spectrum of bosonic 
systems \cite{pitaevskii2003}. Early and spectacular examples of this 
Bose-Einstein condensation (BEC) include superfluidity in liquid $^4$He and 
superconductivity in metals, where attractive interactions allow the electrons, 
which are fermions, to form Cooper pairs, which are bosons. Field-induced gap 
closure and magnetic order in quantum disordered magnetic materials is 
conventionally described as a BEC of magnons, which are spin-1 excitations 
\cite{giamarchi2008,zapf2014}; examples of this phenomenon include the 
dimerized quantum antiferromagnets TlCuCl$_3$ \cite{nikuni2000} and 
BaCuSi$_2$O$_6$ \cite{jaime2004,allenspach2022}, and, under quite different 
experimental conditions, yttrium iron garnet \cite{demokritov2006,sun2017}.

\begin{figure*}
\centering
\includegraphics[width = 0.96\textwidth]{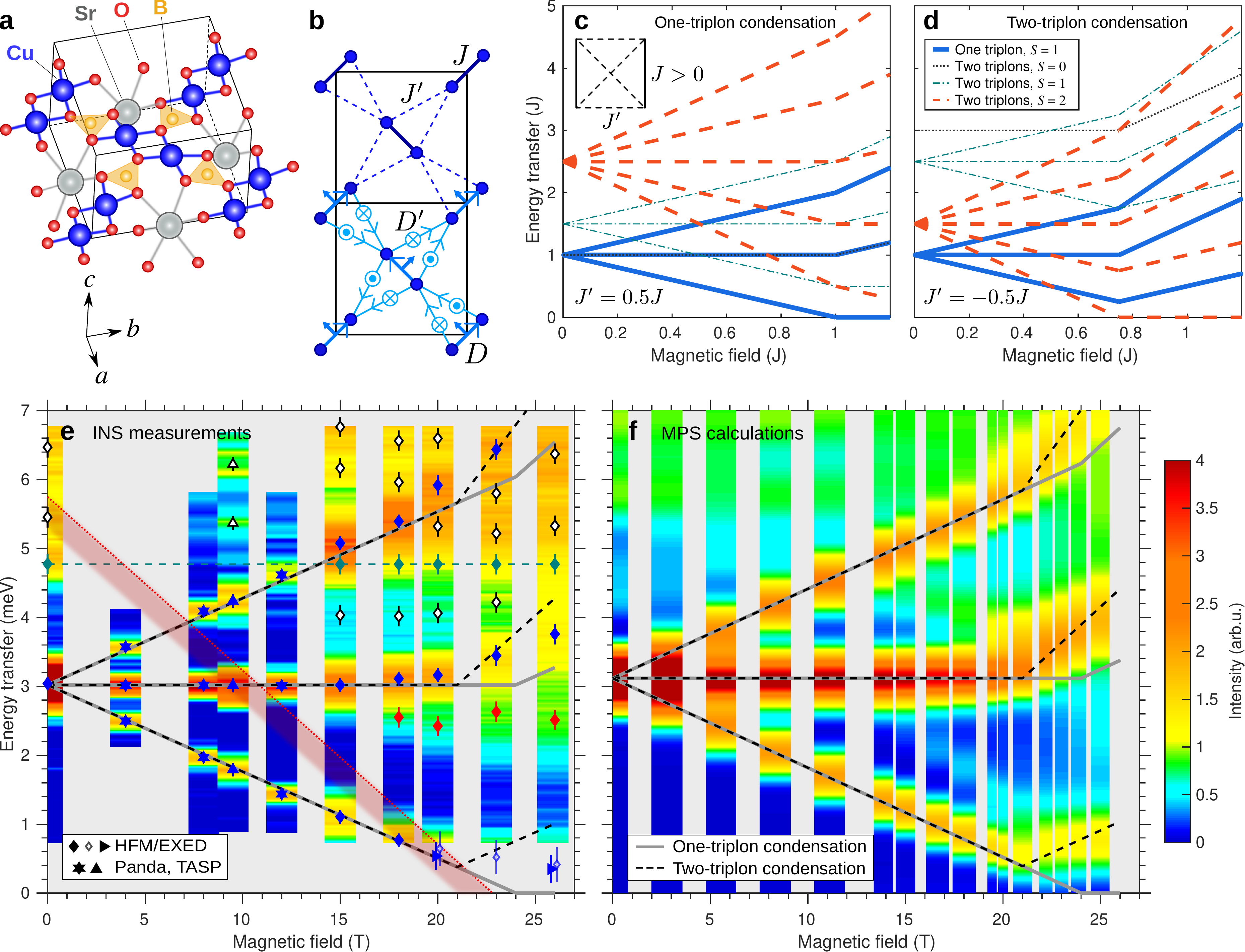}
\caption{{\bf a} Crystal structure of \SCBO~showing one layer of Cu$^{2+}$ 
ions (half a unit cell along ${\hat c}$).
{\bf b} Representation of the superexchange couplings within ($J$) and between 
($J'$) the Cu-Cu dimers, which reproduce the Shastry-Sutherland model (SSM), 
and of the corresponding Dzyaloshinskii-Moriya (DM) interactions (${\vec D}$ 
and ${\vec D}'$).
{\bf c-d} Energy levels of a simple $J$-$J'$ plaquette system (inset in panel 
{\bf c}) shown as a function of magnetic field. For $J'/J > 0$ ({\bf c}), the 
$t_{+}$ one-triplon branch condenses first, whereas when $J'/J < 0$ ({\bf d}), 
the two-triplon bound state condenses first. The two scenarios result in 
different gradients after condensation.
{\bf e} Excitation spectra measured by inelastic neutron scattering (INS) 
over a wide range of fields. Data at low fields (4, 8, 9.5 and 12~T) were 
measured at ${\bf Q} = (1.5,0.5,0)$ as described in the Methods section. 
Data from our HFM/EXED measurements (0, 15, 18, 20, 23 and 25.9~T) are 
presented with full momentum integration, as described in Sec.~S1 of the SI 
\cite{si}. Colour contours represent the neutron intensity and symbols 
show the fitted positions of individual excitations (Figs.~\ref{fig:Fig2}g-l).  
Solid grey lines show the projected locations of the one-triplon branches for 
a scenario of one-triplon condensation and no DM interactions (panel {\bf c}), 
dashed black lines their locations for two-triplon condensation (panel {\bf d}).
The dashed green line shows the $S_z = 0$ branch of the $S = 1$ multiplet 
within the two-triplon bound state. The dotted red line up to 15~T shows 
the $S_z = 2$ branch of the $S = 2$ multiplet as measured by electron spin 
resonance (ESR) \cite{nojiri2003}, i.e.~at ${\bf Q} = (0,0)$, where the 
dispersion of this branch reaches its maximum \cite{momoi2000}. The red 
shading represents the estimated bandwidth of this branch, whose minimum 
determines the binding energy, and our linear extrapolation beyond 15~T 
indicates its effect on gap closure.
{\bf f} Total momentum-integrated spectral functions obtained from our 
cylinder MPS calculations on the SSM with additional DM interactions over 
the same wide field range.}
\label{fig:Fig1}
\end{figure*}

By contrast, the field-induced phase diagram of the highly frustrated quantum 
magnet \SCBO~(Fig.~\ref{fig:Fig1}a) \cite{kageyama1999b} shows neither clear 
gap closure nor induced magnetic order, but instead a spectacular series of 
magnetization plateaux \cite{kageyama1999b,jaime2012,takigawa2013,matsuda2013,
nomura2022}. \SCBO~presents a remarkably faithful realization of the 
Shastry-Sutherland model (SSM), which is a paradigm for ideal frustration 
in a two-dimensional (2D) spin system \cite{shastry1981}. This system 
consists of $S = 1/2$ dimer units arranged orthogonally on a square 
lattice (Fig.~\ref{fig:Fig1}b), and its properties are governed by the 
ratio between the intra-dimer interaction, $J$, and inter-dimer one, $J'$. 
For $J'/J < 0.675$, the ground state is an exact dimer-product singlet 
state \cite{corboz2013}, and the ideally frustrated geometry leads to many 
unconventional effects, including near-dispersionless one-magnon excitations 
(to which we refer henceforth as ``triplons'') and very strongly bound 
multi-triplon states \cite{kageyama2000a,gaulin2004,kakurai2005,zayed2014a}. 
At ambient pressure and zero field, \SCBO~is very well described 
by a SSM with $J'/J \simeq 0.63$ \cite{miyahara2000} and shows anomalous 
thermodynamic behaviour \cite{wietek2019,larrea2021} due to the high spectral 
density of highly localized spin excitations. Under an applied hydrostatic 
pressure \cite{zayed2017,guo2020,larrea2021}, its behaviour mirrors the phases 
of the SSM as $J'/J$ is changed.

Most remarkable of all is the appearance of field-induced magnetization 
plateaux in \SCBO: extremely strong magnetic fields applied along the $c$-axis 
(Fig.~\ref{fig:Fig1}a) induce plateaux at $1/8$, $2/15$, $1/6$, $1/4$, $1/3$, 
$2/5$ and $1/2$ of the saturation magnetization, which is reached around 139~T 
\cite{nomura2022}. In direct analogy with charge-density-wave (CDW) order in 
electronic systems, these plateaux can be seen as bosonic CDW phases of 
magnetic entities. The predicted spin superstructure on the lowest (1/8) 
plateau is a crystal of $S = 2$ two-triplon bound states, as opposed to a 
crystal of unbound $S = 1$ triplons \cite{corboz2014}; this prediction awaits 
experimental verification. Although the transition to the 1/8 plateau takes 
place with a jump at $\mu_0 H_c = 27$~T, the magnetization is clearly finite 
above approximately 16~T, and above 21~T it increases linearly until the jump 
\cite{takigawa2013}. In this crossover region, the gap to the lowest triplon 
branch does not close as the field is increased, but shows an avoided crossing 
previously ascribed to the presence of weak Dzyaloshinskii-Moriya (DM) 
interactions \cite{cepas2001,nojiri2003,kodama2005} 

\begin{figure*}
\centering
\includegraphics[width = 1.7\columnwidth]{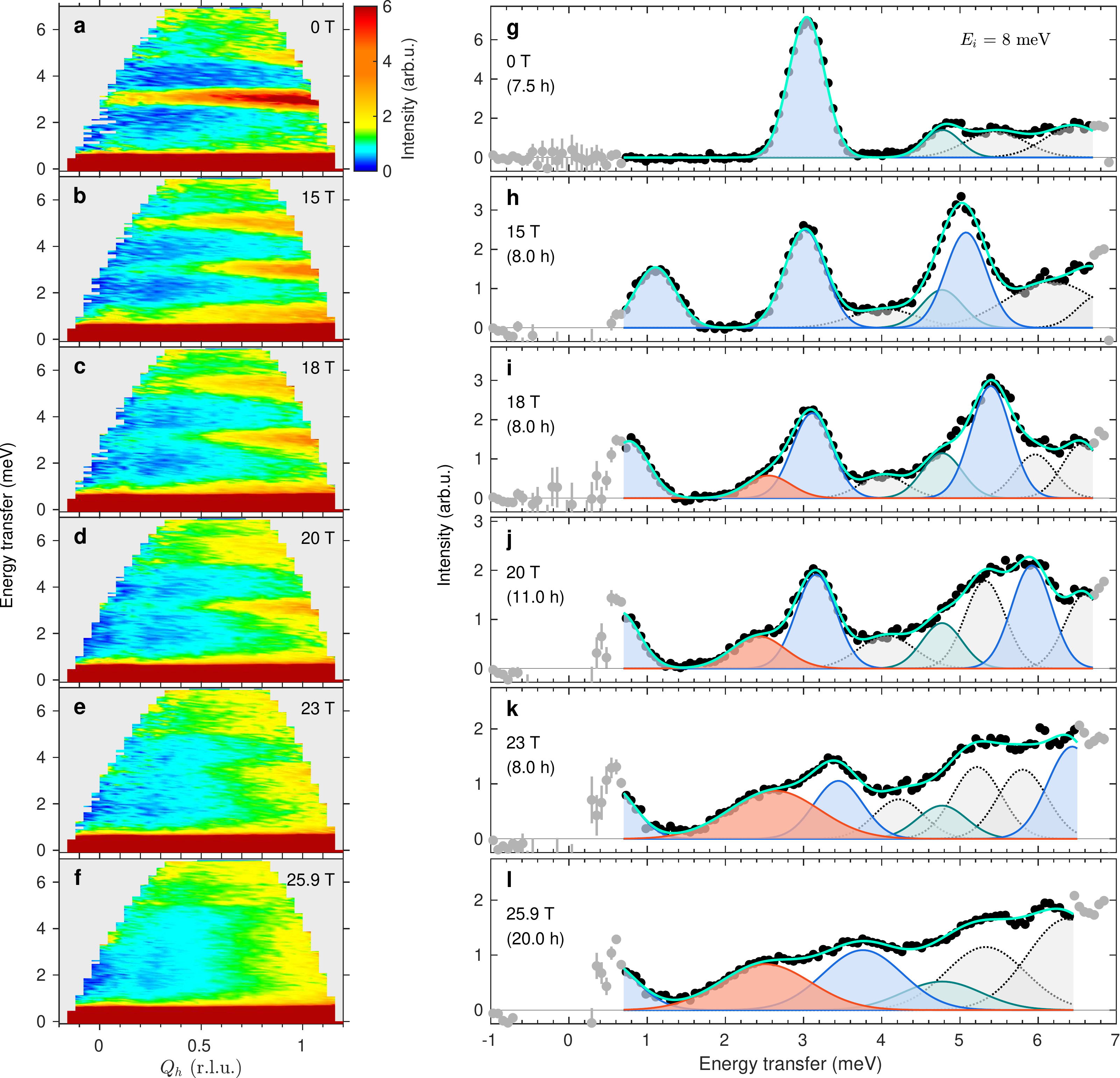}
\caption{{\bf a-f} Measured neutron intensity (colour contours) shown as a 
function of energy transfer and of momentum $Q_h$ for six different field 
strengths. These data were integrated over the intervals $Q_k = [-0.75,0.75]$ 
and $Q_l = [-1.5,2.5]$ and no background was subtracted. It is clear that all 
excitations have very little dispersion, justifying the integration of our 
data over a large region of reciprocal space. {\bf g-l} Neutron intensity 
obtained by further integration over $Q_h = [-0.25,1.25]$ (black symbols), 
shown as a function of energy transfer at the same six field strengths. A 
single background was subtracted as detailed in Sec.~S1B of the SI \cite{si}. 
The one-triplon excitation (solid blue lines with blue shading) exhibits 
Zeeman-type splitting in the applied field and its branches were fitted by 
related Gaussians (Sec.~S1C of the SI \cite{si}). The $S_z = 0$ branch of 
the $S = 1$ two-triplon bound state (dark turquoise lines and shading) 
appears at a constant energy. Additional intensity develops below the $t_0$ 
one-triplon branch at $\mu_0 H \ge 18$~T (solid red lines with red shading) 
and was fitted with a single Gaussian. Further additional intensity visible 
below the $t_-$ one-triplon branch at $\mu_0 H \ge 15$~T (dotted black lines 
with grey shading) was fitted with multiple Gaussians.}
\label{fig:Fig2}
\end{figure*}

It was suggested very early that the field-induced gap closure in \SCBO~could 
be anomalous \cite{bendjama2005}. The zero-field gap of the nearly immobile 
triplons is $\Delta \simeq 3.0$ meV, but triplon pairs gain kinetic energy 
from correlated hopping, such that two-triplon states with $S = 0$ and 1 lie 
far below $2 \Delta$ and hence are strongly bound (with respective energies 
4.0 \cite{lemmens2000} and 4.8~meV \cite{kageyama2000a}). For the $S = 2$ 
multiplet, perturbative (in $J'/J$) calculations on the SSM indicate that 
its minimum, reached at momentum $(\pi,\pi)$, also lies below $2 \Delta$ 
\cite{momoi2000}. In this situation, as the minimal model of 
Figs.~\ref{fig:Fig1}c-d makes clear, increasing the field should cause the 
$S = 2$ bound state to meet the singlet ground state before the lowest 
one-triplon branch does. The result would be a BEC of two-triplon bound 
states, which is a spin-nematic phase \cite{blume1969,andreev1984}, instead 
of field-induced magnetic order. However, a conclusive demonstration has not 
to date been possible: standard momentum-resolved experimental probes do not 
observe $S = 2$ excitations, light-scattering methods probe them only at 
momentum ${\bf Q} = (0,0)$ (depicted in Figs~\ref{fig:Fig1}e) and the DM 
interactions obscure the situation.

The definitive experiment would be to determine the dynamical structure 
factor at all ${\bf Q}$ by inelastic neutron scattering (INS), and to 
compare this with model calculations. Although INS has historically been 
limited to fields below 15~T, for a limited period neutron scattering 
was possible up to 25.9~T (DC) on a specialized facility at the 
Helmholtz-Zentrum 
Berlin (HZB) \cite{prokhnenko2015,smeibidl2016,prokhnenko2017}, 
thus enabling the investigation of \SCBO~over most of the formerly ``dark'' 
field regime below $H_c$. Model calculations for frustrated 2D quantum systems 
have also been a historical impossibility, but recent progress in numerical 
methods based on matrix-product states (MPS) has made it possible to calculate 
hitherto unavailable spectral functions with quantitative accuracy. These 
parallel developments render high-field INS the ideal tool to decode the 
spin dynamics in \SCBO~and in a range of other systems \cite{allenspach2022}.

Here we harness this progress to demonstrate bound-state condensation forming 
a spin nematic in \SCBO. We measure the excitation spectrum using INS at 
200~mK with magnetic field strengths up to 25.9~T (Fig.~\ref{fig:Fig1}e), and 
we perform MPS calculations of the dynamical structure factor of the SSM with 
weak DM interactions (Fig.~\ref{fig:Fig1}f). By tracking the Zeeman splitting 
of the one-triplon mode, we find field-induced changes of the gradients of 
the triplon branches and a persisting, DM-independent gap that are fully 
consistent with bound-state condensation. Our INS and MPS spectra also 
reveal a rapid, field-induced transfer of weight to multiple novel excitations, 
which we show are two- and three-triplon composites that can only exist if the 
condensate contains bound states.

\medskip
\noindent
{\large {\bf Results}}

\noindent
{{\bf Inelastic neutron scattering.}} We performed INS experiments at the 
High-Field Magnet (HFM) facility that operated recently at the HZB. The 
time-of-flight EXED instrument offered the unique capability of performing 
neutron diffraction and spectroscopy at fields up to 25.9~T. A further key 
was a dilution refrigerator enabling a sample temperature of 200~mK, well 
below the smallest energy scales in \SCBO. Unless otherwise stated, all 
the INS data we show were collected with incoming energy $E_i = 8$~meV, 
yielding an energy resolution of 0.64(1)~meV. Further information 
concerning our experiments is summarized in the Methods section and in 
Sec.~S1 of the Supplementary Information (SI) \cite{si} we provide a 
detailed description of the HFM/EXED experimental geometry, our data 
preprocessing and background subtraction. 

Figure \ref{fig:Fig1}e shows the fully ${\bf Q}$-integrated neutron 
intensity as a function of applied field and energy transfer. In 
Figs.~\ref{fig:Fig2}a-f we show the neutron intensity measured as a 
function of $Q_h$, with partial integration around $Q_k = 0$. All the 
excitations we observe are rather flat over the full reciprocal space, 
and this justifies integrating the intensity data over a large ${\bf Q}$ 
volume, which improves the counting statistics significantly. However, a 
weak dispersion is visible in most modes (Figs.~\ref{fig:Fig2}a-f), which 
on integration appears as an additional contribution to the linewidth. 
The intensity in every branch remains concentrated at the largest values 
of $Q_h$ for all fields, while both the intensities and the relative 
intensities between branches show weak changes with $Q_h$, which we 
analyse  in Sec.~1D of the SI \cite{si}.

Integrating over the entire available ${\bf Q}$ range leads to the results 
shown in Figs.~\ref{fig:Fig2}g-l. At zero field (Fig.~\ref{fig:Fig2}g), the 
one-triplon excitation is found around 3.0~meV, with a number of higher-lying 
excitations at 4.8, 5.5 and 6.5~meV, all in good agreement with previous INS 
measurements \cite{kageyama2000a,gaulin2004,kakurai2005,zayed2014a}. 
In an applied field, the one-triplon excitation shows Zeeman splitting into 
three branches, which are clearly visible at 15~T (Fig.~\ref{fig:Fig2}h). We 
label the bottom, middle and top branches respectively as $t_+$, $t_0$ and 
$t_-$. Their splitting increases linearly with the field until approximately 
18~T (Fig.~\ref{fig:Fig2}i), as in ESR measurements \cite{nojiri2003}. Above 
this field, the $t_+$ branch curves towards a rather constant value around 
0.5~meV, suggesting an avoided crossing, with no indication that the 
one-triplon gap closes. In addition, we observe the development of a shoulder 
on the low-energy side of the $t_0$ triplon, which becomes progressively 
stronger as the field is raised (Figs.~\ref{fig:Fig2}i-l). Around the $t_-$ 
triplon we observe further intensity contributions over a broad energy range, 
which become stronger at high fields and can be fitted reasonably by three 
separate Gaussians. Within our experimental resolution, the energies of these 
new excitations remain largely constant as the magnetic field is increased 
(Figs.~\ref{fig:Fig1}e and \ref{fig:Fig2}i-l).

\begin{figure}
\centering
\includegraphics[width = 0.96\columnwidth]{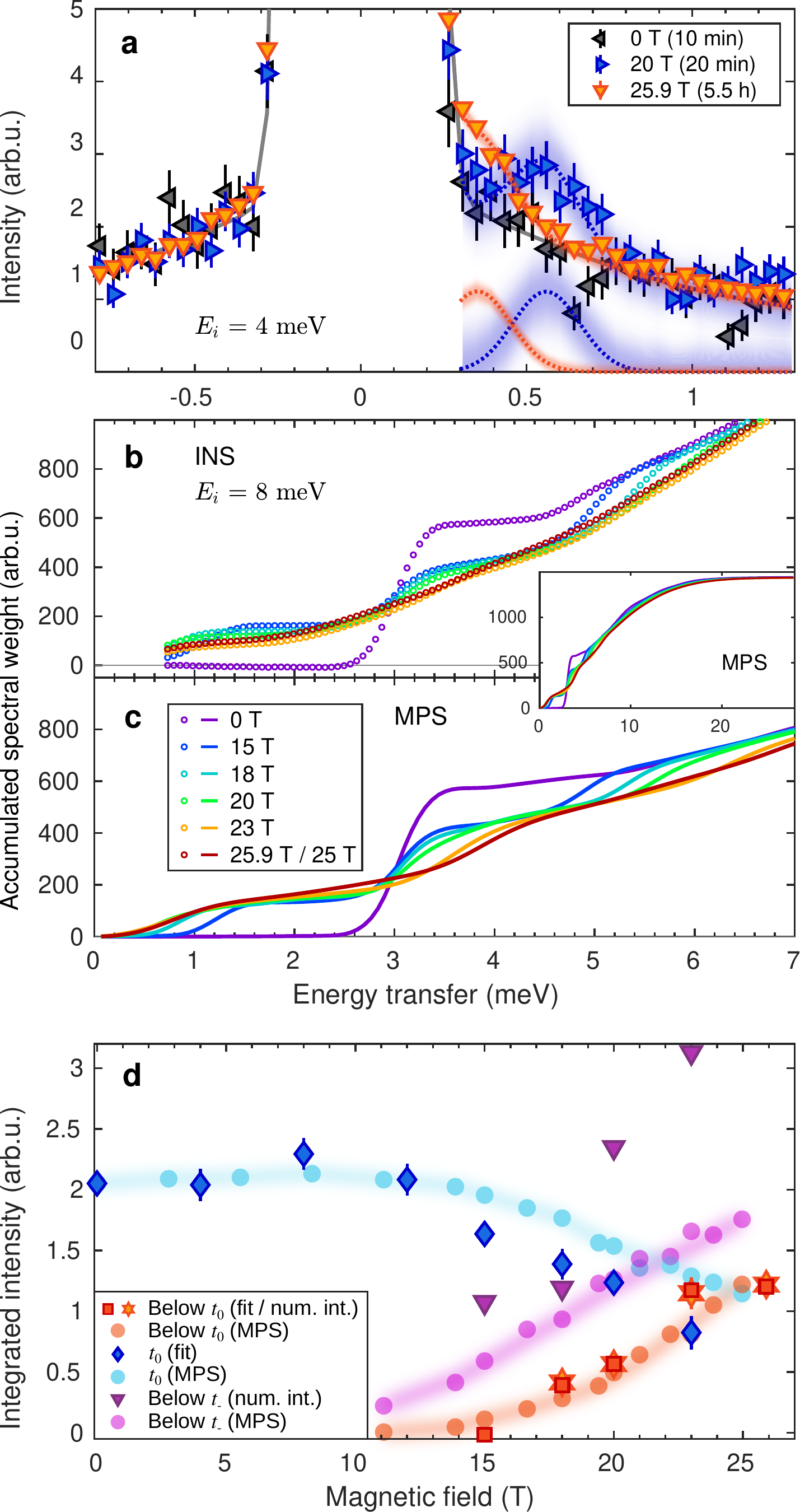}
\caption{{\bf a} Data collected at zero field (black), 20~T (blue) and 25.9~T 
(red) with incident energy $E_i = 4$~meV to resolve the low-energy excitations 
(solid blue and red lines). The data integration range and determination of 
the background (solid black line) are detailed in Sec.~S1B of the SI \cite{si}. 
The mode positions were determined by fitting to a single Gaussian (dotted 
blue and red lines and shading). {\bf b}-{\bf c} Accumulated spectral weight 
shown as a function of energy transfer at each field for our INS ({\bf b}) 
and MPS data ({\bf c}). The inset shows an extended energy range calculated 
by MPS. {\bf d} Integrated intensity of different low-energy spectral 
contributions, shown as a function of field for both the INS and MPS data.}
\label{fig:Fig3}
\end{figure}

As Figs.~\ref{fig:Fig1}c-d showed, the field-induced evolution of the lowest 
($t_+$) triplon mode is important for identifying the nature of condensation. 
The fact that its gap remains finite throughout the crossover regime improves 
the prospects for resolving its position close to the elastic line. We 
performed measurements with an incoming energy of $E_i = 4$ meV, at which 
the resolution improves to 0.24(1) meV. In Fig.~\ref{fig:Fig3}a we show 
the neutron spectra measured at 20 and 25.9~T, where a coherent low-energy 
feature remains discernible, and we extract the respective positions 0.54 
and 0.35~meV (also marked in Fig.~\ref{fig:Fig1}e), reinforcing our result 
that the gap does not close at any field.

To analyse our intensity data in more detail, in Fig.~\ref{fig:Fig3}b 
we show the spectral weights accumulated at energy transfers above a 
0.7~meV cut-off for each measurement field. At zero field, this weight is 
zero until the 3.0~meV triplet branch is encountered, after which it is 
constant again until the $S = 1$ branch of the two-triplon bound state 
at 4.8~meV, and beyond this it increases linearly as multiple higher-lying 
states are encountered. At 15~T, the one-triplon sector is split into three 
clearly defined branches, each of which gives a significant jump in the 
accumulated spectral weight. At higher fields, these jumps become 
increasingly broad as additional states appear at multiple different 
energies, particularly those directly below the $t_0$ and $t_-$ branches 
(Figs.~\ref{fig:Fig1}e and Figs.~\ref{fig:Fig2}i-l). We defer an 
interpretation of these results (Fig.~\ref{fig:Fig3}c) to our numerical 
analysis below. To investigate the physics of the different spectral-weight 
contributions identified in Figs.~\ref{fig:Fig2}g-l, in Fig.~\ref{fig:Fig3}d 
we inspect their evolution as a function of the field. It is clear that the 
intensity of the additional states (i.e.~those below $t_0$ and below $t_-$) 
scales linearly beyond the crossover region, meaning that it follows the 
magnetization, $m$, while the one-triplon weight is proportional to $1 - m$. 

\begin{figure*}
\includegraphics[width = 0.96\textwidth]{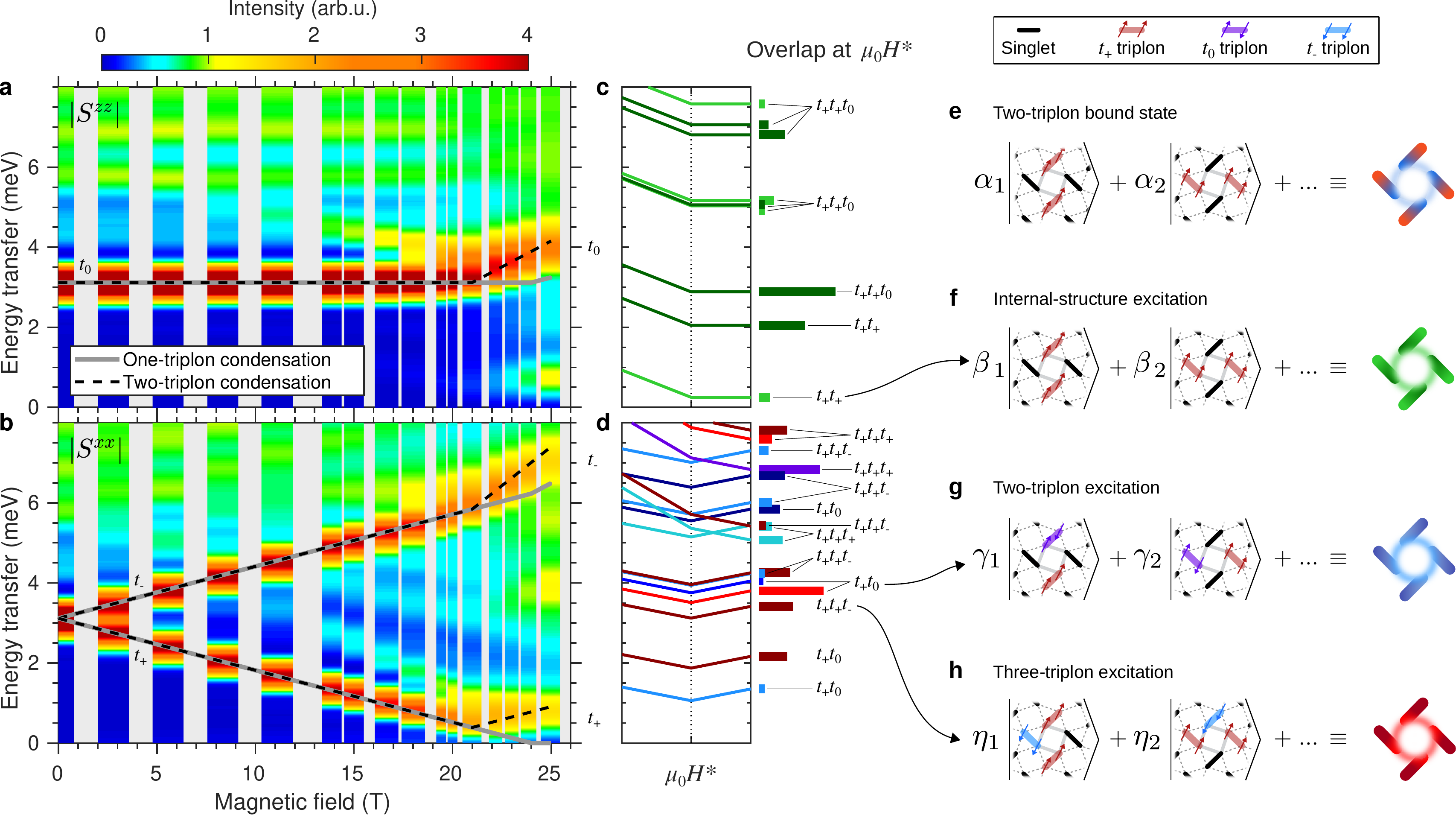}
\caption{{\bf a-b} Spectral functions $S^{zz} (\omega, H)$ and $S^{xx} 
(\omega,H)$ obtained by cylinder MPS calculations and compared with the 
one-triplon branches obtained from the one- and two-triplon-condensation 
scenarios represented in Fig.~\ref{fig:Fig1}. {\bf c-d} Energies and 
ground-state overlaps of states obtained in illustrative ED calculations 
performed around $H^*$ on a 4$\times$4 cluster. {\bf e-h} Schematic 
representations of the ``pinwheel'' structure of the two-triplon bound 
state ({\bf e}), an internal excitation of the pinwheel ({\bf f}), which 
retains the $t_+ t_+$ character, a two-triplon excitation consisting of a 
$t_+$ and a $t_0$ ({\bf g}) and a $t_+ t_+ t_-$ three-triplon excitation 
({\bf h}). In each case illustrated, the coefficients of the basis states 
shown have the same amplitude and differ only by their phases.} 
\label{fig:Fig4}
\end{figure*}

\smallskip
\noindent
{{\bf Numerical modelling.}} To model the experimental results, we have 
performed cylinder MPS calculations of the dynamical spectral function 
for the SSM supplemented by the weak (3\%) DM interactions deduced from 
experiment \cite{kodama2005}. To interpret the broadest features of the 
spectrum we appeal to the plaquette model of Figs.~\ref{fig:Fig1}c-d and 
for a detailed identification of the excitation types we perform exact 
diagonalization (ED) calculations for the SSM on clusters of 4$\times$4 
spins. A summary of our MPS calculations is provided in the Methods 
section, and in Sec.~S2A of the SI \cite{si} we present additional 
details including the effects of the MPS cylinder size. MPS results 
corresponding to the INS spectrum are shown in Fig.~\ref{fig:Fig1}f, 
which makes clear that the calculations contain all the features found 
in the measurements, most notably the additional excitations appearing 
below the $t_0$ and $t_-$ one-triplon branches beyond 18 T, at energy 
transfers in quantitative agreement with experiment.  

The first striking feature of our MPS results is the minimum in the $t_+$ 
branch at 21~T, with a finite gap of 0.6~meV, beyond which the mode energy 
increases again with the field. These low energies are difficult to access 
by INS because of the elastic peak (Fig.~\ref{fig:Fig3}a), but our MPS 
calculations confirm the absence of low-energy spectral weight here. The 
persistence of a gap is not a generic feature of field-induced magnon BEC 
\cite{giamarchi2008,zapf2014}, where the gap closes, the lowest magnon 
condenses and spectral weight appears down to zero energy at all higher 
fields. This spectral weight should extend up to the magnon bandwidth, 
which in \SCBO~is only 0.3~meV \cite{mcclarty2017}, and thus a minimum 
around 0.6~meV suggests different physics. 

It was assumed previously that this avoided closure is a consequence of the 
DM interactions in \SCBO, and indeed DM terms are known to preserve a gap in 
systems undergoing magnon BEC \cite{miyahara2007}. However, the minimum should 
then appear at the field where one-magnon condensation occurs, which by 
extrapolating both our INS and MPS results (Figs.~\ref{fig:Fig1}e-f) is 
clearly 24~T rather than the value $\mu_0 H^* = 21$~T that we observe. To 
confirm that this gap is not caused by the DM interactions, we repeated 
our MPS calculations with different values of $D$ and $D'$, as we show in 
Sec.~S2B of the SI \cite{si}. A robust gap exceeding 0.45~meV remains if 
the DM interactions are removed, and hence is an intrinsic property of the 
pure (Heisenberg) SSM. 

The natural explanation is the condensation of the $S = 2$ bound state. 
The minimal two-dimer model (inset, Fig.~\ref{fig:Fig1}c) is sufficient 
to illustrate some very clear effects in the one-triplon branches. With 
antiferromagnetic intra-dimer interactions ($J > 0$), the field-induced 
condensation process is readily controlled using the inter-dimer bonds: if 
$J' > 0$, the $t_+$ branch condenses (Fig.~\ref{fig:Fig1}c) in a conventional 
magnon BEC, whereas for $J' < 0$, the lowest branch of the two-triplon 
multiplet condenses (Fig.~\ref{fig:Fig1}d). In the latter case, the 
one-triplon branches show several distinctive features: (i) the minimum 
energy of the $t_+$ branch is half of the binding energy, and the 0.6~meV 
observed by MPS is consistent with half the binding energy predicted by 
perturbation theory and iPEPS calculations \cite{corboz2014}; (ii) this 
minimum occurs at a field, $H^*$, lower than the extrapolated $t_+$ 
condensation field; (iii) beyond $H^*$, the $t_+$ energy increases linearly 
with the field; (iv) the $t_0$ branch has a kink at $H^*$ and increases with 
a gradient that is twice as large; (v) the $t_-$ branch also has a kink at 
$H^*$ and increases with a gradient three times as large. These observations 
are completely generic for the condensation of $S = 2$ bound states and are 
fully consistent with both the experimental and the numerical results 
(Figs.~\ref{fig:Fig1}e-f).

Turning now to the novel field-induced excitations, we consider first their 
spectral weight. In Fig.~\ref{fig:Fig3}c we show the accumulated spectral 
weight computed by MPS in the same format as the experimental results of 
Fig.~\ref{fig:Fig3}b, also extending the MPS results to higher energies 
(inset). The agreement is excellent, with both datasets reflecting the 
field-induced smoothing of the steps as increasing amounts of weight 
are shifted to new excitations that appear at intermediate energies. By 
benchmarking the total weight at high energies, it is clear that the 
spectral weight lost from the INS energy window (0.7-7.0~meV) is pushed 
lower (Fig.~\ref{fig:Fig3}a). To quantify these weight-shifts, in 
Fig.~\ref{fig:Fig3}d we show the MPS intensities integrated over energy 
ranges spanning the $t_0$ branch, below this branch and below the $t_-$ 
branch. Beyond the obvious agreement with experiment, the MPS results 
confirm the linear growth of weight in multi-triplon excitations with 
fields beyond approximately 16~T and the remarkable (approximately 2/3) 
loss of one-triplon weight over the field range from there to the 26~T 
edge of the critical region around $H_c$. 

To go further in identifying these multi-triplon states, it is useful to 
separate the spectral function into its longitudinal [$S^{zz}(\omega, H)$] 
and transverse [$S^{xx}(\omega, H)$] components, which we show in 
Figs.~\ref{fig:Fig4}a-b. The $t_0$ branch appears in the $S^{zz}(\omega,H)$ 
channel and the $t^{\pm}$ modes in the $S^{xx}(\omega,H)$ channel, as we 
observe at low fields. However, above 21~T there is a rapid growth of 
low-energy spectral weight in $S^{zz}(\omega,H)$, specifically over a 
broad range centred near 0.7~meV, and of weight around the $t_0$ energy 
in $S^{xx}(\omega, H)$. In Figs.~\ref{fig:Fig4}c-d we show the energies and 
intensities of selected high-weight states obtained around $H^*$ in our 
cluster ED calculations for the SSM (meaning with no DM interactions), 
which as detailed in Sec.~S2D of the SI \cite{si} allow us to make an 
unambiguous identification of the nature ($S$ and $S^z$ quantum numbers) 
of these states. It is clear that two-triplon composite states appear at 
low energies in both sectors and that many three-triplon composites are 
present at energies extending down to the $t_0$ branch. 

\begin{figure}
\includegraphics[width = 0.96\columnwidth]{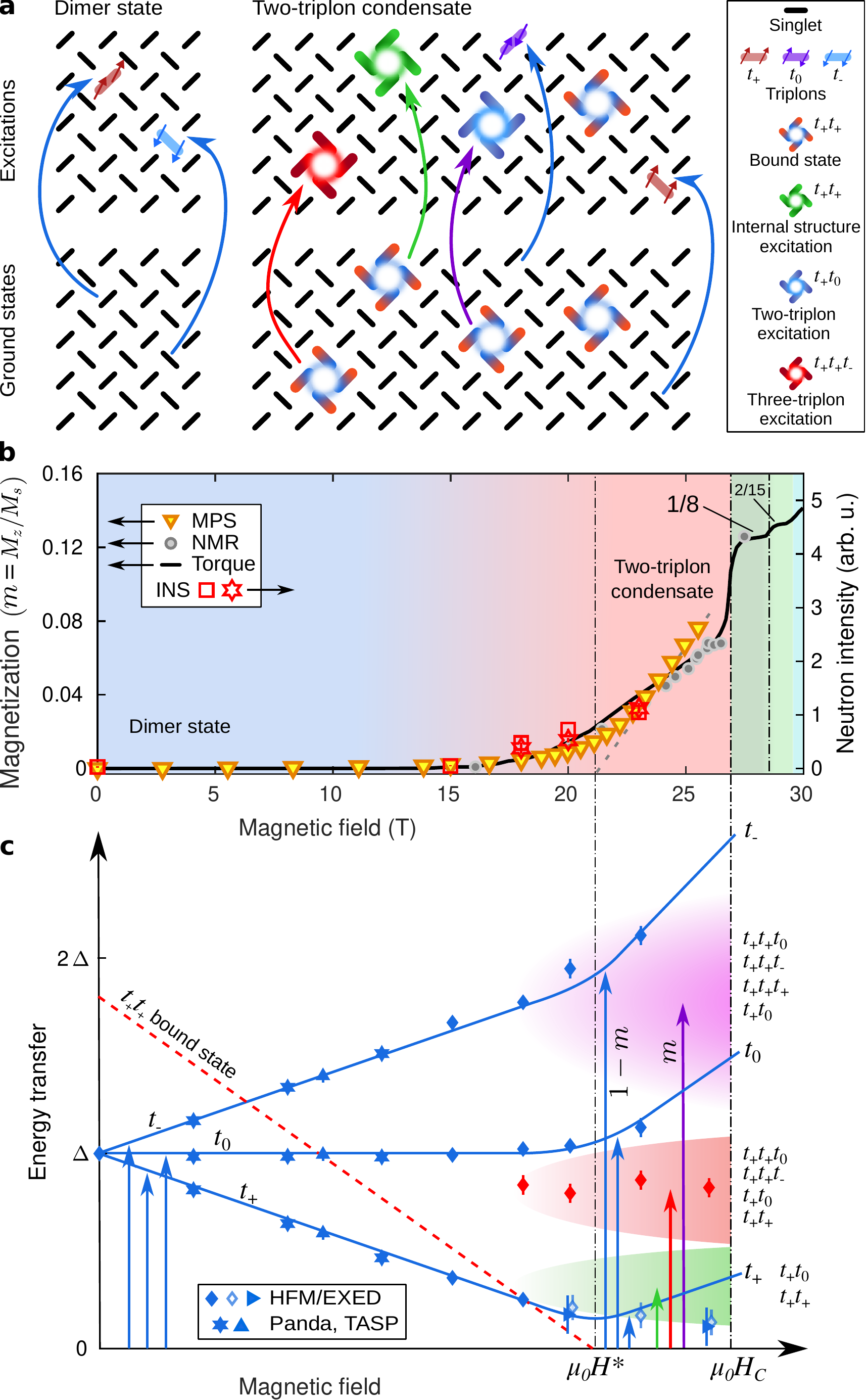}
\caption{{\bf a} Schematic representations of the ground states (lower 
panels) and all associated excitations (upper panels) at low fields in 
the dimer-product phase (left) and at fields beyond $H^*$ (right). {\bf b} 
Magnetization calculated by MPS in the field regime below the 1/8-plateau 
and compared with results obtained from NMR and magnetic torque measurements 
\cite{takigawa2013} (left axis); for comparison we show the measured 
intensity of the multi-triplon excitations below $t_0$ (right axis). 
{\bf c} Schematic illustration of the field-induced evolution of the primary 
spectral features identified by combining our neutron spectroscopy data with 
our MPS and ED results. $\Delta = 3.0$~meV is the one-triplon energy gap at 
zero field. $\mu_0 H^* = 21$~T is the condensation field for two-triplon 
bound states (red dashed line). Green, red and magenta shading represent the 
field-induced increase of spectral weight in novel multi-triplon excitations 
of the types represented in Figs.~\ref{fig:Fig4}f-h and panel {\bf a}.}
\label{fig:Fig5}
\end{figure}

The low-energy spectral weight in $S^{zz} (\omega, H)$ (Figs.~\ref{fig:Fig4}a 
and \ref{fig:Fig4}c) is another strong statement for BEC of bound states: the 
operator $S^z$ can excite either a singlet to a $t_0$ triplet, at relatively 
high energy, or a transition inside the band of condensed excitations. As 
noted above, one-triplon condensation in \SCBO~would be accompanied by weight 
at and below the triplon bandwidth, 0.3~meV, which is too small to explain 
the MPS result. By contrast, the internal structure of the $S = 2$ bound 
state has four bands, of which the lowest can be considered as the 
``pinwheel'' represented in Fig.~\ref{fig:Fig4}e \cite{corboz2014}, and 
$S^z$ can induce ``internal excitations'' within this structure, a schematic 
example being shown in Fig.~\ref{fig:Fig4}f. Using perturbation theory 
\cite{totsuka2001} with the parameters of \SCBO, the energies of the lowest 
internal excitations should cover the range from 0 to approximately 1.8~meV, 
in agreement with the energy range and $|t_+ t_+ \rangle$ character we obtain 
(Figs.~\ref{fig:Fig4}a and \ref{fig:Fig4}c). Unlike the $t_+$ mode, the band 
centre of these excitations does not increase linearly with the field beyond 
$H^*$, and the position and increasing intensity of this additional 
scattering contribution suggest an origin for the low-energy intensity 
observed at 25.9~T (Figs.~\ref{fig:Fig1}e and \ref{fig:Fig3}a).

One of the most striking consequences of two-triplon condensation is the 
presence of low-lying composite three-triplon excitations. The linear 
increase of spectral weight away from the $t_0$ and $t_-$ branches is not 
specific to two-triplon condensation, as several composite two-triplon 
excitations would be observed if single triplons condense \cite{nayak2020}. A 
leading example is the $|t_+ t_0 \rangle$ state represented schematically in 
Fig.~\ref{fig:Fig4}g, but in a scenario of one-triplon ($t_+$) condensation 
this would appear in $S^{zz} (\omega, H)$, whereas our discovery of low-energy 
$|t_+ t_0 \rangle$ weight in $S^{xx} (\omega, H)$ (Figs.~\ref{fig:Fig4}b and 
\ref{fig:Fig4}d) underlines its origin in $S = 2$ bound-state condensation. 
Quite simply, this scenario makes the number of excitations accessible by INS 
much larger, and adding three-triplon composites results in the complex 
spectra we observe both in INS and MPS. Figure \ref{fig:Fig4}h shows a 
schematic representation of a $| t_+ t_+ t_- \rangle$ composite obtained by 
exciting a $t_-$ triplon in close proximity to a $|t_+ t_+ \rangle$ bound 
state, where in general we find that the low-$S$, low-$S_z$ members of the 
multiplet benefit from sizeable net binding energies. The bound-state 
population is reflected in the magnetization, and hence the intensity of 
the two- and three-triplon excitations is also expected to increase linearly 
with the applied field, as we observed in Fig.~\ref{fig:Fig3}d. 

\medskip
\noindent
{\large {\bf Discussion}}

To summarize, we have combined high-field INS experiments and cylinder MPS 
calculations to reach the unambiguous conclusion that, in the field range 
between $\mu_0 H^* = 21$~T and $\mu_0 H_c = 27$~T, the ground state of 
\SCBO~is not a condensate of single triplets, but a condensate of 
two-triplon bound states. The consequences of this result, summarized in 
Fig.~\ref{fig:Fig5}, are a distinctive one-triplon spectrum with altered 
gradients beyond a premature kink at $H^*$, a one-triplon gap that persists 
at all fields and a high field-induced spectral weight of composite two- 
and three-triplon excitations both internal and external to the two-triplon 
bound states. We have shown that the weak DM interactions in \SCBO~play no 
qualitative role in determining this physics, but note nevertheless that 
they do affect the appearance of our results by creating a broad crossover 
regime below $H^*$. Figure \ref{fig:Fig5}a compares the magnetization 
computed from the MPS ground state with different experimental measurements, 
all of which indicate that the crossover from near-zero to linearly 
increasing magnetization spans the range from 16 to 21~T, consistent 
with Figs.~\ref{fig:Fig2}h-k and \ref{fig:Fig3}d.

Placing this result in perspective, the bound-state condensate is an example 
of a spin nematic \cite{blume1969,andreev1984}. This phase has a spontaneously 
broken rotational symmetry around the magnetic field, but its order parameter 
is a two-spin operator, and not simply a transverse magnetization. Spin-nematic 
phases have been predicted to occur just below the saturation field in systems 
where mixed ferro-antiferromagnetic interactions lead to magnon bound states 
\cite{chubukov1991,shannon2006,zhitomirsky2010}, and some evidence for this 
phenomenon has been obtained both by high-field NMR in LiCuVO$_4$ 
\cite{orlova2017} and by high-field calorimetry in 
Cu$_3$V$_2$O$_7$(OH)$_2$.2H$_2$O (volborthite) \cite{kohama2018}. 
What is new in \SCBO~is that the spin-nematic phase occurs below the first 
magnetization jump, i.e.~at low fields, with dilute triplons binding in a sea 
of singlet dimers. One primary hallmark of this phase is a gap to one-triplet 
excitations that persists at all fields and another is the presence of 
low-lying $S = 2$ excitations around $H^*$. While NMR or INS provide 
experimental access to the one-triplet excitations, the weak DM terms ensure 
that ESR can couple to the $S = 2$ excitations (at ${\bf Q} = {\bf 0}$). We 
note that high-field Raman spectroscopy, which is sensitive primarily to 
$S = 0$ states and $\Delta S = 0$ processes, has also provided a complementary 
view of the two-triplon condensation process through the behaviour of those 
excitations \cite{wulferding2021}. 

The spin-nematic phase has a close analogy with the physics of 
superconductivity. In a superconductor, Cooper pairs of fermions undergo 
a BEC, resulting in a state with a single-particle gap and collective 
excitations. In \SCBO~and the SSM, the single particle is a triplon, a 
bosonic particle with an infinite on-site repulsion (a ``hard-core boson''). 
The spin nematic realized in \SCBO~beyond 21~T can be seen as the BEC of 
Cooper pairs of bosons. This analogy establishes a direct connection 
between the superconducting gap of BCS theory and the one-triplon gap 
in \SCBO, which is equal to half the binding energy at the condensation 
field ($H^*$). Building further on the analogy, the BCS-BEC crossover 
\cite{randeria2014} links superconductors with large Cooper pairs that 
extend over many lattice sites to those with compact pairs, both developing 
coherence through the condensate. In \SCBO, the pairs are the strongly 
localized pinwheel objects of Fig.~\ref{fig:Fig4}e and the system 
represents the extreme BEC limit. Also in analogy with unconventional 
superconductivity, which generically competes with other instabilities 
of the interacting Fermi sea such as magnetic and charge order, the 
spin nematic in \SCBO~clearly competes with the bosonic CDW order that 
is favoured above $H_c$, where the 1/8 magnetization-plateau phase is 
established. 

Finally, our results represent new physics found by combining new 
capabilities in high-field neutron scattering with state-of-the-art 
numerical methods, and thus have direct implications in a number of 
disciplines. They demonstrate the profound importance to correlated 
condensed matter of scattering facilities able to operate at high 
magnetic fields, and on the technical side the HZB facility provided 
valuable information for future generations of high-field experiments. 
They illustrate directly the value of modern MPS methods in overcoming 
the long-standing barrier of computing dynamical properties in frustrated 
systems. For quantum matter, our results reinforce the message that novel 
composite states and novel forms of many-body order remain to be found at 
the confluence of strong frustration and controlled extreme conditions.

\medskip
\noindent
{\large {\bf Methods}}

\noindent
{\bf Neutron spectroscopy.} The inelastic neutron scattering (INS) experiment 
was performed at the High-Field Magnet (HFM) facility of the HZB. The Extreme 
Environment Diffractometer (EXED) was a time-of-flight neutron instrument 
that could be operated either in diffraction and small-angle-scattering modes 
or as a direct-geometry spectrometer. It was designed 
specifically to function optimally in combination with the HFM, which provided 
horizontal magnetic fields up to 25.9~T. This magnet was a hybrid solenoid 
with a $30^{\circ}$ conical opening and could be rotated in its entirety with 
respect to the incoming beam in order to extend the accessible region of 
reciprocal space.

The sample was a single-crystalline rod with a mass of 2.5~g, which was grown 
by a floating-zone method with its $(100)$ direction orientated approximately 
along the growth axis \cite{kageyama1999c,jorge2004}. The magnetic field was 
applied along the $(001)$ direction in the horizontal scattering plane, a 
geometry chosen to give access to part of the $(Q_h,Q_k,0)$ plane and to 
avoid the combination of the applied magnetic field with the DM interactions 
breaking any further symmetries of the system. The incident neutron energies 
were $E_i = 4$ and 8~meV and the rotation angle of the magnet relative to the 
incoming neturon beam was $\varphi = - 10^{\circ}$. A dilution refrigerator 
purpose-built by HZB in collaboration with the University of Birmingham 
allowed the measurements to be performed at 200~mK. Data were collected for 
periods between 7 and 20 hours for each selected field strength with $E_i = 
8$~meV and between 10 minutes and 6 hours with $E_i = 4$~meV. 

Supporting INS data were collected on the triple-axis spectrometers TASP, at 
the Paul Scherrer Institute (PSI), and Panda, at the Heinz Maier-Leibnitz 
Zentrum. On TASP, the same single crystal was orientated with $(Q_h,Q_k,0)$ 
in the horizontal scattering plane and a vertical magnetic field of 9.5~T 
was applied. The measurement was performed at 2.3~K, with a constant final 
neutron momentum $k_f = 1.5$~\AA$^{-1}$ and with a liquid-nitrogen-cooled Be 
filter to suppress the $\lambda/2$ contribution. On Panda \cite{zayed2014a},
 measurements were performed in the same geometry, at 1.5~K and with final 
neutron momentum $k_f = 1.55$~\AA$^{-1}$, for fields of 4, 8 and 12~T. 

\medskip
\noindent 
{\bf MPS Calculations.}
Matrix-Product States (MPS) are a variational Ansatz that can be used to 
represent the wavefunction of 1D quantum systems \cite{schollwoeck2011}. 
The representation is formulated in terms of rank-3 tensors and its accuracy 
controlled by the tensor bond dimension, $\chi$. A 2D system can be described 
by wrapping the lattice onto a cylinder and reformulating the model as a 1D 
Hamiltonian with further-neighbour interactions \cite{stoudenmire2012}. The 
ground state in the MPS representation is then obtained using the 
Density-Matrix Renormalization-Group (DMRG) algorithm \cite{schollwoeck2011}. 

The spin physics in \SCBO~is well described by the SSM Hamiltonian with 
Heisenberg interactions $J = 81.5$~K and $J' = 0.63 J$ \cite{shi2022}, 
supplemented by DM interactions $D = 0.034 J$ on the intra-dimer bonds 
\cite{kodama2005} and $D' = - 0.02J$ on the inter-dimer bonds \cite{cepas2001}. 
Finite DM interactions were also required to obtain uniform magnetization 
distributions in our calculations of the ground state in an applied field.
All of the spectral functions shown in the main text were computed with a 
cylinder of length $L = 20$ and circumference $W = 4$ sites, as shown in 
Sec.~S2A of the SI \cite{si}, where we benchmark the effect of $W$ by 
computing the energy per site, magnetization distribution and spectral 
function at two selected fields for a cylinder with $W = 6$. 

To calculate the dynamical properties of the system, we employed the 
time-dependent variational principle (TDVP) \cite{haegeman2016}. This method 
makes use of Lie-Suzuki Trotterization not of the Hamiltonian but of projectors 
to the tangent space of the MPS, which are constructed from the MPS with an 
accuracy dependent on $\chi$. We used the two-site variant of the TDVP 
algorithm, which allows the bond dimension of the MPS Ansatz to be readjusted 
as the entanglement grows upon time-evolution, with a time-step of $0.16/J$ and 
keeping a maximum $\chi$ of 600. The excitations of the SSM are localized as 
a consequence of the strong frustration and thus the time-dependence appears 
as a slowly expanding time-cone. This property allowed us to continue the 
real-time evolution to rather long times even on short cylinders. 

In the magnetized system, $\langle S^z \rangle$ is uniform while $\langle 
S^x \rangle$ and $\langle S^y \rangle$ are staggered, and hence the magnetic 
unit cell contains $n_s = 4$ sites. To compute the dynamical structure factor, 
we therefore started from each of these sites individually and took the 
Fourier transform 
\begin{eqnarray}
S^{\alpha \tilde{\alpha}} (\mathbf{k},\omega) = \frac{1}{n_s} \sum_{a,b} 
e^{-i\mathbf{k} \cdot \mathbf{\left( x_a - x_b \right)}} S^{\alpha 
\tilde{\alpha}}_{ab} (\mathbf{k},\omega), 
\nonumber
\end{eqnarray}
where $\mathbf{x}_a$ and $\mathbf{x}_b$ are the relative positions of sites 
$a$ and $b$ within the unit cell and 
\begin{eqnarray}
S^{\alpha \tilde{\alpha}}_{ab} (\mathbf{k},\omega) = \frac{1}{N} 
\sum_{\mathbf{R}} \! \int_{-\infty}^{\infty} \!\!\! e^{i(\omega t
 - \mathbf{k} \cdot \mathbf{R})} \langle S^{\alpha}_{a,\mathbf{R}}(t) 
S^{\tilde{\alpha}}_{b,\mathbf{0}}(0) \rangle dt, \nonumber
\end{eqnarray}
with $N$ the number of unit cells, $\mathbf{R}$ the spatial position of 
each cell and spin components $\alpha = \tilde{\alpha} \in \lbrace z, x 
\rbrace$. To obtain the results shown in Figs.~\ref{fig:Fig1}f and 
\ref{fig:Fig4}a-b, we integrated these functions over the full reciprocal 
space.

\medskip
\noindent
{\bf Data availability}
The data that support the findings of this study are available from the 
corresponding author upon reasonable request.

\smallskip
\noindent
{\bf Code availability}
The code that supports the findings of this study is available from the 
corresponding author upon reasonable request.


\begin{thebibliography}{54}%
\makeatletter
\providecommand \@ifxundefined [1]{%
 \@ifx{#1\undefined}
}%
\providecommand \@ifnum [1]{%
 \ifnum #1\expandafter \@firstoftwo
 \else \expandafter \@secondoftwo
 \fi
}%
\providecommand \@ifx [1]{%
 \ifx #1\expandafter \@firstoftwo
 \else \expandafter \@secondoftwo
 \fi
}%
\providecommand \natexlab [1]{#1}%
\providecommand \enquote  [1]{``#1''}%
\providecommand \bibnamefont  [1]{#1}%
\providecommand \bibfnamefont [1]{#1}%
\providecommand \citenamefont [1]{#1}%
\providecommand \href@noop [0]{\@secondoftwo}%
\providecommand \href [0]{\begingroup \@sanitize@url \@href}%
\providecommand \@href[1]{\@@startlink{#1}\@@href}%
\providecommand \@@href[1]{\endgroup#1\@@endlink}%
\providecommand \@sanitize@url [0]{\catcode `\\12\catcode `\$12\catcode
  `\&12\catcode `\#12\catcode `\^12\catcode `\_12\catcode `\%12\relax}%
\providecommand \@@startlink[1]{}%
\providecommand \@@endlink[0]{}%
\providecommand \url  [0]{\begingroup\@sanitize@url \@url }%
\providecommand \@url [1]{\endgroup\@href {#1}{\urlprefix }}%
\providecommand \urlprefix  [0]{URL }%
\providecommand \Eprint [0]{\href }%
\providecommand \doibase [0]{http://dx.doi.org/}%
\providecommand \selectlanguage [0]{\@gobble}%
\providecommand \bibinfo  [0]{\@secondoftwo}%
\providecommand \bibfield  [0]{\@secondoftwo}%
\providecommand \translation [1]{[#1]}%
\providecommand \BibitemOpen [0]{}%
\providecommand \bibitemStop [0]{}%
\providecommand \bibitemNoStop [0]{.\EOS\space}%
\providecommand \EOS [0]{\spacefactor3000\relax}%
\providecommand \BibitemShut  [1]{\csname bibitem#1\endcsname}%
\let\auto@bib@innerbib\@empty
\bibitem [{\citenamefont {Pitaevskii}\ and\ \citenamefont
  {Stringari}(2003)}]{pitaevskii2003}%
  \BibitemOpen
  \bibfield  {author} {\bibinfo {author} {\bibfnamefont {L.}~\bibnamefont
  {Pitaevskii}}\ and\ \bibinfo {author} {\bibfnamefont {S.}~\bibnamefont
  {Stringari}},\ }\href@noop {} {\emph {\bibinfo {title} {{Bose-Einstein
  Condensation}}}}\ (\bibinfo  {publisher} {Oxford University Press},\ \bibinfo
  {address} {New York},\ \bibinfo {year} {2003})\BibitemShut {NoStop}%
\bibitem [{\citenamefont {Giamarchi}\ \emph {et~al.}(2008)\citenamefont
  {Giamarchi}, \citenamefont {R\"uegg},\ and\ \citenamefont
  {Tchernyshyov}}]{giamarchi2008}%
  \BibitemOpen
  \bibfield  {author} {\bibinfo {author} {\bibfnamefont {T.}~\bibnamefont
  {Giamarchi}}, \bibinfo {author} {\bibfnamefont {Ch.}\ \bibnamefont
  {R\"uegg}}, \ and\ \bibinfo {author} {\bibfnamefont {O.}~\bibnamefont
  {Tchernyshyov}},\ }\bibfield  {title} {\enquote {\bibinfo {title}
  {{Bose-Einstein condensation in magnetic insulators}},}\ }\href
  {https://doi.org/10.1038/nphys893} {\bibfield  {journal} {\bibinfo  {journal}
  {Nature Phys.}\ }\textbf {\bibinfo {volume} {4}},\ \bibinfo {pages}
  {198--204} (\bibinfo {year} {2008})}\BibitemShut {NoStop}%
\bibitem [{\citenamefont {Zapf}\ \emph {et~al.}(2014)\citenamefont {Zapf},
  \citenamefont {Jaime},\ and\ \citenamefont {Batista}}]{zapf2014}%
  \BibitemOpen
  \bibfield  {author} {\bibinfo {author} {\bibfnamefont {V.}~\bibnamefont
  {Zapf}}, \bibinfo {author} {\bibfnamefont {M.}~\bibnamefont {Jaime}}, \ and\
  \bibinfo {author} {\bibfnamefont {C.~D.}\ \bibnamefont {Batista}},\
  }\bibfield  {title} {\enquote {\bibinfo {title} {{Bose-Einstein condensation
  in quantum magnets}},}\ }\href {https://doi.org/10.1103/RevModPhys.86.563}
  {\bibfield  {journal} {\bibinfo  {journal} {Rev. Mod. Phys.}\ }\textbf
  {\bibinfo {volume} {86}},\ \bibinfo {pages} {563} (\bibinfo {year}
  {2014})}\BibitemShut {NoStop}%
\bibitem [{\citenamefont {Nikuni}\ \emph {et~al.}(2000)\citenamefont {Nikuni},
  \citenamefont {Oshikawa}, \citenamefont {Oosawa},\ and\ \citenamefont
  {Tanaka}}]{nikuni2000}%
  \BibitemOpen
  \bibfield  {author} {\bibinfo {author} {\bibfnamefont {T.}~\bibnamefont
  {Nikuni}}, \bibinfo {author} {\bibfnamefont {M.}~\bibnamefont {Oshikawa}},
  \bibinfo {author} {\bibfnamefont {A.}~\bibnamefont {Oosawa}}, \ and\ \bibinfo
  {author} {\bibfnamefont {H.}~\bibnamefont {Tanaka}},\ }\bibfield  {title}
  {\enquote {\bibinfo {title} {{Bose-Einstein Condensation of Diluted Magnons
  in TlCuCl$_3$}},}\ }\href {https://doi.org/10.1103/PhysRevLett.84.5868}
  {\bibfield  {journal} {\bibinfo  {journal} {Phys. Rev. Lett.}\ }\textbf
  {\bibinfo {volume} {84}},\ \bibinfo {pages} {5868} (\bibinfo {year}
  {2000})}\BibitemShut {NoStop}%
\bibitem [{\citenamefont {Jaime}\ \emph {et~al.}(2004)\citenamefont {Jaime},
  \citenamefont {Correa}, \citenamefont {Harrison}, \citenamefont {Batista},
  \citenamefont {Kawashima}, \citenamefont {Kazuma}, \citenamefont {Jorge},
  \citenamefont {Stern}, \citenamefont {Heinmaa}, \citenamefont {Zvyagin},
  \citenamefont {Sasago},\ and\ \citenamefont {Uchinokura}}]{jaime2004}%
  \BibitemOpen
  \bibfield  {author} {\bibinfo {author} {\bibfnamefont {M.}~\bibnamefont
  {Jaime}}, \bibinfo {author} {\bibfnamefont {V.~F.}\ \bibnamefont {Correa}},
  \bibinfo {author} {\bibfnamefont {N.}~\bibnamefont {Harrison}}, \bibinfo
  {author} {\bibfnamefont {C.~D.}\ \bibnamefont {Batista}}, \bibinfo {author}
  {\bibfnamefont {N.}~\bibnamefont {Kawashima}}, \bibinfo {author}
  {\bibfnamefont {Y.}~\bibnamefont {Kazuma}}, \bibinfo {author} {\bibfnamefont
  {G.~A.}\ \bibnamefont {Jorge}}, \bibinfo {author} {\bibfnamefont
  {R.}~\bibnamefont {Stern}}, \bibinfo {author} {\bibfnamefont
  {I.}~\bibnamefont {Heinmaa}}, \bibinfo {author} {\bibfnamefont {S.~A.}\
  \bibnamefont {Zvyagin}}, \bibinfo {author} {\bibfnamefont {Y.}~\bibnamefont
  {Sasago}}, \ and\ \bibinfo {author} {\bibfnamefont {K.}~\bibnamefont
  {Uchinokura}},\ }\bibfield  {title} {\enquote {\bibinfo {title}
  {{Magnetic-Field-Induced Condensation of Triplons in Han Purple Pigment
  BaCuSi$_2$O$_6$}},}\ }\href {\doibase 10.1103/PhysRevLett.93.087203}
  {\bibfield  {journal} {\bibinfo  {journal} {Phys. Rev. Lett.}\ }\textbf
  {\bibinfo {volume} {93}},\ \bibinfo {pages} {087203} (\bibinfo {year}
  {2004})}\BibitemShut {NoStop}%
\bibitem [{\citenamefont {Allenspach}\ \emph {et~al.}(2022)\citenamefont
  {Allenspach}, \citenamefont {Madsen}, \citenamefont {Biffin}, \citenamefont
  {Bartkowiak}, \citenamefont {Prokhnenko}, \citenamefont {Gazizulina},
  \citenamefont {Liu}, \citenamefont {Wahle}, \citenamefont {Gerischer},
  \citenamefont {Kempfer}, \citenamefont {Heller}, \citenamefont {Smeibidl},
  \citenamefont {Mira}, \citenamefont {Laflorencie}, \citenamefont {Mila},
  \citenamefont {Normand},\ and\ \citenamefont {R\"uegg}}]{allenspach2022}%
  \BibitemOpen
  \bibfield  {author} {\bibinfo {author} {\bibfnamefont {S.}~\bibnamefont
  {Allenspach}}, \bibinfo {author} {\bibfnamefont {A.}~\bibnamefont {Madsen}},
  \bibinfo {author} {\bibfnamefont {A.}~\bibnamefont {Biffin}}, \bibinfo
  {author} {\bibfnamefont {M.}~\bibnamefont {Bartkowiak}}, \bibinfo {author}
  {\bibfnamefont {O.}~\bibnamefont {Prokhnenko}}, \bibinfo {author}
  {\bibfnamefont {A.}~\bibnamefont {Gazizulina}}, \bibinfo {author}
  {\bibfnamefont {X.}~\bibnamefont {Liu}}, \bibinfo {author} {\bibfnamefont
  {R.}~\bibnamefont {Wahle}}, \bibinfo {author} {\bibfnamefont
  {S.}~\bibnamefont {Gerischer}}, \bibinfo {author} {\bibfnamefont
  {S.}~\bibnamefont {Kempfer}}, \bibinfo {author} {\bibfnamefont
  {P.}~\bibnamefont {Heller}}, \bibinfo {author} {\bibfnamefont
  {P.}~\bibnamefont {Smeibidl}}, \bibinfo {author} {\bibfnamefont
  {A.}~\bibnamefont {Mira}}, \bibinfo {author} {\bibfnamefont {N.}~\bibnamefont
  {Laflorencie}}, \bibinfo {author} {\bibfnamefont {F.}~\bibnamefont {Mila}},
  \bibinfo {author} {\bibfnamefont {B.}~\bibnamefont {Normand}}, \ and\
  \bibinfo {author} {\bibfnamefont {Ch.}\ \bibnamefont {R\"uegg}},\ }\bibfield
  {title} {\enquote {\bibinfo {title} {{Investigating field-induced magnetic
  order in Han purple by neutron scattering up to 25.9 T}},}\ }\href {\doibase
  10.1103/PhysRevB.106.104418} {\bibfield  {journal} {\bibinfo  {journal}
  {Phys. Rev. B}\ }\textbf {\bibinfo {volume} {106}},\ \bibinfo {pages}
  {104418} (\bibinfo {year} {2022})}\BibitemShut {NoStop}%
\bibitem [{\citenamefont {Demokritov}\ \emph {et~al.}(2006)\citenamefont
  {Demokritov}, \citenamefont {Demidov}, \citenamefont {Dzyapko}, \citenamefont
  {Melkov}, \citenamefont {Serga}, \citenamefont {Hillebrands},\ and\
  \citenamefont {Slavin}}]{demokritov2006}%
  \BibitemOpen
  \bibfield  {author} {\bibinfo {author} {\bibfnamefont {S.~O.}\ \bibnamefont
  {Demokritov}}, \bibinfo {author} {\bibfnamefont {V.~E.}\ \bibnamefont
  {Demidov}}, \bibinfo {author} {\bibfnamefont {O.}~\bibnamefont {Dzyapko}},
  \bibinfo {author} {\bibfnamefont {G.~A.}\ \bibnamefont {Melkov}}, \bibinfo
  {author} {\bibfnamefont {A.~A.}\ \bibnamefont {Serga}}, \bibinfo {author}
  {\bibfnamefont {B.}~\bibnamefont {Hillebrands}}, \ and\ \bibinfo {author}
  {\bibfnamefont {A.~N.}\ \bibnamefont {Slavin}},\ }\bibfield  {title}
  {\enquote {\bibinfo {title} {{Bose-Einstein condensation of quasi-equilibrium
  magnons at room temperature under pumping}},}\ }\href
  {https://doi.org/10.1038/nature05117} {\bibfield  {journal} {\bibinfo
  {journal} {Nature}\ }\textbf {\bibinfo {volume} {443}},\ \bibinfo {pages}
  {430--433} (\bibinfo {year} {2006})}\BibitemShut {NoStop}%
\bibitem [{\citenamefont {Sun}\ \emph {et~al.}(2017)\citenamefont {Sun},
  \citenamefont {Nattermann},\ and\ \citenamefont {Pokrovsky}}]{sun2017}%
  \BibitemOpen
  \bibfield  {author} {\bibinfo {author} {\bibfnamefont {C.}~\bibnamefont
  {Sun}}, \bibinfo {author} {\bibfnamefont {T.}~\bibnamefont {Nattermann}}, \
  and\ \bibinfo {author} {\bibfnamefont {V.~L.}\ \bibnamefont {Pokrovsky}},\
  }\bibfield  {title} {\enquote {\bibinfo {title} {{Topical Review:
  Bose-Einstein condensation and superfluidity of magnons in yttrium iron
  garnet films}},}\ }\href {\doibase 10.1088/1361-6463/aa5cfc} {\bibfield
  {journal} {\bibinfo  {journal} {J. Phys. D: Appl. Phys.}\ }\textbf {\bibinfo
  {volume} {50}},\ \bibinfo {pages} {143002} (\bibinfo {year}
  {2017})}\BibitemShut {NoStop}%
\bibitem [{si()}]{si}%
  \BibitemOpen
  \href@noop {} {}\bibinfo {howpublished} {See the Supplementary Information at
  http://www.xxx.yyy, for a full exposition of the HFM/EXED experimental
  geometry, our data reduction and fitting, background subtraction, cylinder
  MPS calculations and the indicative ED calculations we performed on small
  clusters to identify the states found by MPS.}\BibitemShut {Stop}%
\bibitem [{\citenamefont {Nojiri}\ \emph {et~al.}(2003)\citenamefont {Nojiri},
  \citenamefont {Kageyama}, \citenamefont {Ueda},\ and\ \citenamefont
  {Motokawa}}]{nojiri2003}%
  \BibitemOpen
  \bibfield  {author} {\bibinfo {author} {\bibfnamefont {H.}~\bibnamefont
  {Nojiri}}, \bibinfo {author} {\bibfnamefont {H.}~\bibnamefont {Kageyama}},
  \bibinfo {author} {\bibfnamefont {Y.}~\bibnamefont {Ueda}}, \ and\ \bibinfo
  {author} {\bibfnamefont {M.}~\bibnamefont {Motokawa}},\ }\bibfield  {title}
  {\enquote {\bibinfo {title} {{ESR} study on the excited state energy spectrum
  of {SrCu$_2$(BO$_3$)$_2$} $-$ a central role of multiple-triplet bound
  states},}\ }\href {\doibase 10.1143/JPSJ.72.3243} {\bibfield  {journal}
  {\bibinfo  {journal} {J. Phys. Soc. Jpn.}\ }\textbf {\bibinfo {volume}
  {72}},\ \bibinfo {pages} {3243} (\bibinfo {year} {2003})}\BibitemShut
  {NoStop}%
\bibitem [{\citenamefont {Momoi}\ and\ \citenamefont
  {Totsuka}(2000)}]{momoi2000}%
  \BibitemOpen
  \bibfield  {author} {\bibinfo {author} {\bibfnamefont {T.}~\bibnamefont
  {Momoi}}\ and\ \bibinfo {author} {\bibfnamefont {K.}~\bibnamefont
  {Totsuka}},\ }\bibfield  {title} {\enquote {\bibinfo {title} {{Magnetization
  plateaus of the Shastry-Sutherland model for
  ${\mathrm{SrCu}}_{2}({\mathrm{BO}}_{3}{)}_{2}$: Spin-density wave,
  supersolid, and bound states}},}\ }\href {\doibase 10.1103/PhysRevB.62.15067}
  {\bibfield  {journal} {\bibinfo  {journal} {Phys. Rev. B}\ }\textbf {\bibinfo
  {volume} {62}},\ \bibinfo {pages} {15067--15078} (\bibinfo {year}
  {2000})}\BibitemShut {NoStop}%
\bibitem [{\citenamefont {Kageyama}\ \emph
  {et~al.}(1999{\natexlab{a}})\citenamefont {Kageyama}, \citenamefont
  {Yoshimura}, \citenamefont {Stern}, \citenamefont {Mushnikov}, \citenamefont
  {Onizuka}, \citenamefont {Kato}, \citenamefont {Kosuge}, \citenamefont
  {Slichter}, \citenamefont {Goto},\ and\ \citenamefont
  {Ueda}}]{kageyama1999b}%
  \BibitemOpen
  \bibfield  {author} {\bibinfo {author} {\bibfnamefont {H.}~\bibnamefont
  {Kageyama}}, \bibinfo {author} {\bibfnamefont {K.}~\bibnamefont {Yoshimura}},
  \bibinfo {author} {\bibfnamefont {R.}~\bibnamefont {Stern}}, \bibinfo
  {author} {\bibfnamefont {N.~V.}\ \bibnamefont {Mushnikov}}, \bibinfo {author}
  {\bibfnamefont {K.}~\bibnamefont {Onizuka}}, \bibinfo {author} {\bibfnamefont
  {M.}~\bibnamefont {Kato}}, \bibinfo {author} {\bibfnamefont {K.}~\bibnamefont
  {Kosuge}}, \bibinfo {author} {\bibfnamefont {C.~P.}\ \bibnamefont
  {Slichter}}, \bibinfo {author} {\bibfnamefont {T.}~\bibnamefont {Goto}}, \
  and\ \bibinfo {author} {\bibfnamefont {Y.}~\bibnamefont {Ueda}},\ }\bibfield
  {title} {\enquote {\bibinfo {title} {Exact dimer ground state and quantized
  magnetization plateaus in the two-dimensional spin system
  {S}r{C}u$_2$({BO}$_3$)$_2$},}\ }\href {\doibase 10.1103/PhysRevLett.82.3168}
  {\bibfield  {journal} {\bibinfo  {journal} {Phys. Rev. Lett.}\ }\textbf
  {\bibinfo {volume} {82}},\ \bibinfo {pages} {3168} (\bibinfo {year}
  {1999}{\natexlab{a}})}\BibitemShut {NoStop}%
\bibitem [{\citenamefont {Jaime}\ \emph {et~al.}(2012)\citenamefont {Jaime},
  \citenamefont {Daou}, \citenamefont {Crooker}, \citenamefont {Weickert},
  \citenamefont {Uchida}, \citenamefont {Feiguin}, \citenamefont {Batista},
  \citenamefont {Dabkowska},\ and\ \citenamefont {Gaulin}}]{jaime2012}%
  \BibitemOpen
  \bibfield  {author} {\bibinfo {author} {\bibfnamefont {Marcelo}\ \bibnamefont
  {Jaime}}, \bibinfo {author} {\bibfnamefont {Ramzy}\ \bibnamefont {Daou}},
  \bibinfo {author} {\bibfnamefont {Scott~A.}\ \bibnamefont {Crooker}},
  \bibinfo {author} {\bibfnamefont {Franziska}\ \bibnamefont {Weickert}},
  \bibinfo {author} {\bibfnamefont {Atsuko}\ \bibnamefont {Uchida}}, \bibinfo
  {author} {\bibfnamefont {Adrian~E.}\ \bibnamefont {Feiguin}}, \bibinfo
  {author} {\bibfnamefont {Cristian~D.}\ \bibnamefont {Batista}}, \bibinfo
  {author} {\bibfnamefont {Hanna~A.}\ \bibnamefont {Dabkowska}}, \ and\
  \bibinfo {author} {\bibfnamefont {Bruce~D.}\ \bibnamefont {Gaulin}},\
  }\bibfield  {title} {\enquote {\bibinfo {title} {{Magnetostriction and
  magnetic texture to 100.75 Tesla in frustrated SrCu$_2$(BO$_3$)$_2$}},}\
  }\href {https://doi.org/10.1073/pnas.1200743109} {\bibfield  {journal}
  {\bibinfo  {journal} {Proc. Natl. Acad. Sci.}\ }\textbf {\bibinfo {volume}
  {109}},\ \bibinfo {pages} {12404--12407} (\bibinfo {year}
  {2012})}\BibitemShut {NoStop}%
\bibitem [{\citenamefont {Takigawa}\ \emph {et~al.}(2013)\citenamefont
  {Takigawa}, \citenamefont {Horvati\'{c}}, \citenamefont {Waki}, \citenamefont
  {Kr\"amer}, \citenamefont {Berthier}, \citenamefont {L\'evy-Bertrand},
  \citenamefont {Sheikin}, \citenamefont {Kageyama}, \citenamefont {Ueda},\
  and\ \citenamefont {Mila}}]{takigawa2013}%
  \BibitemOpen
  \bibfield  {author} {\bibinfo {author} {\bibfnamefont {M.}~\bibnamefont
  {Takigawa}}, \bibinfo {author} {\bibfnamefont {M.}~\bibnamefont
  {Horvati\'{c}}}, \bibinfo {author} {\bibfnamefont {T.}~\bibnamefont {Waki}},
  \bibinfo {author} {\bibfnamefont {S.}~\bibnamefont {Kr\"amer}}, \bibinfo
  {author} {\bibfnamefont {C.}~\bibnamefont {Berthier}}, \bibinfo {author}
  {\bibfnamefont {F.}~\bibnamefont {L\'evy-Bertrand}}, \bibinfo {author}
  {\bibfnamefont {I.}~\bibnamefont {Sheikin}}, \bibinfo {author} {\bibfnamefont
  {H.}~\bibnamefont {Kageyama}}, \bibinfo {author} {\bibfnamefont
  {Y.}~\bibnamefont {Ueda}}, \ and\ \bibinfo {author} {\bibfnamefont
  {F.}~\bibnamefont {Mila}},\ }\bibfield  {title} {\enquote {\bibinfo {title}
  {{Incomplete Devil's Staircase in the Magnetization Curve of
  SrCu$_{2}$(BO$_{3}$)$_{2}$}},}\ }\href {\doibase
  10.1103/PhysRevLett.110.067210} {\bibfield  {journal} {\bibinfo  {journal}
  {Phys. Rev. Lett.}\ }\textbf {\bibinfo {volume} {110}},\ \bibinfo {pages}
  {067210} (\bibinfo {year} {2013})}\BibitemShut {NoStop}%
\bibitem [{\citenamefont {Matsuda}\ \emph {et~al.}(2013)\citenamefont
  {Matsuda}, \citenamefont {Abe}, \citenamefont {Takeyama}, \citenamefont
  {Kageyama}, \citenamefont {Corboz}, \citenamefont {Honecker}, \citenamefont
  {Manmana}, \citenamefont {Foltin}, \citenamefont {Schmidt},\ and\
  \citenamefont {Mila}}]{matsuda2013}%
  \BibitemOpen
  \bibfield  {author} {\bibinfo {author} {\bibfnamefont {Y.~H.}\ \bibnamefont
  {Matsuda}}, \bibinfo {author} {\bibfnamefont {N.}~\bibnamefont {Abe}},
  \bibinfo {author} {\bibfnamefont {S.}~\bibnamefont {Takeyama}}, \bibinfo
  {author} {\bibfnamefont {H.}~\bibnamefont {Kageyama}}, \bibinfo {author}
  {\bibfnamefont {P.}~\bibnamefont {Corboz}}, \bibinfo {author} {\bibfnamefont
  {A.}~\bibnamefont {Honecker}}, \bibinfo {author} {\bibfnamefont {S.~R.}\
  \bibnamefont {Manmana}}, \bibinfo {author} {\bibfnamefont {G.~R.}\
  \bibnamefont {Foltin}}, \bibinfo {author} {\bibfnamefont {K.~P.}\
  \bibnamefont {Schmidt}}, \ and\ \bibinfo {author} {\bibfnamefont
  {F.}~\bibnamefont {Mila}},\ }\bibfield  {title} {\enquote {\bibinfo {title}
  {{Magnetization of SrCu$_{2}$(BO$_{3}$)$_{2}$ in Ultrahigh Magnetic Fields up
  to 118 T}},}\ }\href {\doibase 10.1103/PhysRevLett.111.137204} {\bibfield
  {journal} {\bibinfo  {journal} {Phys. Rev. Lett.}\ }\textbf {\bibinfo
  {volume} {111}},\ \bibinfo {pages} {137204} (\bibinfo {year}
  {2013})}\BibitemShut {NoStop}%
\bibitem [{\citenamefont {Nomura}\ \emph {et~al.}(2022)\citenamefont {Nomura},
  \citenamefont {Corboz}, \citenamefont {Miyata}, \citenamefont {Zherlitsyn},
  \citenamefont {Ishii}, \citenamefont {Kohama}, \citenamefont {Matsuda},
  \citenamefont {Ikeda}, \citenamefont {Zhong}, \citenamefont {Kageyama},\ and\
  \citenamefont {Mila}}]{nomura2022}%
  \BibitemOpen
  \bibfield  {author} {\bibinfo {author} {\bibfnamefont {T.}~\bibnamefont
  {Nomura}}, \bibinfo {author} {\bibfnamefont {P.}~\bibnamefont {Corboz}},
  \bibinfo {author} {\bibfnamefont {A.}~\bibnamefont {Miyata}}, \bibinfo
  {author} {\bibfnamefont {S.}~\bibnamefont {Zherlitsyn}}, \bibinfo {author}
  {\bibfnamefont {Y.}~\bibnamefont {Ishii}}, \bibinfo {author} {\bibfnamefont
  {Y.}~\bibnamefont {Kohama}}, \bibinfo {author} {\bibfnamefont {Y.~H.}\
  \bibnamefont {Matsuda}}, \bibinfo {author} {\bibfnamefont {A.}~\bibnamefont
  {Ikeda}}, \bibinfo {author} {\bibfnamefont {C.}~\bibnamefont {Zhong}},
  \bibinfo {author} {\bibfnamefont {H.}~\bibnamefont {Kageyama}}, \ and\
  \bibinfo {author} {\bibfnamefont {F.}~\bibnamefont {Mila}},\ }\bibfield
  {title} {\enquote {\bibinfo {title} {{The Shastry-Sutherland Compound
  SrCu$_2$(BO$_3$)$_2$ Studied up to the Saturation Magnetic Field}},}\ }\href
  {https://doi.org/10.48550/arXiv.2209.07652} {\bibfield  {journal} {\bibinfo
  {journal} {arXiv:2209.07652}\ } (\bibinfo {year} {2022})}\BibitemShut
  {NoStop}%
\bibitem [{\citenamefont {Shastry}\ and\ \citenamefont
  {Sutherland}(1981)}]{shastry1981}%
  \BibitemOpen
  \bibfield  {author} {\bibinfo {author} {\bibfnamefont {B.~S.}\ \bibnamefont
  {Shastry}}\ and\ \bibinfo {author} {\bibfnamefont {B.}~\bibnamefont
  {Sutherland}},\ }\bibfield  {title} {\enquote {\bibinfo {title} {Exact ground
  state of a quantum mechanical antiferromagnet},}\ }\href {\doibase
  10.1016/0378-4363(81)90838-X} {\bibfield  {journal} {\bibinfo  {journal}
  {Physica B+C}\ }\textbf {\bibinfo {volume} {108}},\ \bibinfo {pages} {1069}
  (\bibinfo {year} {1981})}\BibitemShut {NoStop}%
\bibitem [{\citenamefont {Corboz}\ and\ \citenamefont
  {Mila}(2013)}]{corboz2013}%
  \BibitemOpen
  \bibfield  {author} {\bibinfo {author} {\bibfnamefont {P.}~\bibnamefont
  {Corboz}}\ and\ \bibinfo {author} {\bibfnamefont {F.}~\bibnamefont {Mila}},\
  }\bibfield  {title} {\enquote {\bibinfo {title} {Tensor network study of the
  {S}hastry-{S}utherland model in zero magnetic field},}\ }\href {\doibase
  10.1103/PhysRevB.87.115144} {\bibfield  {journal} {\bibinfo  {journal} {Phys.
  Rev. B}\ }\textbf {\bibinfo {volume} {87}},\ \bibinfo {pages} {115144}
  (\bibinfo {year} {2013})}\BibitemShut {NoStop}%
\bibitem [{\citenamefont {Kageyama}\ \emph {et~al.}(2000)\citenamefont
  {Kageyama}, \citenamefont {Nishi}, \citenamefont {Aso}, \citenamefont
  {Onizuka}, \citenamefont {Yosihama}, \citenamefont {Nukui}, \citenamefont
  {Kodama}, \citenamefont {Kakurai},\ and\ \citenamefont
  {Ueda}}]{kageyama2000a}%
  \BibitemOpen
  \bibfield  {author} {\bibinfo {author} {\bibfnamefont {H.}~\bibnamefont
  {Kageyama}}, \bibinfo {author} {\bibfnamefont {M.}~\bibnamefont {Nishi}},
  \bibinfo {author} {\bibfnamefont {N.}~\bibnamefont {Aso}}, \bibinfo {author}
  {\bibfnamefont {K.}~\bibnamefont {Onizuka}}, \bibinfo {author} {\bibfnamefont
  {T.}~\bibnamefont {Yosihama}}, \bibinfo {author} {\bibfnamefont
  {K.}~\bibnamefont {Nukui}}, \bibinfo {author} {\bibfnamefont
  {K.}~\bibnamefont {Kodama}}, \bibinfo {author} {\bibfnamefont
  {K.}~\bibnamefont {Kakurai}}, \ and\ \bibinfo {author} {\bibfnamefont
  {Y.}~\bibnamefont {Ueda}},\ }\bibfield  {title} {\enquote {\bibinfo {title}
  {Direct evidence for the localized single-triplet excitations and the
  dispersive multitriplet excitations in {S}r{C}u$_2$({BO}$_3$)$_2$},}\ }\href
  {\doibase 10.1103/PhysRevLett.84.5876} {\bibfield  {journal} {\bibinfo
  {journal} {Phys. Rev. Lett.}\ }\textbf {\bibinfo {volume} {84}},\ \bibinfo
  {pages} {5876} (\bibinfo {year} {2000})}\BibitemShut {NoStop}%
\bibitem [{\citenamefont {Gaulin}\ \emph {et~al.}(2004)\citenamefont {Gaulin},
  \citenamefont {Lee}, \citenamefont {Haravifard}, \citenamefont {Castellan},
  \citenamefont {Berlinsky}, \citenamefont {Dabkowska}, \citenamefont {Qiu},\
  and\ \citenamefont {Copley}}]{gaulin2004}%
  \BibitemOpen
  \bibfield  {author} {\bibinfo {author} {\bibfnamefont {B.~D.}\ \bibnamefont
  {Gaulin}}, \bibinfo {author} {\bibfnamefont {S.~H.}\ \bibnamefont {Lee}},
  \bibinfo {author} {\bibfnamefont {S.}~\bibnamefont {Haravifard}}, \bibinfo
  {author} {\bibfnamefont {J.~P.}\ \bibnamefont {Castellan}}, \bibinfo {author}
  {\bibfnamefont {A.~J.}\ \bibnamefont {Berlinsky}}, \bibinfo {author}
  {\bibfnamefont {H.~A.}\ \bibnamefont {Dabkowska}}, \bibinfo {author}
  {\bibfnamefont {Y.}~\bibnamefont {Qiu}}, \ and\ \bibinfo {author}
  {\bibfnamefont {J.~R.~D.}\ \bibnamefont {Copley}},\ }\bibfield  {title}
  {\enquote {\bibinfo {title} {{High-Resolution Study of Spin Excitations in
  the Singlet Ground State of SrCu$_2$(BO$_3$)$_2$}},}\ }\href {\doibase
  10.1103/PhysRevLett.93.267202} {\bibfield  {journal} {\bibinfo  {journal}
  {Phys. Rev. Lett.}\ }\textbf {\bibinfo {volume} {93}},\ \bibinfo {pages}
  {267202} (\bibinfo {year} {2004})}\BibitemShut {NoStop}%
\bibitem [{\citenamefont {Kakurai}\ \emph {et~al.}(2005)\citenamefont
  {Kakurai}, \citenamefont {Nukui}, \citenamefont {Aso}, \citenamefont {Nishi},
  \citenamefont {Kadowaki}, \citenamefont {Kageyama}, \citenamefont {Ueda},
  \citenamefont {Regnault},\ and\ \citenamefont {C\'epas}}]{kakurai2005}%
  \BibitemOpen
  \bibfield  {author} {\bibinfo {author} {\bibfnamefont {K.}~\bibnamefont
  {Kakurai}}, \bibinfo {author} {\bibfnamefont {K.}~\bibnamefont {Nukui}},
  \bibinfo {author} {\bibfnamefont {N.}~\bibnamefont {Aso}}, \bibinfo {author}
  {\bibfnamefont {M.}~\bibnamefont {Nishi}}, \bibinfo {author} {\bibfnamefont
  {H.}~\bibnamefont {Kadowaki}}, \bibinfo {author} {\bibfnamefont
  {H.}~\bibnamefont {Kageyama}}, \bibinfo {author} {\bibfnamefont
  {Y.}~\bibnamefont {Ueda}}, \bibinfo {author} {\bibfnamefont {L-P.}\
  \bibnamefont {Regnault}}, \ and\ \bibinfo {author} {\bibfnamefont
  {O.}~\bibnamefont {C\'epas}},\ }\bibfield  {title} {\enquote {\bibinfo
  {title} {{Neutron Scattering Investigation on Quantum Spin System
  SrCu$_2$(BO$_3$)$_2$}},}\ }\href {\doibase 10.1143/PTPS.159.22} {\bibfield
  {journal} {\bibinfo  {journal} {Prog. Theor. Phys. Suppl.}\ }\textbf
  {\bibinfo {volume} {159}},\ \bibinfo {pages} {22} (\bibinfo {year}
  {2005})}\BibitemShut {NoStop}%
\bibitem [{\citenamefont {Zayed}\ \emph {et~al.}(2014)\citenamefont {Zayed},
  \citenamefont {R\"uegg}, \citenamefont {Str\"assle}, \citenamefont {Stuhr},
  \citenamefont {Roessli}, \citenamefont {Ay}, \citenamefont {Mesot},
  \citenamefont {Link}, \citenamefont {Pomjakushina}, \citenamefont
  {Stingaciu}, \citenamefont {Conder},\ and\ \citenamefont
  {R\o{}nnow}}]{zayed2014a}%
  \BibitemOpen
  \bibfield  {author} {\bibinfo {author} {\bibfnamefont {M.~E.}\ \bibnamefont
  {Zayed}}, \bibinfo {author} {\bibfnamefont {Ch.}\ \bibnamefont {R\"uegg}},
  \bibinfo {author} {\bibfnamefont {Th.}\ \bibnamefont {Str\"assle}}, \bibinfo
  {author} {\bibfnamefont {U.}~\bibnamefont {Stuhr}}, \bibinfo {author}
  {\bibfnamefont {B.}~\bibnamefont {Roessli}}, \bibinfo {author} {\bibfnamefont
  {M.}~\bibnamefont {Ay}}, \bibinfo {author} {\bibfnamefont {J.}~\bibnamefont
  {Mesot}}, \bibinfo {author} {\bibfnamefont {P.}~\bibnamefont {Link}},
  \bibinfo {author} {\bibfnamefont {E.}~\bibnamefont {Pomjakushina}}, \bibinfo
  {author} {\bibfnamefont {M.}~\bibnamefont {Stingaciu}}, \bibinfo {author}
  {\bibfnamefont {K.}~\bibnamefont {Conder}}, \ and\ \bibinfo {author}
  {\bibfnamefont {H.~M.}\ \bibnamefont {R\o{}nnow}},\ }\bibfield  {title}
  {\enquote {\bibinfo {title} {Correlated decay of triplet excitations in the
  {Shastry-Sutherland} compound {SrCu$_2$(BO$_3$)$_2$}},}\ }\href {\doibase
  10.1103/PhysRevLett.113.067201} {\bibfield  {journal} {\bibinfo  {journal}
  {Phys. Rev. Lett.}\ }\textbf {\bibinfo {volume} {113}},\ \bibinfo {pages}
  {067201} (\bibinfo {year} {2014})}\BibitemShut {NoStop}%
\bibitem [{\citenamefont {Miyahara}\ and\ \citenamefont
  {Ueda}(2000)}]{miyahara2000}%
  \BibitemOpen
  \bibfield  {author} {\bibinfo {author} {\bibfnamefont {S.}~\bibnamefont
  {Miyahara}}\ and\ \bibinfo {author} {\bibfnamefont {K.}~\bibnamefont
  {Ueda}},\ }\bibfield  {title} {\enquote {\bibinfo {title} {Thermodynamic
  properties of three-dimensional orthogonal dimer model for
  {SrCu$_2$(BO$_3$)$_2$}},}\ }\href@noop {} {\bibfield  {journal} {\bibinfo
  {journal} {J. Phys. Soc. Jpn. (Suppl.) B}\ }\textbf {\bibinfo {volume}
  {69}},\ \bibinfo {pages} {72} (\bibinfo {year} {2000})}\BibitemShut {NoStop}%
\bibitem [{\citenamefont {Wietek}\ \emph {et~al.}(2019)\citenamefont {Wietek},
  \citenamefont {Corboz}, \citenamefont {Wessel}, \citenamefont {Normand},
  \citenamefont {Mila},\ and\ \citenamefont {Honecker}}]{wietek2019}%
  \BibitemOpen
  \bibfield  {author} {\bibinfo {author} {\bibfnamefont {A.}~\bibnamefont
  {Wietek}}, \bibinfo {author} {\bibfnamefont {P.}~\bibnamefont {Corboz}},
  \bibinfo {author} {\bibfnamefont {S.}~\bibnamefont {Wessel}}, \bibinfo
  {author} {\bibfnamefont {B.}~\bibnamefont {Normand}}, \bibinfo {author}
  {\bibfnamefont {F.}~\bibnamefont {Mila}}, \ and\ \bibinfo {author}
  {\bibfnamefont {A.}~\bibnamefont {Honecker}},\ }\bibfield  {title} {\enquote
  {\bibinfo {title} {Thermodynamic properties of the {S}hastry-{S}utherland
  model throughout the dimer-product phase},}\ }\href
  {https://doi.org/10.1103/PhysRevResearch.1.033038} {\bibfield  {journal}
  {\bibinfo  {journal} {Phys. Rev. Research}\ }\textbf {\bibinfo {volume}
  {1}},\ \bibinfo {pages} {033038} (\bibinfo {year} {2019})}\BibitemShut
  {NoStop}%
\bibitem [{\citenamefont {{Larrea Jim\'enez}}\ \emph
  {et~al.}(2021)\citenamefont {{Larrea Jim\'enez}}, \citenamefont {Crone},
  \citenamefont {Fogh}, \citenamefont {Zayed}, \citenamefont {Lortz},
  \citenamefont {Pomjakushina}, \citenamefont {Conder}, \citenamefont
  {L\"auchli}, \citenamefont {Weber}, \citenamefont {Wessel}, \citenamefont
  {Honecker}, \citenamefont {Normand}, \citenamefont {R\"uegg}, \citenamefont
  {Corboz}, \citenamefont {R{\o}nnow},\ and\ \citenamefont
  {Mila}}]{larrea2021}%
  \BibitemOpen
  \bibfield  {author} {\bibinfo {author} {\bibfnamefont {J.}~\bibnamefont
  {{Larrea Jim\'enez}}}, \bibinfo {author} {\bibfnamefont {S.~P.~G.}\
  \bibnamefont {Crone}}, \bibinfo {author} {\bibfnamefont {E.}~\bibnamefont
  {Fogh}}, \bibinfo {author} {\bibfnamefont {M.~E.}\ \bibnamefont {Zayed}},
  \bibinfo {author} {\bibfnamefont {R.}~\bibnamefont {Lortz}}, \bibinfo
  {author} {\bibfnamefont {E.}~\bibnamefont {Pomjakushina}}, \bibinfo {author}
  {\bibfnamefont {K.}~\bibnamefont {Conder}}, \bibinfo {author} {\bibfnamefont
  {A.~M.}\ \bibnamefont {L\"auchli}}, \bibinfo {author} {\bibfnamefont
  {L.}~\bibnamefont {Weber}}, \bibinfo {author} {\bibfnamefont
  {S.}~\bibnamefont {Wessel}}, \bibinfo {author} {\bibfnamefont
  {A.}~\bibnamefont {Honecker}}, \bibinfo {author} {\bibfnamefont
  {B.}~\bibnamefont {Normand}}, \bibinfo {author} {\bibfnamefont {Ch.}\
  \bibnamefont {R\"uegg}}, \bibinfo {author} {\bibfnamefont {P.}~\bibnamefont
  {Corboz}}, \bibinfo {author} {\bibfnamefont {H.~M.}\ \bibnamefont
  {R{\o}nnow}}, \ and\ \bibinfo {author} {\bibfnamefont {F.}~\bibnamefont
  {Mila}},\ }\bibfield  {title} {\enquote {\bibinfo {title} {A quantum magnetic
  analogue to the critical point of water},}\ }\href
  {https://doi.org/10.1038/s41586-021-03411-8} {\bibfield  {journal} {\bibinfo
  {journal} {Nature}\ }\textbf {\bibinfo {volume} {592}},\ \bibinfo {pages}
  {370--375} (\bibinfo {year} {2021})}\BibitemShut {NoStop}%
\bibitem [{\citenamefont {Zayed}\ \emph {et~al.}(2017)\citenamefont {Zayed},
  \citenamefont {R{\"u}egg}, \citenamefont {Larrea~J.}, \citenamefont
  {L{\"a}uchli}, \citenamefont {Panagopoulos}, \citenamefont {Saxena},
  \citenamefont {Ellerby}, \citenamefont {McMorrow}, \citenamefont
  {Str{\"a}ssle}, \citenamefont {Klotz}, \citenamefont {Hamel}, \citenamefont
  {Sadykov}, \citenamefont {Pomjakushin}, \citenamefont {Boehm}, \citenamefont
  {Jim{\'e}nez-Ruiz}, \citenamefont {Schneidewind}, \citenamefont
  {Pomjakushina}, \citenamefont {Stingaciu}, \citenamefont {Conder},\ and\
  \citenamefont {R{\o}nnow}}]{zayed2017}%
  \BibitemOpen
  \bibfield  {author} {\bibinfo {author} {\bibfnamefont {M.~E.}\ \bibnamefont
  {Zayed}}, \bibinfo {author} {\bibfnamefont {Ch.}\ \bibnamefont {R{\"u}egg}},
  \bibinfo {author} {\bibfnamefont {J.}~\bibnamefont {Larrea~J.}}, \bibinfo
  {author} {\bibfnamefont {A.~M.}\ \bibnamefont {L{\"a}uchli}}, \bibinfo
  {author} {\bibfnamefont {C.}~\bibnamefont {Panagopoulos}}, \bibinfo {author}
  {\bibfnamefont {S.~S.}\ \bibnamefont {Saxena}}, \bibinfo {author}
  {\bibfnamefont {M.}~\bibnamefont {Ellerby}}, \bibinfo {author} {\bibfnamefont
  {D.~F.}\ \bibnamefont {McMorrow}}, \bibinfo {author} {\bibfnamefont
  {Th}~\bibnamefont {Str{\"a}ssle}}, \bibinfo {author} {\bibfnamefont
  {S.}~\bibnamefont {Klotz}}, \bibinfo {author} {\bibfnamefont
  {G.}~\bibnamefont {Hamel}}, \bibinfo {author} {\bibfnamefont {R.~A.}\
  \bibnamefont {Sadykov}}, \bibinfo {author} {\bibfnamefont {V.}~\bibnamefont
  {Pomjakushin}}, \bibinfo {author} {\bibfnamefont {M.}~\bibnamefont {Boehm}},
  \bibinfo {author} {\bibfnamefont {M.}~\bibnamefont {Jim{\'e}nez-Ruiz}},
  \bibinfo {author} {\bibfnamefont {A.}~\bibnamefont {Schneidewind}}, \bibinfo
  {author} {\bibfnamefont {E.}~\bibnamefont {Pomjakushina}}, \bibinfo {author}
  {\bibfnamefont {M.}~\bibnamefont {Stingaciu}}, \bibinfo {author}
  {\bibfnamefont {K.}~\bibnamefont {Conder}}, \ and\ \bibinfo {author}
  {\bibfnamefont {H.~M.}\ \bibnamefont {R{\o}nnow}},\ }\bibfield  {title}
  {\enquote {\bibinfo {title} {{4-spin plaquette singlet state in the
  Shastry-Sutherland compound SrCu$_2$(BO$_3$)$_2$}},}\ }\href
  {https://doi.org/10.1038/nphys4190} {\bibfield  {journal} {\bibinfo
  {journal} {Nature Phys.}\ }\textbf {\bibinfo {volume} {13}},\ \bibinfo
  {pages} {962} (\bibinfo {year} {2017})}\BibitemShut {NoStop}%
\bibitem [{\citenamefont {Guo}\ \emph {et~al.}(2020)\citenamefont {Guo},
  \citenamefont {Sun}, \citenamefont {Zhao}, \citenamefont {Wang},
  \citenamefont {Hong}, \citenamefont {Sidorov}, \citenamefont {Ma},
  \citenamefont {Wu}, \citenamefont {Li}, \citenamefont {Meng}, \citenamefont
  {Sandvik},\ and\ \citenamefont {Sun}}]{guo2020}%
  \BibitemOpen
  \bibfield  {author} {\bibinfo {author} {\bibfnamefont {J.}~\bibnamefont
  {Guo}}, \bibinfo {author} {\bibfnamefont {G.}~\bibnamefont {Sun}}, \bibinfo
  {author} {\bibfnamefont {B.}~\bibnamefont {Zhao}}, \bibinfo {author}
  {\bibfnamefont {L.}~\bibnamefont {Wang}}, \bibinfo {author} {\bibfnamefont
  {W.}~\bibnamefont {Hong}}, \bibinfo {author} {\bibfnamefont {V.~A.}\
  \bibnamefont {Sidorov}}, \bibinfo {author} {\bibfnamefont {N.}~\bibnamefont
  {Ma}}, \bibinfo {author} {\bibfnamefont {Q.}~\bibnamefont {Wu}}, \bibinfo
  {author} {\bibfnamefont {S.}~\bibnamefont {Li}}, \bibinfo {author}
  {\bibfnamefont {Z.~Y.}\ \bibnamefont {Meng}}, \bibinfo {author}
  {\bibfnamefont {A.~W.}\ \bibnamefont {Sandvik}}, \ and\ \bibinfo {author}
  {\bibfnamefont {L.}~\bibnamefont {Sun}},\ }\bibfield  {title} {\enquote
  {\bibinfo {title} {{Quantum phases of SrCu$_2$(BO$_3$)$_2$ from high-pressure
  thermodynamics}},}\ }\href {\doibase 10.1103/PhysRevLett.124.206602}
  {\bibfield  {journal} {\bibinfo  {journal} {Phys. Rev. Lett.}\ }\textbf
  {\bibinfo {volume} {124}},\ \bibinfo {pages} {206602} (\bibinfo {year}
  {2020})}\BibitemShut {NoStop}%
\bibitem [{\citenamefont {Corboz}\ and\ \citenamefont
  {Mila}(2014)}]{corboz2014}%
  \BibitemOpen
  \bibfield  {author} {\bibinfo {author} {\bibfnamefont {P.}~\bibnamefont
  {Corboz}}\ and\ \bibinfo {author} {\bibfnamefont {F.}~\bibnamefont {Mila}},\
  }\bibfield  {title} {\enquote {\bibinfo {title} {Crystals of bound states in
  the magnetization plateaus of the {Shastry-Sutherland} model},}\ }\href
  {\doibase 10.1103/PhysRevLett.112.147203} {\bibfield  {journal} {\bibinfo
  {journal} {Phys. Rev. Lett.}\ }\textbf {\bibinfo {volume} {112}},\ \bibinfo
  {pages} {147203} (\bibinfo {year} {2014})}\BibitemShut {NoStop}%
\bibitem [{\citenamefont {C\'epas}\ \emph {et~al.}(2001)\citenamefont
  {C\'epas}, \citenamefont {Kakurai}, \citenamefont {Regnault}, \citenamefont
  {Ziman}, \citenamefont {Boucher}, \citenamefont {Aso}, \citenamefont {Nishi},
  \citenamefont {Kageyama},\ and\ \citenamefont {Ueda}}]{cepas2001}%
  \BibitemOpen
  \bibfield  {author} {\bibinfo {author} {\bibfnamefont {O.}~\bibnamefont
  {C\'epas}}, \bibinfo {author} {\bibfnamefont {K.}~\bibnamefont {Kakurai}},
  \bibinfo {author} {\bibfnamefont {L.~P.}\ \bibnamefont {Regnault}}, \bibinfo
  {author} {\bibfnamefont {T.}~\bibnamefont {Ziman}}, \bibinfo {author}
  {\bibfnamefont {J.~P.}\ \bibnamefont {Boucher}}, \bibinfo {author}
  {\bibfnamefont {N.}~\bibnamefont {Aso}}, \bibinfo {author} {\bibfnamefont
  {M.}~\bibnamefont {Nishi}}, \bibinfo {author} {\bibfnamefont
  {H.}~\bibnamefont {Kageyama}}, \ and\ \bibinfo {author} {\bibfnamefont
  {Y.}~\bibnamefont {Ueda}},\ }\bibfield  {title} {\enquote {\bibinfo {title}
  {{Dzyaloshinskii-Moriya Interaction in the 2D Spin Gap System
  SrCu$_2$(BO$_3$)$_2$}},}\ }\href {\doibase 10.1103/PhysRevLett.87.167205}
  {\bibfield  {journal} {\bibinfo  {journal} {Phys. Rev. Lett.}\ }\textbf
  {\bibinfo {volume} {87}},\ \bibinfo {pages} {167205} (\bibinfo {year}
  {2001})}\BibitemShut {NoStop}%
\bibitem [{\citenamefont {Kodama}\ \emph {et~al.}(2005)\citenamefont {Kodama},
  \citenamefont {Miyahara}, \citenamefont {Takigawa}, \citenamefont
  {Horvati\'c}, \citenamefont {Berthier}, \citenamefont {Mila}, \citenamefont
  {Kageyama},\ and\ \citenamefont {Ueda}}]{kodama2005}%
  \BibitemOpen
  \bibfield  {author} {\bibinfo {author} {\bibfnamefont {K.}~\bibnamefont
  {Kodama}}, \bibinfo {author} {\bibfnamefont {S.}~\bibnamefont {Miyahara}},
  \bibinfo {author} {\bibfnamefont {M.}~\bibnamefont {Takigawa}}, \bibinfo
  {author} {\bibfnamefont {M.}~\bibnamefont {Horvati\'c}}, \bibinfo {author}
  {\bibfnamefont {C.}~\bibnamefont {Berthier}}, \bibinfo {author}
  {\bibfnamefont {F.}~\bibnamefont {Mila}}, \bibinfo {author} {\bibfnamefont
  {H.}~\bibnamefont {Kageyama}}, \ and\ \bibinfo {author} {\bibfnamefont
  {Y.}~\bibnamefont {Ueda}},\ }\bibfield  {title} {\enquote {\bibinfo {title}
  {Field-induced effects of anisotropic magnetic interactions in
  {SrCu$_2$(BO$_3$)$_2$}},}\ }\href {\doibase 10.1088/0953-8984/17/4/L02}
  {\bibfield  {journal} {\bibinfo  {journal} {J. Phys. Condens. Matter}\
  }\textbf {\bibinfo {volume} {17}},\ \bibinfo {pages} {L61} (\bibinfo {year}
  {2005})}\BibitemShut {NoStop}%
\bibitem [{\citenamefont {Bendjama}\ \emph {et~al.}(2005)\citenamefont
  {Bendjama}, \citenamefont {Kumar},\ and\ \citenamefont
  {Mila}}]{bendjama2005}%
  \BibitemOpen
  \bibfield  {author} {\bibinfo {author} {\bibfnamefont {R.}~\bibnamefont
  {Bendjama}}, \bibinfo {author} {\bibfnamefont {B.}~\bibnamefont {Kumar}}, \
  and\ \bibinfo {author} {\bibfnamefont {F.}~\bibnamefont {Mila}},\ }\bibfield
  {title} {\enquote {\bibinfo {title} {{Absence of Single-Particle
  Bose-Einstein Condensation at Low Densitiesfor Bosons with Correlated
  Hopping}},}\ }\href {\doibase 10.1103/PhysRevLett.95.110406} {\bibfield
  {journal} {\bibinfo  {journal} {Phys. Rev. Lett.}\ }\textbf {\bibinfo
  {volume} {95}},\ \bibinfo {pages} {110406} (\bibinfo {year}
  {2005})}\BibitemShut {NoStop}%
\bibitem [{\citenamefont {Lemmens}\ \emph {et~al.}(2000)\citenamefont
  {Lemmens}, \citenamefont {Grove}, \citenamefont {Fischer}, \citenamefont
  {G\"untherodt}, \citenamefont {Kotov}, \citenamefont {Kageyama},
  \citenamefont {Onizuka},\ and\ \citenamefont {Ueda}}]{lemmens2000}%
  \BibitemOpen
  \bibfield  {author} {\bibinfo {author} {\bibfnamefont {P.}~\bibnamefont
  {Lemmens}}, \bibinfo {author} {\bibfnamefont {M.}~\bibnamefont {Grove}},
  \bibinfo {author} {\bibfnamefont {M.}~\bibnamefont {Fischer}}, \bibinfo
  {author} {\bibfnamefont {G.}~\bibnamefont {G\"untherodt}}, \bibinfo {author}
  {\bibfnamefont {V.N.}\ \bibnamefont {Kotov}}, \bibinfo {author}
  {\bibfnamefont {H.}~\bibnamefont {Kageyama}}, \bibinfo {author}
  {\bibfnamefont {K.}~\bibnamefont {Onizuka}}, \ and\ \bibinfo {author}
  {\bibfnamefont {Y.}~\bibnamefont {Ueda}},\ }\bibfield  {title} {\enquote
  {\bibinfo {title} {Collective singlet excitations and evolution of {R}aman
  spectral weights in the {2D} spin dimer compound
  {S}r{C}u$_2$({BO}$_3$)$_2$},}\ }\href {\doibase 10.1103/PhysRevLett.85.2605}
  {\bibfield  {journal} {\bibinfo  {journal} {Phys. Rev. Lett.}\ }\textbf
  {\bibinfo {volume} {85}},\ \bibinfo {pages} {2605} (\bibinfo {year}
  {2000})}\BibitemShut {NoStop}%
\bibitem [{\citenamefont {Blume}\ and\ \citenamefont
  {Hsieh}(1969)}]{blume1969}%
  \BibitemOpen
  \bibfield  {author} {\bibinfo {author} {\bibfnamefont {M.}~\bibnamefont
  {Blume}}\ and\ \bibinfo {author} {\bibfnamefont {Y.~Y.}\ \bibnamefont
  {Hsieh}},\ }\bibfield  {title} {\enquote {\bibinfo {title} {{Biquadratic
  Exchange and Quadrupolar Ordering}},}\ }\href {\doibase
  doi.org/10.1063/1.1657616} {\bibfield  {journal} {\bibinfo  {journal} {J.
  Appl. Phys.}\ }\textbf {\bibinfo {volume} {40}},\ \bibinfo {pages} {1249}
  (\bibinfo {year} {1969})}\BibitemShut {NoStop}%
\bibitem [{\citenamefont {Andreev}\ and\ \citenamefont
  {Grishchuk}(1984)}]{andreev1984}%
  \BibitemOpen
  \bibfield  {author} {\bibinfo {author} {\bibfnamefont {A.~F.}\ \bibnamefont
  {Andreev}}\ and\ \bibinfo {author} {\bibfnamefont {I.~A.}\ \bibnamefont
  {Grishchuk}},\ }\bibfield  {title} {\enquote {\bibinfo {title} {{Spin
  Nematics}},}\ }\href
  {http://www.jetp.ras.ru/cgi-bin/e/index/e/60/2/p267?a=list} {\bibfield
  {journal} {\bibinfo  {journal} {Zh. Eksp. Teor. Fiz.}\ }\textbf {\bibinfo
  {volume} {87}},\ \bibinfo {pages} {467} (\bibinfo {year} {1984})},\ \bibinfo
  {note} {[Sov. Phys. JETP {\bf 60}, 267 (1984)]}\BibitemShut {NoStop}%
\bibitem [{\citenamefont {Prokhnenko}\ \emph {et~al.}(2015)\citenamefont
  {Prokhnenko}, \citenamefont {Stein}, \citenamefont {Bleif}, \citenamefont
  {Fromme}, \citenamefont {Bartkowiak},\ and\ \citenamefont
  {Wilpert}}]{prokhnenko2015}%
  \BibitemOpen
  \bibfield  {author} {\bibinfo {author} {\bibfnamefont {O.}~\bibnamefont
  {Prokhnenko}}, \bibinfo {author} {\bibfnamefont {W.-D.}\ \bibnamefont
  {Stein}}, \bibinfo {author} {\bibfnamefont {H.-J.}\ \bibnamefont {Bleif}},
  \bibinfo {author} {\bibfnamefont {M.}~\bibnamefont {Fromme}}, \bibinfo
  {author} {\bibfnamefont {M.}~\bibnamefont {Bartkowiak}}, \ and\ \bibinfo
  {author} {\bibfnamefont {T.}~\bibnamefont {Wilpert}},\ }\bibfield  {title}
  {\enquote {\bibinfo {title} {{Time-of-flight Extreme Environment
  Diffractometer at the Helmholtz-Zentrum Berlin}},}\ }\href
  {https://doi.org/10.1063/1.4913656} {\bibfield  {journal} {\bibinfo
  {journal} {Rev. Sci. Instrum.}\ }\textbf {\bibinfo {volume} {86}},\ \bibinfo
  {pages} {033102} (\bibinfo {year} {2015})}\BibitemShut {NoStop}%
\bibitem [{\citenamefont {Smeibidl}\ \emph {et~al.}(2016)\citenamefont
  {Smeibidl}, \citenamefont {Bird}, \citenamefont {Ehmler}, \citenamefont
  {Dixon}, \citenamefont {Heinrich}, \citenamefont {Hoffmann}, \citenamefont
  {Kempfer}, \citenamefont {Bole}, \citenamefont {Toth}, \citenamefont
  {Prokhnenko},\ and\ \citenamefont {Lake}}]{smeibidl2016}%
  \BibitemOpen
  \bibfield  {author} {\bibinfo {author} {\bibfnamefont {P.}~\bibnamefont
  {Smeibidl}}, \bibinfo {author} {\bibfnamefont {M.}~\bibnamefont {Bird}},
  \bibinfo {author} {\bibfnamefont {H.}~\bibnamefont {Ehmler}}, \bibinfo
  {author} {\bibfnamefont {I.}~\bibnamefont {Dixon}}, \bibinfo {author}
  {\bibfnamefont {J.}~\bibnamefont {Heinrich}}, \bibinfo {author}
  {\bibfnamefont {M.}~\bibnamefont {Hoffmann}}, \bibinfo {author}
  {\bibfnamefont {S.}~\bibnamefont {Kempfer}}, \bibinfo {author} {\bibfnamefont
  {S.}~\bibnamefont {Bole}}, \bibinfo {author} {\bibfnamefont {J.}~\bibnamefont
  {Toth}}, \bibinfo {author} {\bibfnamefont {O.}~\bibnamefont {Prokhnenko}}, \
  and\ \bibinfo {author} {\bibfnamefont {B.}~\bibnamefont {Lake}},\ }\bibfield
  {title} {\enquote {\bibinfo {title} {{First Hybrid Magnet for Neutron
  Scattering at Helmholtz-Zentrum Berlin}},}\ }\href
  {http://dx.doi.org/10.1109/TASC.2016.2525773} {\bibfield  {journal} {\bibinfo
   {journal} {IEEE Trans. Appl. Supercond.}\ }\textbf {\bibinfo {volume}
  {26}},\ \bibinfo {pages} {4301606} (\bibinfo {year} {2016})}\BibitemShut
  {NoStop}%
\bibitem [{\citenamefont {Prokhnenko}\ \emph {et~al.}(2017)\citenamefont
  {Prokhnenko}, \citenamefont {Smeibidl}, \citenamefont {Stein}, \citenamefont
  {Bartkowiak},\ and\ \citenamefont {Stuesser}}]{prokhnenko2017}%
  \BibitemOpen
  \bibfield  {author} {\bibinfo {author} {\bibfnamefont {O.}~\bibnamefont
  {Prokhnenko}}, \bibinfo {author} {\bibfnamefont {P.}~\bibnamefont
  {Smeibidl}}, \bibinfo {author} {\bibfnamefont {W.~D.}\ \bibnamefont {Stein}},
  \bibinfo {author} {\bibfnamefont {M.}~\bibnamefont {Bartkowiak}}, \ and\
  \bibinfo {author} {\bibfnamefont {N.}~\bibnamefont {Stuesser}},\ }\bibfield
  {title} {\enquote {\bibinfo {title} {{HFM/EXED: The High Magnetic Field
  Facility for Neutron Scattering at BER II}},}\ }\href
  {https://doi.org/10.17815/jlsrf-3-111} {\bibfield  {journal} {\bibinfo
  {journal} {JLSRF}\ }\textbf {\bibinfo {volume} {3}},\ \bibinfo {pages} {A115}
  (\bibinfo {year} {2017})}\BibitemShut {NoStop}%
\bibitem [{\citenamefont {McClarty}\ \emph {et~al.}(2017)\citenamefont
  {McClarty}, \citenamefont {Kr\"uger}, \citenamefont {Guidi}, \citenamefont
  {Parker}, \citenamefont {Refson}, \citenamefont {Parker}, \citenamefont
  {Prabhakaran},\ and\ \citenamefont {Coldea}}]{mcclarty2017}%
  \BibitemOpen
  \bibfield  {author} {\bibinfo {author} {\bibfnamefont {P.~A.}\ \bibnamefont
  {McClarty}}, \bibinfo {author} {\bibfnamefont {F.}~\bibnamefont {Kr\"uger}},
  \bibinfo {author} {\bibfnamefont {T.}~\bibnamefont {Guidi}}, \bibinfo
  {author} {\bibfnamefont {S.~F.}\ \bibnamefont {Parker}}, \bibinfo {author}
  {\bibfnamefont {K.}~\bibnamefont {Refson}}, \bibinfo {author} {\bibfnamefont
  {A.~W.}\ \bibnamefont {Parker}}, \bibinfo {author} {\bibfnamefont
  {D.}~\bibnamefont {Prabhakaran}}, \ and\ \bibinfo {author} {\bibfnamefont
  {R.}~\bibnamefont {Coldea}},\ }\bibfield  {title} {\enquote {\bibinfo {title}
  {{Topological triplon modes and bound states in a Shastry-Sutherland
  magnet}},}\ }\href {\doibase 10.1038/nphys4117} {\bibfield  {journal}
  {\bibinfo  {journal} {Nature Phys.}\ }\textbf {\bibinfo {volume} {13}},\
  \bibinfo {pages} {736} (\bibinfo {year} {2017})}\BibitemShut {NoStop}%
\bibitem [{\citenamefont {Miyahara}\ \emph {et~al.}(2007)\citenamefont
  {Miyahara}, \citenamefont {Fouet}, \citenamefont {Manmana}, \citenamefont
  {Noack}, \citenamefont {Mayaffre}, \citenamefont {Sheikin}, \citenamefont
  {Berthier},\ and\ \citenamefont {Mila}}]{miyahara2007}%
  \BibitemOpen
  \bibfield  {author} {\bibinfo {author} {\bibfnamefont {S.}~\bibnamefont
  {Miyahara}}, \bibinfo {author} {\bibfnamefont {J.-B.}\ \bibnamefont {Fouet}},
  \bibinfo {author} {\bibfnamefont {S.~R.}\ \bibnamefont {Manmana}}, \bibinfo
  {author} {\bibfnamefont {R.~M.}\ \bibnamefont {Noack}}, \bibinfo {author}
  {\bibfnamefont {H.}~\bibnamefont {Mayaffre}}, \bibinfo {author}
  {\bibfnamefont {I.}~\bibnamefont {Sheikin}}, \bibinfo {author} {\bibfnamefont
  {C.}~\bibnamefont {Berthier}}, \ and\ \bibinfo {author} {\bibfnamefont
  {F.}~\bibnamefont {Mila}},\ }\bibfield  {title} {\enquote {\bibinfo {title}
  {{Uniform and staggered magnetizations induced by Dzyaloshinskii-Moriya
  interactions in isolated and coupled spin-1$/$2 dimers in a magnetic
  field}},}\ }\href {\doibase 10.1103/PhysRevB.75.184402} {\bibfield  {journal}
  {\bibinfo  {journal} {Phys. Rev. B}\ }\textbf {\bibinfo {volume} {75}},\
  \bibinfo {pages} {184402} (\bibinfo {year} {2007})}\BibitemShut {NoStop}%
\bibitem [{\citenamefont {Totsuka}\ \emph {et~al.}(2001)\citenamefont
  {Totsuka}, \citenamefont {Miyahara},\ and\ \citenamefont
  {Ueda}}]{totsuka2001}%
  \BibitemOpen
  \bibfield  {author} {\bibinfo {author} {\bibfnamefont {K.}~\bibnamefont
  {Totsuka}}, \bibinfo {author} {\bibfnamefont {S.}~\bibnamefont {Miyahara}}, \
  and\ \bibinfo {author} {\bibfnamefont {K.}~\bibnamefont {Ueda}},\ }\bibfield
  {title} {\enquote {\bibinfo {title} {Low-lying magnetic excitation of the
  {S}hastry-{S}utherland model},}\ }\href {\doibase 10.1103/PhysRevLett.86.520}
  {\bibfield  {journal} {\bibinfo  {journal} {Phys. Rev. Lett.}\ }\textbf
  {\bibinfo {volume} {86}},\ \bibinfo {pages} {520} (\bibinfo {year}
  {2001})}\BibitemShut {NoStop}%
\bibitem [{\citenamefont {Nayak}\ \emph {et~al.}(2020)\citenamefont {Nayak},
  \citenamefont {Blosser}, \citenamefont {Zheludev},\ and\ \citenamefont
  {Mila}}]{nayak2020}%
  \BibitemOpen
  \bibfield  {author} {\bibinfo {author} {\bibfnamefont {M.}~\bibnamefont
  {Nayak}}, \bibinfo {author} {\bibfnamefont {D.}~\bibnamefont {Blosser}},
  \bibinfo {author} {\bibfnamefont {A.}~\bibnamefont {Zheludev}}, \ and\
  \bibinfo {author} {\bibfnamefont {F.}~\bibnamefont {Mila}},\ }\bibfield
  {title} {\enquote {\bibinfo {title} {{Magnetic-Field-Induced Bound States in
  Spin-1/2 Ladders}},}\ }\href {\doibase 10.1103/PhysRevLett.124.087203}
  {\bibfield  {journal} {\bibinfo  {journal} {Phys. Rev. Lett.}\ }\textbf
  {\bibinfo {volume} {124}},\ \bibinfo {pages} {087203} (\bibinfo {year}
  {2020})}\BibitemShut {NoStop}%
\bibitem [{\citenamefont {Chubukov}(1991)}]{chubukov1991}%
  \BibitemOpen
  \bibfield  {author} {\bibinfo {author} {\bibfnamefont {A.~V.}\ \bibnamefont
  {Chubukov}},\ }\bibfield  {title} {\enquote {\bibinfo {title} {{Chiral,
  Nematic, and Dimer States in Quantum Spin Chains}},}\ }\href {\doibase
  10.1103/PhysRevB.44.4693} {\bibfield  {journal} {\bibinfo  {journal} {Phys.
  Rev. B}\ }\textbf {\bibinfo {volume} {44}},\ \bibinfo {pages} {4693}
  (\bibinfo {year} {1991})}\BibitemShut {NoStop}%
\bibitem [{\citenamefont {Shannon}\ \emph {et~al.}(2006)\citenamefont
  {Shannon}, \citenamefont {Momoi},\ and\ \citenamefont
  {Sindzingre}}]{shannon2006}%
  \BibitemOpen
  \bibfield  {author} {\bibinfo {author} {\bibfnamefont {N.}~\bibnamefont
  {Shannon}}, \bibinfo {author} {\bibfnamefont {T.}~\bibnamefont {Momoi}}, \
  and\ \bibinfo {author} {\bibfnamefont {P.}~\bibnamefont {Sindzingre}},\
  }\bibfield  {title} {\enquote {\bibinfo {title} {{Nematic Order in Square
  Lattice Frustrated Ferromagnets}},}\ }\href {\doibase
  10.1103/PhysRevLett.96.027213} {\bibfield  {journal} {\bibinfo  {journal}
  {Phys. Rev. Lett.}\ }\textbf {\bibinfo {volume} {96}},\ \bibinfo {pages}
  {027213} (\bibinfo {year} {2006})}\BibitemShut {NoStop}%
\bibitem [{\citenamefont {Zhitomirsky}\ and\ \citenamefont
  {Tsunetsugu}(2010)}]{zhitomirsky2010}%
  \BibitemOpen
  \bibfield  {author} {\bibinfo {author} {\bibfnamefont {M.~E.}\ \bibnamefont
  {Zhitomirsky}}\ and\ \bibinfo {author} {\bibfnamefont {H.}~\bibnamefont
  {Tsunetsugu}},\ }\bibfield  {title} {\enquote {\bibinfo {title} {{Magnon
  Pairing in Quantum Spin Nematic}},}\ }\href {\doibase
  10.1209/0295-5075/92/37001} {\bibfield  {journal} {\bibinfo  {journal}
  {Europhys. Lett.}\ }\textbf {\bibinfo {volume} {92}},\ \bibinfo {pages}
  {37001} (\bibinfo {year} {2010})}\BibitemShut {NoStop}%
\bibitem [{\citenamefont {Orlova}\ \emph {et~al.}(2017)\citenamefont {Orlova},
  \citenamefont {Green}, \citenamefont {Law}, \citenamefont {Gorbunov},
  \citenamefont {Chanda}, \citenamefont {Kr\"amer}, \citenamefont {Horvati\'c},
  \citenamefont {Kremer}, \citenamefont {Wosnitza},\ and\ \citenamefont
  {Rikken}}]{orlova2017}%
  \BibitemOpen
  \bibfield  {author} {\bibinfo {author} {\bibfnamefont {A.}~\bibnamefont
  {Orlova}}, \bibinfo {author} {\bibfnamefont {E.~L.}\ \bibnamefont {Green}},
  \bibinfo {author} {\bibfnamefont {J.~M.}\ \bibnamefont {Law}}, \bibinfo
  {author} {\bibfnamefont {D.~I.}\ \bibnamefont {Gorbunov}}, \bibinfo {author}
  {\bibfnamefont {G.}~\bibnamefont {Chanda}}, \bibinfo {author} {\bibfnamefont
  {S.}~\bibnamefont {Kr\"amer}}, \bibinfo {author} {\bibfnamefont
  {M.}~\bibnamefont {Horvati\'c}}, \bibinfo {author} {\bibfnamefont {R.~K.}\
  \bibnamefont {Kremer}}, \bibinfo {author} {\bibfnamefont {J.}~\bibnamefont
  {Wosnitza}}, \ and\ \bibinfo {author} {\bibfnamefont {G.~L.~J.~A.}\
  \bibnamefont {Rikken}},\ }\bibfield  {title} {\enquote {\bibinfo {title}
  {{Nuclear Magnetic Resonance Signature of the Spin-Nematic Phase in
  LiCuVO$_4$ at High Magnetic Fields}},}\ }\href
  {https://doi.org/10.1103/PhysRevLett.118.247201} {\bibfield  {journal}
  {\bibinfo  {journal} {Phys. Rev. Lett.}\ }\textbf {\bibinfo {volume} {118}},\
  \bibinfo {pages} {247201} (\bibinfo {year} {2017})}\BibitemShut {NoStop}%
\bibitem [{\citenamefont {Kohama}\ \emph {et~al.}(2018)\citenamefont {Kohama},
  \citenamefont {Ishikawa}, \citenamefont {Matsuo}, \citenamefont {Kindo},
  \citenamefont {Shannon},\ and\ \citenamefont {Hiroi}}]{kohama2018}%
  \BibitemOpen
  \bibfield  {author} {\bibinfo {author} {\bibfnamefont {Yoshimitsu}\
  \bibnamefont {Kohama}}, \bibinfo {author} {\bibfnamefont {Hajime}\
  \bibnamefont {Ishikawa}}, \bibinfo {author} {\bibfnamefont {Akira}\
  \bibnamefont {Matsuo}}, \bibinfo {author} {\bibfnamefont {Koichi}\
  \bibnamefont {Kindo}}, \bibinfo {author} {\bibfnamefont {Nic}\ \bibnamefont
  {Shannon}}, \ and\ \bibinfo {author} {\bibfnamefont {Zenji}\ \bibnamefont
  {Hiroi}},\ }\bibfield  {title} {\enquote {\bibinfo {title} {Possible
  observation of quantum spin-nematic phase in a frustrated magnet},}\ }\href
  {https://doi.org/10.1073/pnas.1821969116} {\bibfield  {journal} {\bibinfo
  {journal} {Proc. Natl. Acad. Sci.}\ }\textbf {\bibinfo {volume} {116}},\
  \bibinfo {pages} {10686--10690} (\bibinfo {year} {2018})}\BibitemShut
  {NoStop}%
\bibitem [{\citenamefont {Wulferding}\ \emph {et~al.}(2021)\citenamefont
  {Wulferding}, \citenamefont {Choi}, \citenamefont {Lee}, \citenamefont
  {Prosnikov}, \citenamefont {Gallais}, \citenamefont {Lemmens}, \citenamefont
  {Zhong}, \citenamefont {Kageyama},\ and\ \citenamefont
  {Choi}}]{wulferding2021}%
  \BibitemOpen
  \bibfield  {author} {\bibinfo {author} {\bibfnamefont {D.}~\bibnamefont
  {Wulferding}}, \bibinfo {author} {\bibfnamefont {Y.}~\bibnamefont {Choi}},
  \bibinfo {author} {\bibfnamefont {S.}~\bibnamefont {Lee}}, \bibinfo {author}
  {\bibfnamefont {M.~A.}\ \bibnamefont {Prosnikov}}, \bibinfo {author}
  {\bibfnamefont {Y.}~\bibnamefont {Gallais}}, \bibinfo {author} {\bibfnamefont
  {P.}~\bibnamefont {Lemmens}}, \bibinfo {author} {\bibfnamefont
  {C.}~\bibnamefont {Zhong}}, \bibinfo {author} {\bibfnamefont
  {H.}~\bibnamefont {Kageyama}}, \ and\ \bibinfo {author} {\bibfnamefont
  {K.-Y.}\ \bibnamefont {Choi}},\ }\bibfield  {title} {\enquote {\bibinfo
  {title} {{Thermally populated versus field-induced triplon bound states in
  the Shastry-Sutherland lattice SrCu$_2$(BO$_3$)$_2$}},}\ }\href {\doibase
  10.1038/s41535-021-00405-7} {\bibfield  {journal} {\bibinfo  {journal} {npj
  Quantum Materials}\ }\textbf {\bibinfo {volume} {6}},\ \bibinfo {pages} {102}
  (\bibinfo {year} {2021})}\BibitemShut {NoStop}%
\bibitem [{\citenamefont {Randeria}\ and\ \citenamefont
  {Taylor}(2014)}]{randeria2014}%
  \BibitemOpen
  \bibfield  {author} {\bibinfo {author} {\bibfnamefont {Mohit}\ \bibnamefont
  {Randeria}}\ and\ \bibinfo {author} {\bibfnamefont {Edward}\ \bibnamefont
  {Taylor}},\ }\bibfield  {title} {\enquote {\bibinfo {title} {{Crossover from
  Bardeen-Cooper-Schrieffer to Bose-Einstein Condensation and the Unitary Fermi
  Gas}},}\ }\href {\doibase 10.1146/annurev-conmatphys-031113-133829}
  {\bibfield  {journal} {\bibinfo  {journal} {Annu. Rev. Condens. Matter
  Phys.}\ }\textbf {\bibinfo {volume} {5}},\ \bibinfo {pages} {209--232}
  (\bibinfo {year} {2014})}\BibitemShut {NoStop}%
\bibitem [{\citenamefont {Kageyama}\ \emph
  {et~al.}(1999{\natexlab{b}})\citenamefont {Kageyama}, \citenamefont
  {Onizuka}, \citenamefont {Yamauchi},\ and\ \citenamefont
  {Ueda}}]{kageyama1999c}%
  \BibitemOpen
  \bibfield  {author} {\bibinfo {author} {\bibfnamefont {H.}~\bibnamefont
  {Kageyama}}, \bibinfo {author} {\bibfnamefont {K.}~\bibnamefont {Onizuka}},
  \bibinfo {author} {\bibfnamefont {T.}~\bibnamefont {Yamauchi}}, \ and\
  \bibinfo {author} {\bibfnamefont {Y.}~\bibnamefont {Ueda}},\ }\bibfield
  {title} {\enquote {\bibinfo {title} {Crystal growth of the two-dimensional
  spin gap system {SrCu$_2$(BO$_3$)$_2$}},}\ }\href {\doibase
  10.1016/S0022-0248(99)00313-9} {\bibfield  {journal} {\bibinfo  {journal}
  {Journal of Crystal Growth}\ }\textbf {\bibinfo {volume} {206}},\ \bibinfo
  {pages} {65--67} (\bibinfo {year} {1999}{\natexlab{b}})}\BibitemShut
  {NoStop}%
\bibitem [{\citenamefont {Jorge}\ \emph {et~al.}(2004)\citenamefont {Jorge},
  \citenamefont {Jaime}, \citenamefont {Harrison}, \citenamefont {Stern},
  \citenamefont {Dabkowska},\ and\ \citenamefont {Gaulin}}]{jorge2004}%
  \BibitemOpen
  \bibfield  {author} {\bibinfo {author} {\bibfnamefont {G.}~\bibnamefont
  {Jorge}}, \bibinfo {author} {\bibfnamefont {M.}~\bibnamefont {Jaime}},
  \bibinfo {author} {\bibfnamefont {N.}~\bibnamefont {Harrison}}, \bibinfo
  {author} {\bibfnamefont {R.}~\bibnamefont {Stern}}, \bibinfo {author}
  {\bibfnamefont {H.}~\bibnamefont {Dabkowska}}, \ and\ \bibinfo {author}
  {\bibfnamefont {B.~D.}\ \bibnamefont {Gaulin}},\ }\bibfield  {title}
  {\enquote {\bibinfo {title} {{High magnetic field magnetization and specific
  heat of the 2D spin–dimer system SrCu$_2$(BO$_3$)$_2$}},}\ }\href {\doibase
  10.1016/S0022-0248(99)00313-9} {\bibfield  {journal} {\bibinfo  {journal}
  {Journal of Alloys and Compounds}\ }\textbf {\bibinfo {volume} {369}},\
  \bibinfo {pages} {90--92} (\bibinfo {year} {2004})}\BibitemShut {NoStop}%
\bibitem [{\citenamefont {Schollw{\"o}ck}(2011)}]{schollwoeck2011}%
  \BibitemOpen
  \bibfield  {author} {\bibinfo {author} {\bibfnamefont {U.}~\bibnamefont
  {Schollw{\"o}ck}},\ }\bibfield  {title} {\enquote {\bibinfo {title} {The
  density-matrix renormalization group in the age of matrix product states},}\
  }\href {\doibase https://doi.org/10.1016/j.aop.2010.09.012} {\bibfield
  {journal} {\bibinfo  {journal} {Ann. Phys.}\ }\textbf {\bibinfo {volume}
  {326}},\ \bibinfo {pages} {96 -- 192} (\bibinfo {year} {2011})}\BibitemShut
  {NoStop}%
\bibitem [{\citenamefont {Stoudenmire}\ and\ \citenamefont
  {White}(2012)}]{stoudenmire2012}%
  \BibitemOpen
  \bibfield  {author} {\bibinfo {author} {\bibfnamefont {E.~M.}\ \bibnamefont
  {Stoudenmire}}\ and\ \bibinfo {author} {\bibfnamefont {Steven~R.}\
  \bibnamefont {White}},\ }\bibfield  {title} {\enquote {\bibinfo {title}
  {{Studying Two-Dimensional Systems with the Density Matrix Renormalization
  Group}},}\ }\href {\doibase 10.1146/annurev-conmatphys-020911-125018}
  {\bibfield  {journal} {\bibinfo  {journal} {Annu. Rev. Condens. Matter
  Phys.}\ }\textbf {\bibinfo {volume} {3}},\ \bibinfo {pages} {111--128}
  (\bibinfo {year} {2012})}\BibitemShut {NoStop}%
\bibitem [{\citenamefont {Shi}\ \emph {et~al.}(2022)\citenamefont {Shi},
  \citenamefont {Dissanayake}, \citenamefont {Corboz}, \citenamefont
  {Steinhardt}, \citenamefont {Graf}, \citenamefont {Silevitch}, \citenamefont
  {Dabkowska}, \citenamefont {Rosenbaum}, \citenamefont {Mila},\ and\
  \citenamefont {Haravifard}}]{shi2022}%
  \BibitemOpen
  \bibfield  {author} {\bibinfo {author} {\bibfnamefont {Z.}~\bibnamefont
  {Shi}}, \bibinfo {author} {\bibfnamefont {S.}~\bibnamefont {Dissanayake}},
  \bibinfo {author} {\bibfnamefont {P.}~\bibnamefont {Corboz}}, \bibinfo
  {author} {\bibfnamefont {W.}~\bibnamefont {Steinhardt}}, \bibinfo {author}
  {\bibfnamefont {D.}~\bibnamefont {Graf}}, \bibinfo {author} {\bibfnamefont
  {D.~M.}\ \bibnamefont {Silevitch}}, \bibinfo {author} {\bibfnamefont {H.~A.}\
  \bibnamefont {Dabkowska}}, \bibinfo {author} {\bibfnamefont {T.~F.}\
  \bibnamefont {Rosenbaum}}, \bibinfo {author} {\bibfnamefont {F.}~\bibnamefont
  {Mila}}, \ and\ \bibinfo {author} {\bibfnamefont {S.}~\bibnamefont
  {Haravifard}},\ }\bibfield  {title} {\enquote {\bibinfo {title} {{Discovery
  of quantum phases in the Shastry-Sutherland compound SrCu$_2$(BO$_3$)$_2$
  under extreme conditions of field and pressure}},}\ }\href {\doibase
  10.1038/s41467-022-30036-w} {\bibfield  {journal} {\bibinfo  {journal}
  {Nature Commun.}\ }\textbf {\bibinfo {volume} {13}},\ \bibinfo {pages} {2301}
  (\bibinfo {year} {2022})}\BibitemShut {NoStop}%
\bibitem [{\citenamefont {Haegeman}\ \emph {et~al.}(2016)\citenamefont
  {Haegeman}, \citenamefont {Lubich}, \citenamefont {Oseledets}, \citenamefont
  {Vandereycken},\ and\ \citenamefont {Verstraete}}]{haegeman2016}%
  \BibitemOpen
  \bibfield  {author} {\bibinfo {author} {\bibfnamefont {J.}~\bibnamefont
  {Haegeman}}, \bibinfo {author} {\bibfnamefont {Ch.}\ \bibnamefont {Lubich}},
  \bibinfo {author} {\bibfnamefont {I.}~\bibnamefont {Oseledets}}, \bibinfo
  {author} {\bibfnamefont {B.}~\bibnamefont {Vandereycken}}, \ and\ \bibinfo
  {author} {\bibfnamefont {F.}~\bibnamefont {Verstraete}},\ }\bibfield  {title}
  {\enquote {\bibinfo {title} {Unifying time evolution and optimization with
  matrix product states},}\ }\href {\doibase 10.1103/PhysRevB.94.165116}
  {\bibfield  {journal} {\bibinfo  {journal} {Phys. Rev. B}\ }\textbf {\bibinfo
  {volume} {94}},\ \bibinfo {pages} {165116} (\bibinfo {year}
  {2016})}\BibitemShut {NoStop}%
\end{thebibliography}

%

\smallskip
\noindent
{\bf Acknowledgements}

\noindent
We are especially grateful to S. Gerischer, P. Heller and R. Wahle for their 
assistance with cryogenics and to S. Kempfer and P. Smeibidl for ensuring the 
operation of the high-field magnet. We thank O. Gauth\'e, L. Herviou and S. 
Miyahara for helpful discussions. We acknowledge the financial support of the 
European Research Council through the Synergy network HERO (Grant No.~810451). 
We are grateful to the Helmholtz-Zentrum Berlin for the allocation of neutron 
beam time on EXED, to the Paul Scherrer Institute for beam time on TASP and 
to the Heinz Meier-Leibnitz Zentrum for beam time on Panda. Our numerical 
calculations were performed on the facilities of the Scientific IT and 
Application Support Center of the EPFL. This publication was made possible 
by the generous support of the Qatar Foundation through the Seed Research 
Funding Program of Carnegie Mellon University in Qatar. The statements made 
herein are solely the responsibility of the authors.

\smallskip
\noindent
{\bf Author contributions}

\noindent
The experimental project was conceived by E.F., H.M.R. and K.K., with input 
from H.N. and H.K. The theoretical framework was provided by F.M. Single 
crystals were grown by E.P. INS measurements on HFM/EXED were performed by 
E.F., K.M., O.P. and M.B., on TASP by E.F., J.-R.S. and A.A.T. and on Panda 
by M.E.Z. Data analysis was performed by E.F., M.E.Z. and H.M.R. MPS and ED 
calculations were performed by M.N. and their theoretical interpretation 
was provided by M.N., B.N. and F.M. The manuscript was written by E.F., 
H.M.R., M.N., B.N. and F.M. with contributions from all the authors.

\smallskip
\noindent
{\bf Competing financial interests} The authors declare no competing 
financial interests.

\smallskip
\noindent
{\bf Additional information}

\noindent
{\bf Supplementary information} The online version contains supplementary 
material available at https://doi.org/xxx.yyy.zzz.

\noindent
{\bf Correspondence and requests for materials} should be addressed to E.F.

\end{document}


\renewcommand*{\citenumfont}[1]{S#1}
\renewcommand*{\bibnumfmt}[1]{[S#1]}

\setcounter{figure}{0}
\renewcommand{\thefigure}{S\arabic{figure}}

\setcounter{section}{0}
\renewcommand{\thesection}{S\arabic{section}}

\setcounter{equation}{0}
\renewcommand{\theequation}{S\arabic{equation}}

\setcounter{table}{0}
\renewcommand{\thetable}{S\arabic{table}}

\def\SCBO{SrCu$_2$(BO$_3$)$_2$ }

\onecolumngrid

\noindent
{\large {\bf {Supplementary Information to accompany the article}}}

\vskip4mm

\noindent
{\large {\bf {Field-induced bound-state condensation and spin-nematic phase 
in SrCu$_2$(BO$_3$)$_2$ revealed by neutron scattering up to 25.9 T }}}

\vskip4mm

\noindent
Ellen Fogh, Mithilesh Nayak, Oleksandr Prokhnenko, Maciej Bartkowiak, Koji 
Munakata, Jian-Rui Soh, Alexandra A. Turrini, Mohamed E. Zayed, Ekaterina 
Pomjakushina, Hiroshi Kageyama, Hiroyuki Nojiri, Kazuhisa Kakurai, Bruce 
Normand, Fr\'ed\'eric Mila and Henrik. M. R\o{}nnow

\vskip6mm

\twocolumngrid

\section{Inelastic neutron scattering experiment}

\subsection{Geometric considerations in real and reciprocal space} 

\begin{figure*}
\includegraphics[width = \textwidth]{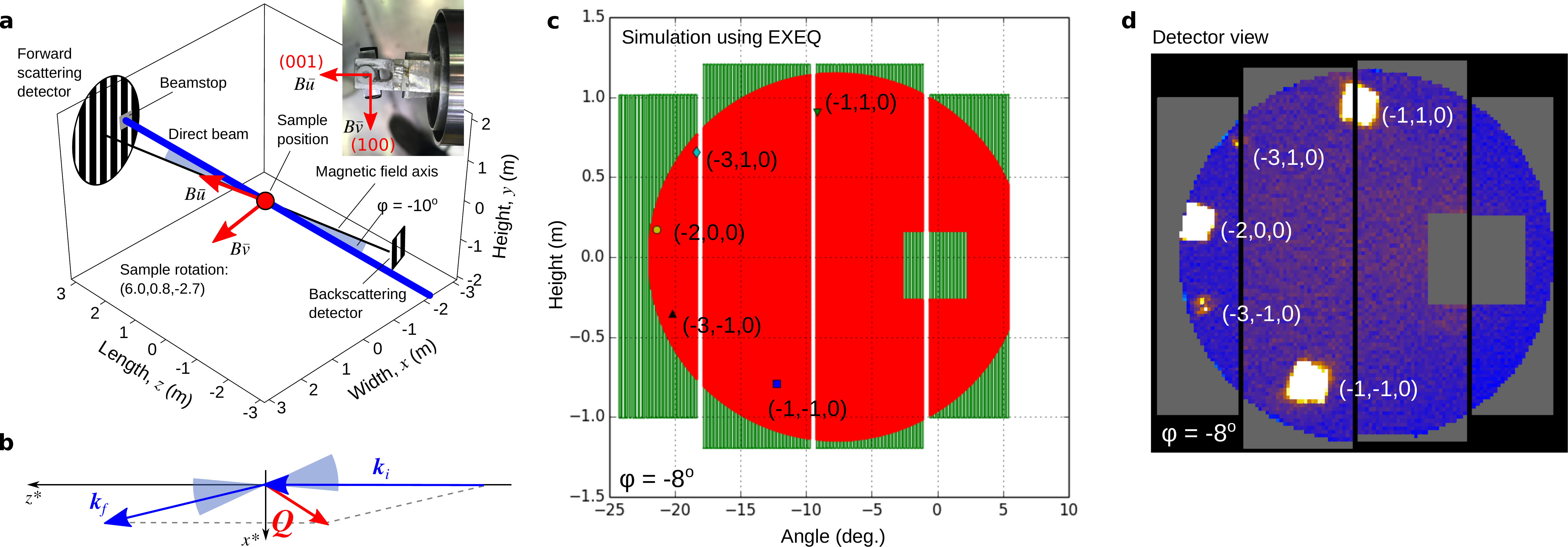}
\caption{{\bf a} Schematic representation of the instrument geometry, showing 
the magnet axis (black line), the neutron beam (thick blue line) and the 
detector positions. The axes $B\overline{u}$ and $B\overline{v}$ (red arrows) 
define the sample orientation with respect to the field. The inset shows a photo of the sample mounted 
in the dilution refrigerator, indicating the orientation. {\bf b} Scattering 
triangle showing the wavevectors ${\bf k}_i$ of the incoming neutrons, ${\bf 
k}_f$ of the outgoing neutrons and the scattering vector, ${\bf Q} = (Q_h,Q_k,
Q_l)$. The cones represent the opening angles of the magnet, whose rotation 
with respect to the incoming beam is $\varphi$. {\bf c} Simulated detector 
view for $\varphi = -8^{\circ}$, showing the predicted positions of five 
different nuclear Bragg peaks. {\bf d} Detector view from the experiment 
with the same settings as in panel {\bf c}. Panels {\bf a} and {\bf c} were 
generated using the EXEQ prediction tool \cite{bartkowiak2020}.}
\label{fig:supp1}
\end{figure*}

The HFM/EXED instrument that was built at the HZB 
is described in a series of detailed publications \cite{bartkowiak2015,prokhnenko2015,
smeibidl2016,prokhnenko2016,prokhnenko2017}. The sample orientation for 
our experiment was established by using EXED in diffraction mode with a 
wavelength range of $\lambda = [0.6,6.3]$ $\mathrm{\AA}$. Three axes are used 
to define this orientation, $B\overline{u}$, $B\overline{v}$ and $B\overline{u} 
\times B\overline{v}$, as represented in Fig.~\ref{fig:supp1}a. The sample 
rotation angles with respect to the three axes, $(6.0^{\circ}, 0.8^{\circ}, - 
2.7^{\circ})$, were determined by indexing a number of nuclear Bragg peaks and 
following their position on the detector at different magnet rotation angles, 
$\varphi$. The scattering geometry, defined in Fig.~\ref{fig:supp1}b, was 
constrained by the cone-shaped $30^{\circ}$ openings of the magnet. The 
predicted detector image, shown in Fig.~\ref{fig:supp1}c, agreed well with 
the measured image shown in Fig.~\ref{fig:supp1}d; these images were made 
using $\varphi = -8^{\circ}$, where five nuclear Bragg peaks were observed. The 
INS experiment was then carried out with $\varphi = -10^{\circ}$, where only 
three Bragg peaks were visible, but the larger rotation angle allowed larger 
scattering angles, which enlarged the accessible portion of reciprocal space.

\begin{figure*}
\includegraphics[width = \textwidth]{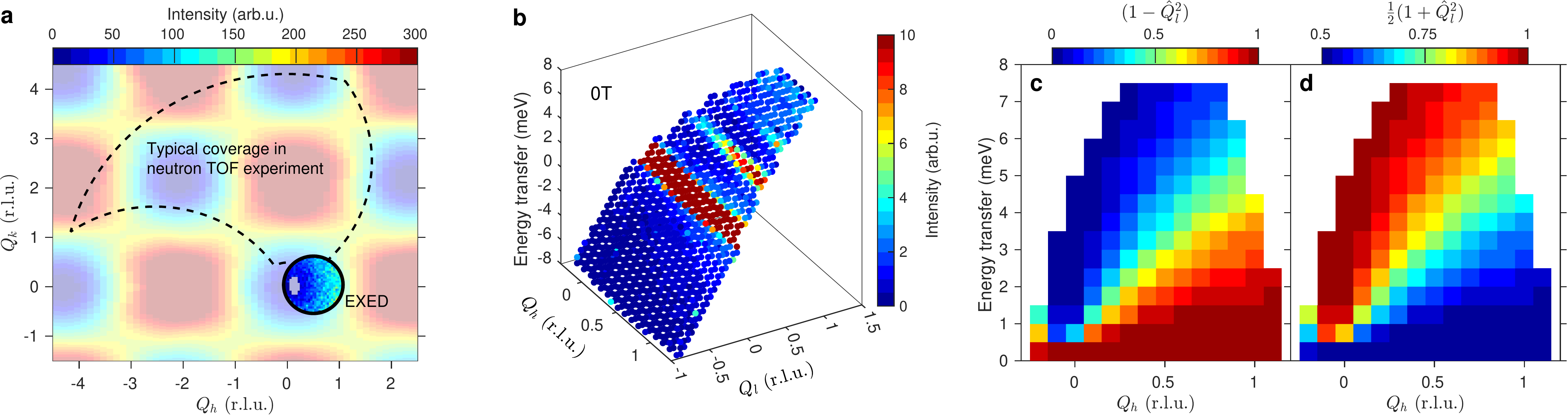}
\caption{{\bf a} The area inside the solid circle shows the portion of 
reciprocal space accessible in our EXED experiment and the area inside the 
dashed lines its comparison with the range covered in a typical time-of-flight 
INS experiment with $E_i = 8$~meV and detector coverage 10-100$^{\circ}$. This 
slice in $(Q_h,Q_k,0)$ was made for energies in the range $[2.79,3.29]$~meV 
around the one-triplon excitation. The pattern shown with faded colour is the 
corresponding structure factor calculated from the zero-field MPS spectrum. 
{\bf b} Representation of the connection between the energy transfer and the 
$Q_l$ component of the scattering vector. The data shown were taken at zero 
field and integrated over the range $Q_k = [-0.75,0.75]$. {\bf c}, {\bf d} 
Structure-factor components $(1 - \hat{Q}^2_l)$ and $\frac{1}{2} (1 + 
\hat{Q}^2_l)$, shown as functions of energy transfer and of $Q_h$, again 
with integration over $Q_k = [-0.75,0.75]$.}
\label{fig:supp2}
\end{figure*}

The range of ${\bf Q}$ accessible on EXED was restricted by the high-field 
magnet, as Fig.~\ref{fig:supp2}a shows for an energy transfer of 3.0~meV, 
where almost one Brillouin zone is covered. This area was larger at lower 
energy transfers, but smaller at higher ones. The comparison in 
Fig.~\ref{fig:supp2}a makes clear that on EXED we could not access the 
high-intensity point at ${\bf Q} = (-0.5,1.5,0)$, or any of its equivalent 
points, found in a previous experiment at zero field \cite{lancon2017}. While 
the structure factors at finite fields are not known, the earlier zero-field 
data showed that different excitations have markedly different structure 
factors, but also that this difference is detectable only at higher absolute 
values of ${\bf Q}$. As a result, the only ${\bf Q}$-dependence we could 
observe on EXED was an overall increase of intensity with increasing 
$|{\bf Q}|$.

The range accessible along $(0,0,Q_l)$ was restricted by the choice of $Q_h$ 
and $Q_k$ as well as by the energy transfer, as shown in Fig.~\ref{fig:supp2}b. 
Thus a given point in the spectrum represented a certain range of $Q_l$ values 
and consequently a specific ratio of the measured dynamical structure factors, 
because $S^{zz} \propto (1 - \hat{Q}^2_l)$ whereas $S^{xx} \propto \frac{1}{2}(1
 + \hat{Q}^2_l)$; these two quantities are illustrated in 
Figs.~\ref{fig:supp2}c-d. For the $Q$-integrated spectra shown in Figs.~2g-l 
of the main text, higher energies contain stronger contributions from $S^{xx} 
({\bf Q}, \omega)$ whereas lower energies are more reflective of $S^{zz} 
({\bf Q}, \omega)$. We comment that $Q_k = 0$ was used when calculating 
$\hat{Q}^2_l$ in Fig.~\ref{fig:supp2}c-d, although in practice the integration 
range along $(0,Q_k,0)$ was rather large, as in Figs.~2g-l of the main text.

\subsection{Data treatment and background subtraction} 

\begin{figure}
\includegraphics[width = \columnwidth]{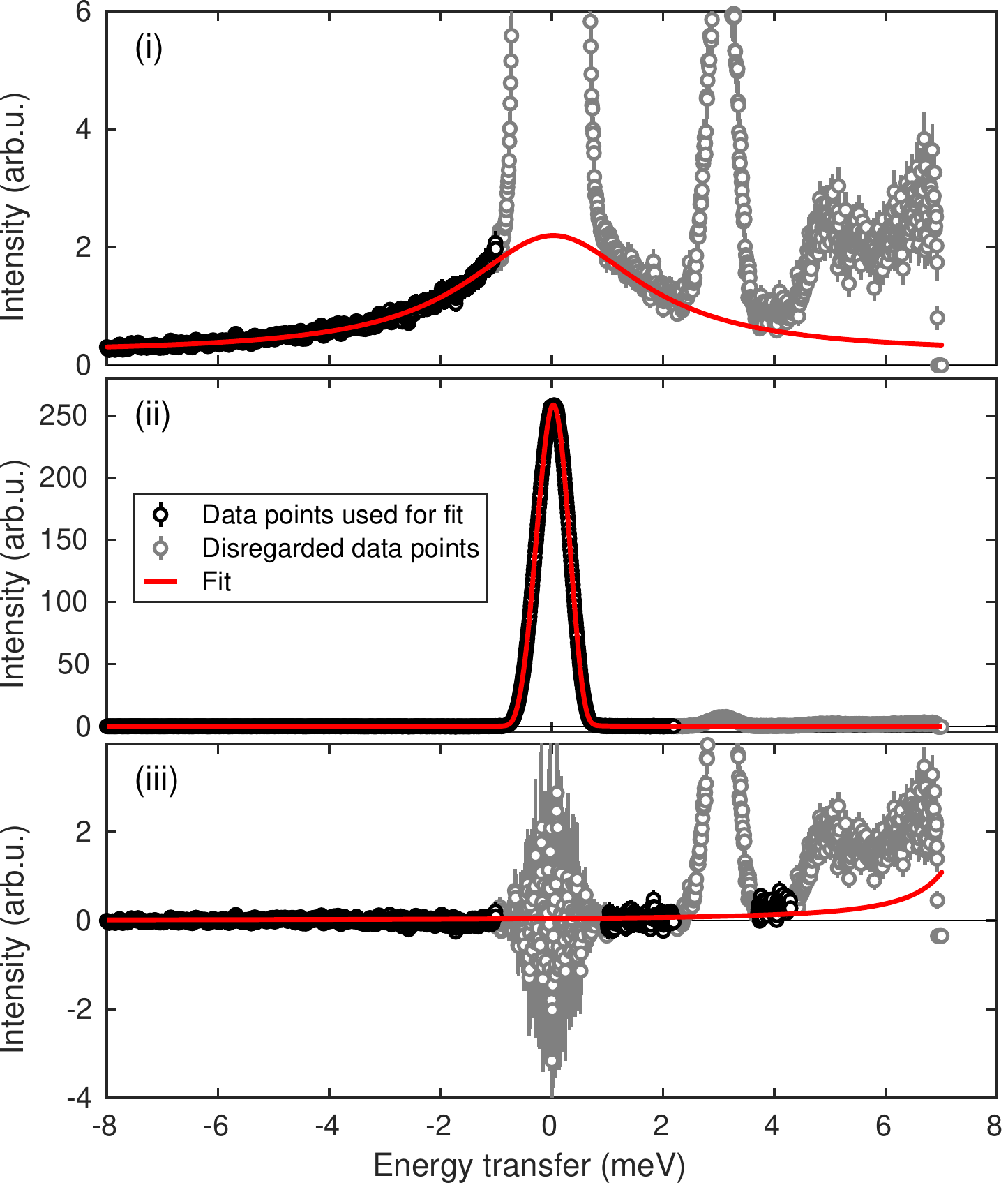}
\caption{Three-step process for background estimation based on the zero-field 
data. The black data points are those used for each step of the fitting 
procedure and the grey ones are those which were disregarded. The red lines 
show the fit at each step and the final background is their sum.}
\label{fig:supp3}
\end{figure}

The INS data were corrected for the detector efficiency following a standard 
procedure where individual detector elements are normalized with respect to a 
vanadium standard. This correction and all subsequent cuts through the data 
were performed using the Mantid software package \cite{arnold2014}. Figures 
2g-l of the main text, and the slices of Fig.~1e at the same fields, show 
the background-subtracted neutron intensity integrated over the entire 
available area of reciprocal space, $Q_h = [-0.25,1.25]$, $Q_k = [-0.75,
0.75]$ and $Q_l = [-0.9,1.3]$, for the data collected with $E_i = 8$ meV. 
This background was based on the zero-field data, where it is known that 
the lowest-lying excitation is the one-triplon mode at 3.0~meV 
\cite{kageyama2000a,gaulin2004,kakurai2005,zayed2014a,lancon2017}, and at 
the low temperature of our experiment one expects a signal only on the neutron 
energy-loss side of the spectrum. The background was defined by a three-step 
fitting procedure that we illustrate in Fig.~\ref{fig:supp3}. 
(i) Quasi-elastic scattering was approximated by a Lorentzian. (ii) After 
subtracting the Lorentzian, the elastic lineshape was fitted to six Gaussians.
This number is purely empirical and was optimized to describe the shape of the
elastic line. (iii) Once both Lorentzian and Gaussian contributions had been 
subtracted, the time-independent part of the background was fitted to the form
$I(E) = I_0 (E_i - E)^{-3/2}$, with $E_i = 8$~meV fixed and the constant $I_0$ 
as the only fitting parameter. The final background is then a sum of the three 
contributions.

For the intensity data measured with $E_i = 4$~meV and shown in Fig.~3a of the 
main text, the integration range was $Q_h = [-0.15,0.85]$, $Q_k = [-0.5,0.5]$ 
and $Q_l = [-1.5,2.5]$. The zero-field results in this case did not have good 
statistics because the count time was only 10 minutes, and instead we based 
our background estimate on the 25.9 T data, which were collected for 5.5 h. 
Because the spectrum is not known at this field, the background estimation 
procedure was based on data points in the energy interval 
$E = [-4.0, 0.1]$~meV.

A spurion was observed around 2.1~meV for data with $E_i = 4$~meV and around 
5.1 meV for data with $E_i = 8$~meV. This feature was field-independent and 
appeared close to ${\bf Q} = 0$. A Gaussian was fitted and subtracted to 
remove the spurion before fitting the spectrum of excitations originating 
in the sample.

\subsection{Peak fitting and intensities} 

Here we describe in detail the procedure used to fit the data shown in 
Figs.~2g-l of the main text. Previous theoretical analysis at zero field and 
our own numerical results at finite fields motivate the presence, positions 
and widths of a significant number of candidate excitations, as a result of 
which we based our fit on multiple Gaussian profiles. At zero field, the peak 
width we obtained for the one-triplon mode around 3.0~meV was used to fix the 
widths of the excitations at 4.8, 5.5 and 6.5~meV, with the ratio of these 
widths taken from Ref.~\cite{kakurai2005}. The zero-field parameters of the 
$S = 1$ two-triplon bound state were used to fix the properties of the $S^z
 = 0$ branch of this multiplet at finite fields. We did not include Gaussians 
for the $S^z = \pm 1$ branches of this multiplet, because their neutron 
intensities are expected to be weak. We note also that the $S = 1$ bound 
state has vanishing intensity at ${\bf Q} = (1.5,0.5,0)$, which is why it 
cannot be observed in some of the data we show in Fig.~1e of the main text, 
or in Fig.~\ref{fig:supp4} below. 

\begin{figure}
\includegraphics[width = \columnwidth]{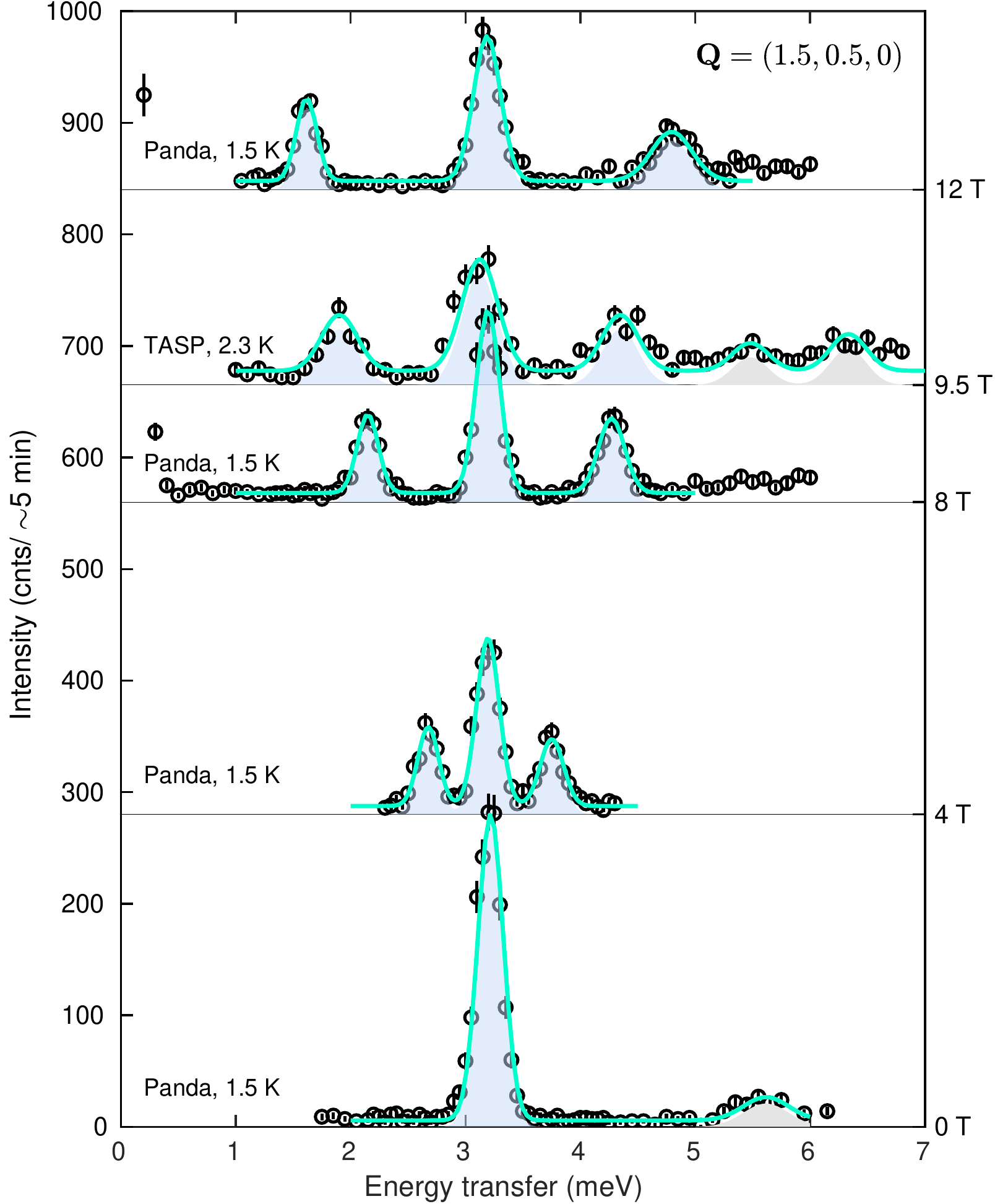}
\caption{Neutron intensities measured as functions of energy transfer at 
constant ${\bf Q} = (1.5,0.5,0)$ on Panda and TASP at selected field strengths. 
The data points are offset along the vertical axis by a distance proportional 
to the corresponding applied field value, as noted at right. The blue shading 
shows fits to the three one-triplon branches. The grey shading shows additional 
excitations at higher energies. The data from TASP were scaled to those from 
Panda by matching the integrated intensities of the measured $t_0$ branches.}
\label{fig:supp4}
\end{figure}

\begin{figure}
\includegraphics[width = 0.9\columnwidth]{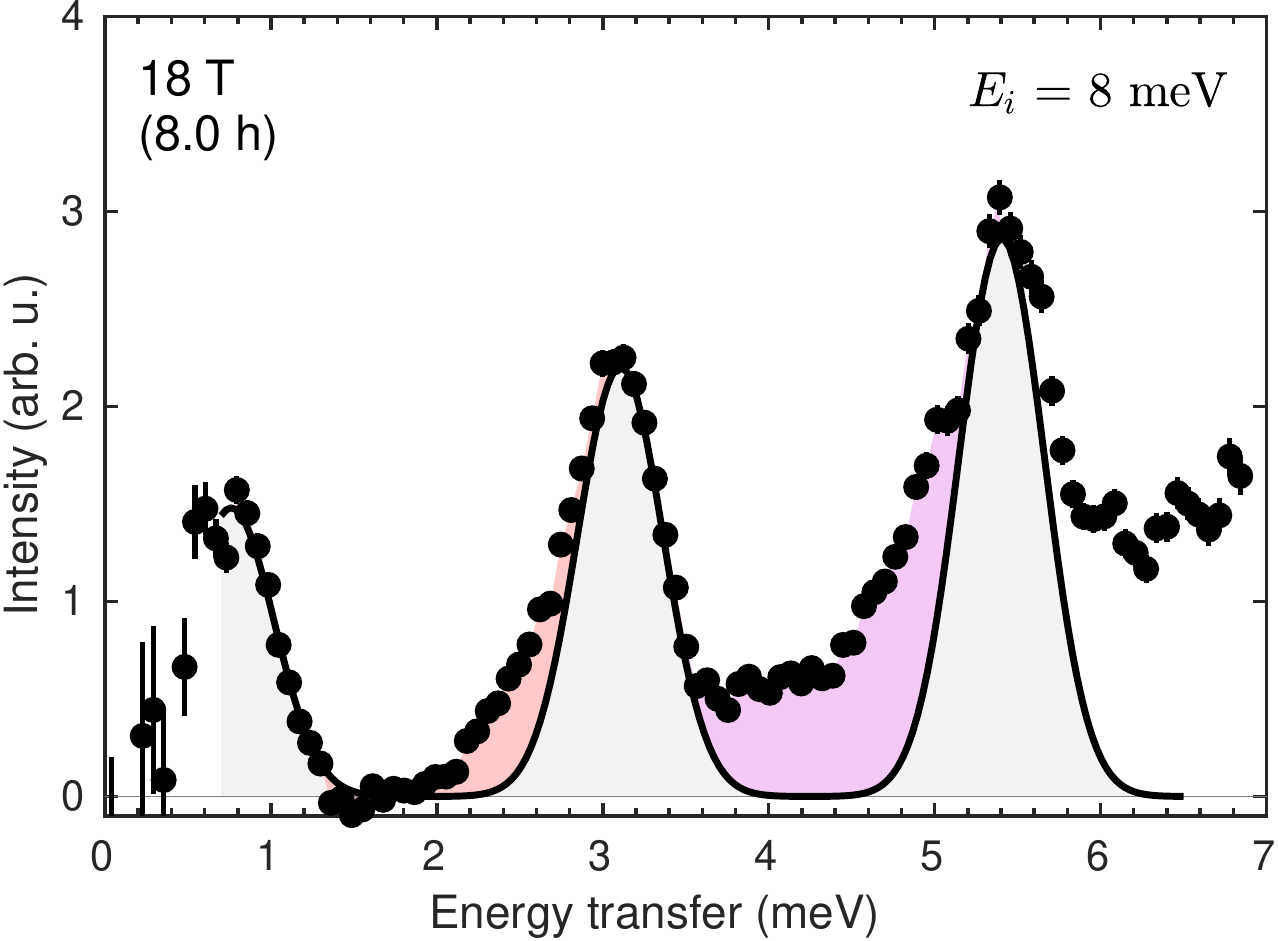}
\caption{Illustration of INS data fitting, shown for an applied field of 18~T. 
The one-triplon branches are fitted by three Gaussians of fixed width (black 
lines and grey shading). Red shading indicates the extra intensity found below 
the $t_0$ branch and purple shading the extra intensity appearing below the 
$t_-$ branch.}
\label{fig:intensities}
\end{figure}

\begin{figure*}
\includegraphics[width = \textwidth]{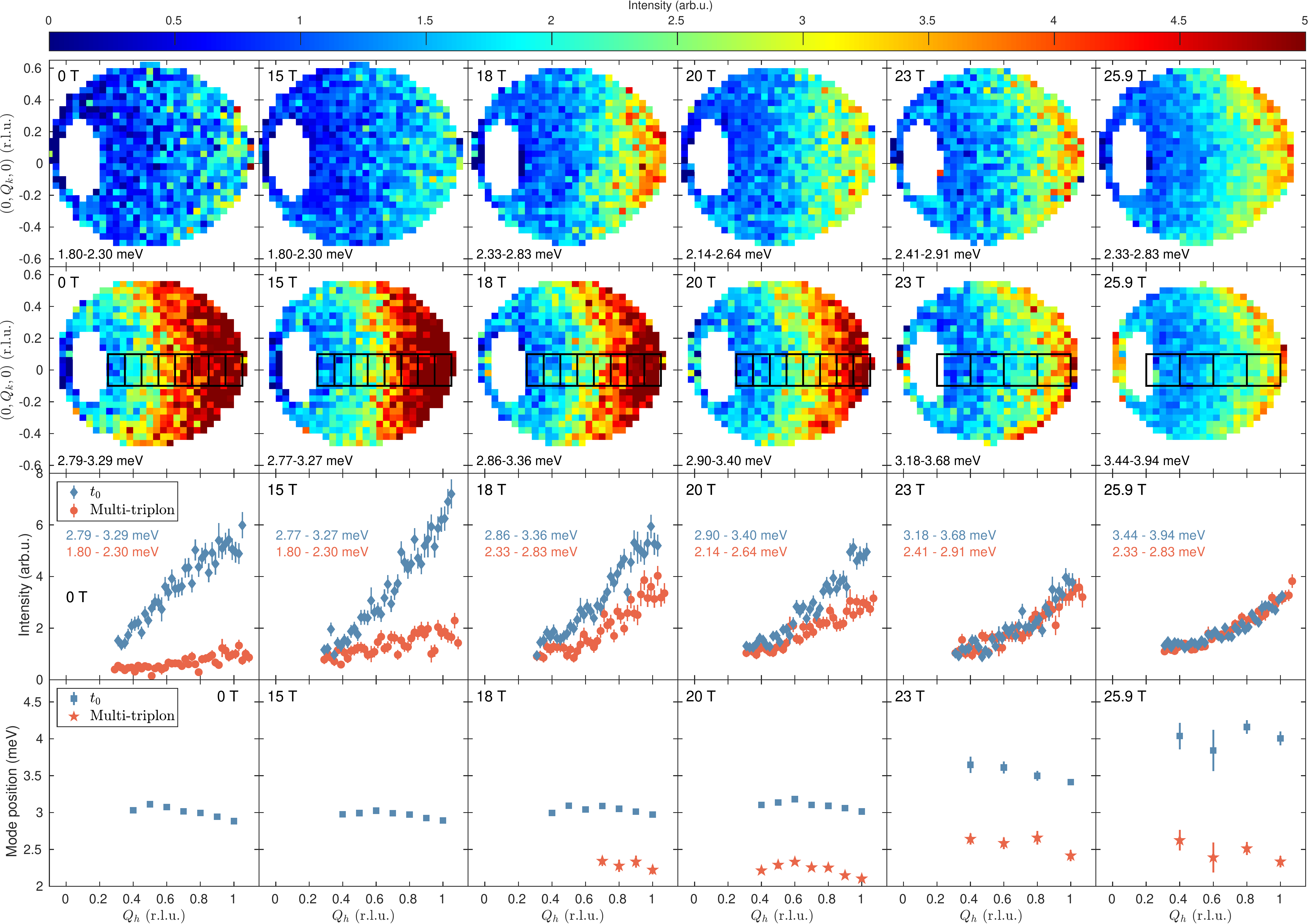}
\caption{Data slices showing the ${\bf Q}$-dependence of the neutron intensity 
below the $t_0$ excitation branch (top row) and around the $t_0$ branch (second 
row). In preparing these slices, the intensity data were integrated over the 
momentum interval $Q_l = [-0.9,1.3]$ and over an energy interval of width 
$\Delta E = 0.5$~meV centred on the fitted positions of the red Gaussians 
(to obtain the top row) and the $t_0$ Gaussians (to obtain the second row) 
in Figs.~2g-l of the main text. The zero-field intensity of the $t_0$ branch 
has been divided by 2 for illustration on the same colour scale. Below 18~T, 
where no extra intensity was detected below $t_0$, the top two colour panels 
show simply an energy window well below the $t_0$ window. The black boxes in 
the second row illustrate the region of integration used to produce the panels 
in the third and fourth rows. These show respectively the integrated 
intensities and mode positions as a function of $Q_h$.} 
\label{fig:supp5}
\end{figure*}

At finite fields, the one-triplon mode undergoes a Zeeman splitting, with the 
$t_-$ and $t_+$ branches positioned at energies $\pm \alpha \, \mu_0 H$ from 
the $t_0$ branch, where $\alpha \simeq 0.131$~meV/T \cite{nojiri2003}. We fixed 
the widths of the $t_+$, $t_-$ and $S = 1$, $S^z = 0$ bound-state branches to 
that fitted for the $t_0$ mode. Only at 15~T, where the $t_0$ peak is close to 
the $S_z = + 1$ branch of the $S = 1$ multiplet, did we use the fitted width of 
the $t_+$ mode, which is well defined at this field, to fix the others. At all 
measurement fields from 18 to 25.9~T, we found that a single additional 
Gaussian placed at an energy below the $t_0$ branch, and up to three additional 
Gaussians placed below the $t_-$ branch, provided a good fit to the intensity 
profiles measured as a function of energy transfer. The positions and widths 
of these Gaussians were allowed to vary freely, except if the widths became 
much greater or smaller than that of the $t_0$ branch, in which case their 
widths were fixed relative to that of this branch. Despite their two-triplon 
nature, the excitations in this energy and field range are not expected to 
display a dispersion significantly larger than that due to the DM term 
\cite{kakurai2005}, and the width of the momentum-integrated $t_0$ excitation 
represents this bandwidth. We stress that our fit does not imply the presence 
of just one additional excitation below the $t_0$ branch, only that a single 
Gaussian is suitable for capturing all of the additional intensity arising 
from these excitations, a quantity we used in Fig.~3d of the main text and 
analyse in detail in Sec.~S2D. For completeness we state that the position 
of the highest-energy mode below the $t_-$ branch at 23~T was fixed.

In Fig.~1e of the main text, the symbols shown at 0, 15, 18, 20, 23 and 25.9~T 
were taken from the Gaussians obtained in Figs.~2g-l. The data slices shown at 
fields between 0 and 15~T were taken from separate experiments: data at 4, 8 
and 12~T were measured on Panda, as noted in the Methods section, at 1.5~K and 
at constant ${\bf Q} = (1.5,0.5,0)$ \cite{zayed2014a,zayed2010}; the data at 
9.5~T were measured on TASP, also at constant ${\bf Q} = (1.5,0.5,0)$. The 
fitting procedure for these data, which are represented with field-scaled 
separation in Fig.~\ref{fig:supp4}, followed a simplified version of the 
procedure applied at high fields. For the Panda data, three Gaussians were 
fitted to the three one-triplon branches, with all parameters allowed to vary. 
For the TASP data, the three one-triplon branches were fitted with three 
Gaussians of the same width, with the positions of the $t_-$ and $t_+$ branches 
constrained to have the same separation from the $t_0$ position and with their 
intensities constrained to have half the intensity of the $t_0$ excitation. Two 
additional Gaussians were fitted at higher energies, with their individual 
positions, widths and intensities taken as free parameters. 

Figure 3a of the main text shows the low-energy parts of three INS spectra 
collected with $E_i = 4$~meV. The momentum-integration ranges and background 
fitting are described in Sec.~S1B. To estimate the peak positions close to the 
elastic line at 20 and 25.9~T, we fitted one Gaussian whose width was fixed to 
the value obtained for the $t_0$ mode at zero field in a separate experiment 
on HFM/EXED.

Figure 3d of the main text shows the integrated spectral weights found in 
energy windows below the $t_0$ and $t_-$ one-triplon branches. To illustrate 
the extraction of these weights, in Fig.~\ref{fig:intensities} we show the 
experimental data at a representative field of 18~T. The one-triplon 
contributions were fitted by three Gaussians following the above procedure. 
The excess spectral weight induced below $t_0$ is then defined as that 
appearing above the one-triplon contribution between $t_+$ and $t_0$, 
highlighted with red shading in Fig.~\ref{fig:intensities} and it is 
integrated numerically. The excess weight below $t_-$ is defined in the 
same way and is shown by the purple shading. 

\subsection{{\bf Q}-dependence} 

As Figs.~2a-f of the main text make clear, the spin excitations of \SCBO~are 
extremely insensitive to ${\bf Q}$, and this was the basis on which we 
integrated our EXED results over the full reciprocal space for further 
analysis. Here we nevertheless inspect the differences discernible over the 
${\bf Q}$ range covered in our measurements (Fig.~\ref{fig:supp2}a). The top 
two rows of Fig.~\ref{fig:supp5} (colour panels) represent the intensities 
measured in the $(Q_h,Q_k,0)$ plane in energy windows selected to cover the 
regions below (top row) and spanning (second row) the $t_0$ excitation branch 
for all six measurement fields. While the shape of the intensity distribution 
is indeed very similar in all panels, with the highest intensities at the 
largest $|Q|$ values, the net intensity reflects the general shift from the 
$t_0$ branch to the novel multi-triplon excitations below it (red Gaussians 
in Figs.~2g-l) as the field is increased, as quantified in Fig.~3d of the 
main text. 

The third row of panels in Fig.~\ref{fig:supp5} quantifies the evolution of 
the two intensity contributions with $Q_h$ by integrating over $Q_k = [-0.1,
0.1]$ (black boxes in second row). At 0 and 15~T, where no statistically 
significant extra intensity could be discerned below the $t_0$ branch 
(Figs.~2g and 2h of the main text), the intensity in the selected 
integration window can be attributed to the sample environment; because 
different energy intervals correspond to different intervals in $Q_l$, we 
do not subtract this background contribution. The intensities measured for 
the $t_0$ branch at each field show a largely linear dependence on $Q_h$ and 
a steady decrease as the field is increased above 18~T; the intensities of 
the multi-triplon contribution below $t_0$ also increase with $|Q_h|$ and 
increase with field until they match the $t_0$ weight. 
The bottom row of panels in Fig.~\ref{fig:supp5} shows how the position of 
the $t_0$ branch, and the centre position of the spectral weight below $t_0$, 
vary weakly with $Q_h$. The data were discretized in units of $\Delta Q_h = 
0.1$ for fields up to 20~T and $\Delta Q_h = 0.2$ at 23 and 25.9~T. The very 
minor dispersion observed in all cases is consistent with the bandwidth 
identified in zero-field studies \cite{gaulin2004,kakurai2005}.

\begin{figure*}
\centering
\includegraphics[width = 0.85\textwidth]{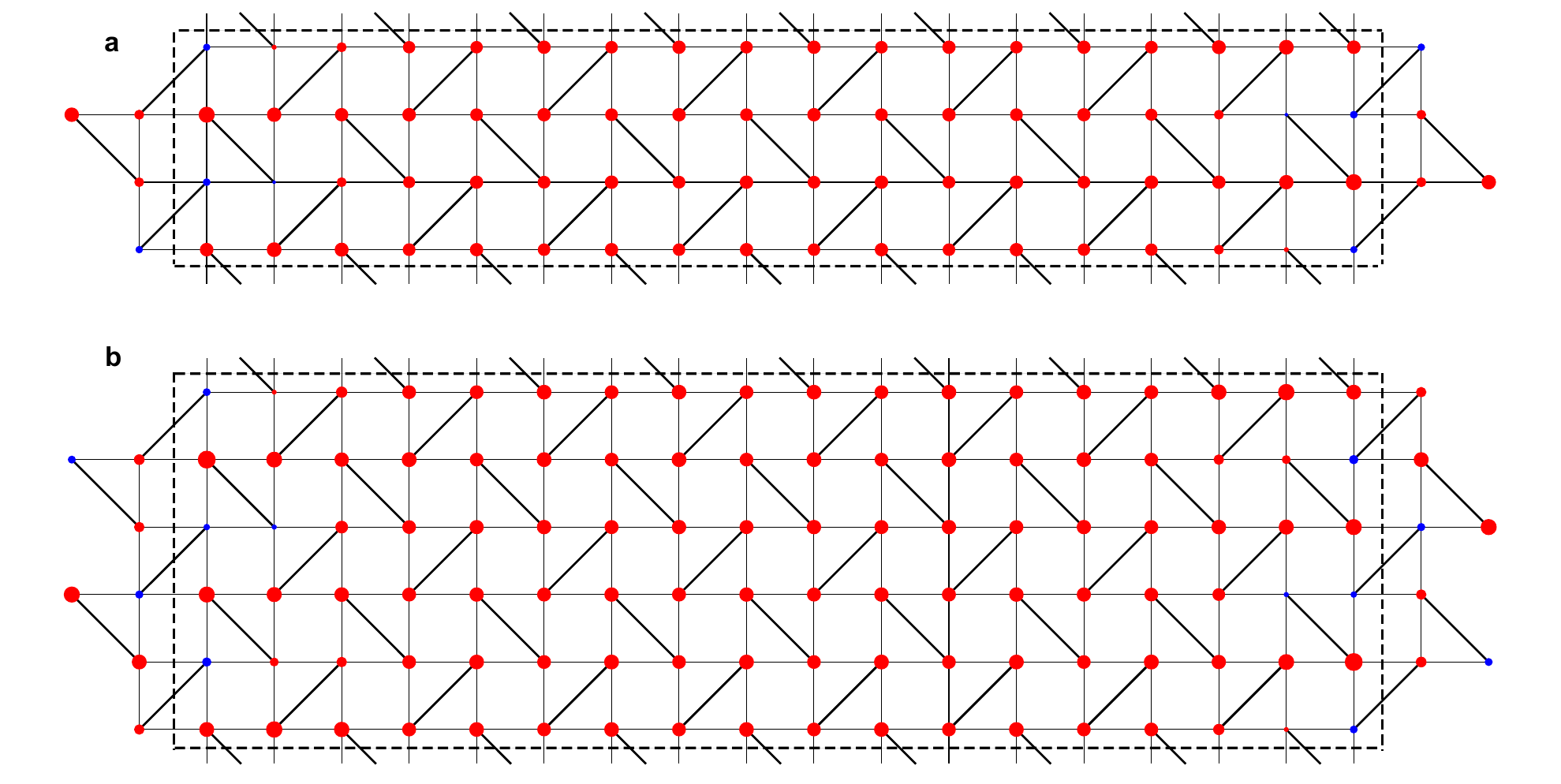}
\caption{Illustration of the cylinder geometries used for our MPS calculations. 
{\bf a} Cylinder width $W = 4$. {\bf b} $W = 6$. The red (positive) and blue 
(negative) circles indicate the magnetization per site obtained at $H = 20$~T, 
and the degree to which it is uniform across the central regions of the 
cylinder. The dashed rectangle indicates the region used for the spatial 
Fourier transform.}
\label{cgvw}
\end{figure*}

\section{Numerical calculations and analysis}

\subsection{Dependence on cylinder size}

An MPS Ansatz describes a one-dimensional (1D) quantum many-body wavefunction. 
A 2D system can be reformulated as a 1D model on a cylinder with long-ranged 
interactions \cite{stoudenmire2012}, such that MPS-based DMRG \cite{white1992,
schollwoeck2011} can be used to find the ground state and compute the 
associated static observables. The dynamical structure factor (DSF) of a 2D 
system can also be computed using MPS-based methods, and for this work we 
applied the TDVP \cite{haegeman2011,haegeman2016,paeckel2019}, as described in 
the Methods section. Here we calibrate how the finite length and circumference 
of the cylinder affects our calculations. All results reported in the main text 
were obtained using a cylinder circumference of $W = 4$ sites and a length of 
$L = 20$ sites, as represented in Fig.~\ref{cgvw}a. As a first step towards 
justifying that these values of $W$ and $L$ return results fully representative 
of the 2D SSM, we computed the ground-state energy and field-induced static 
magnetization with $W = 6$ (Fig.~\ref{cgvw}b), finding that both quantities 
change rather little. 

\begin{figure}
\centering
\includegraphics[width=0.92\columnwidth]{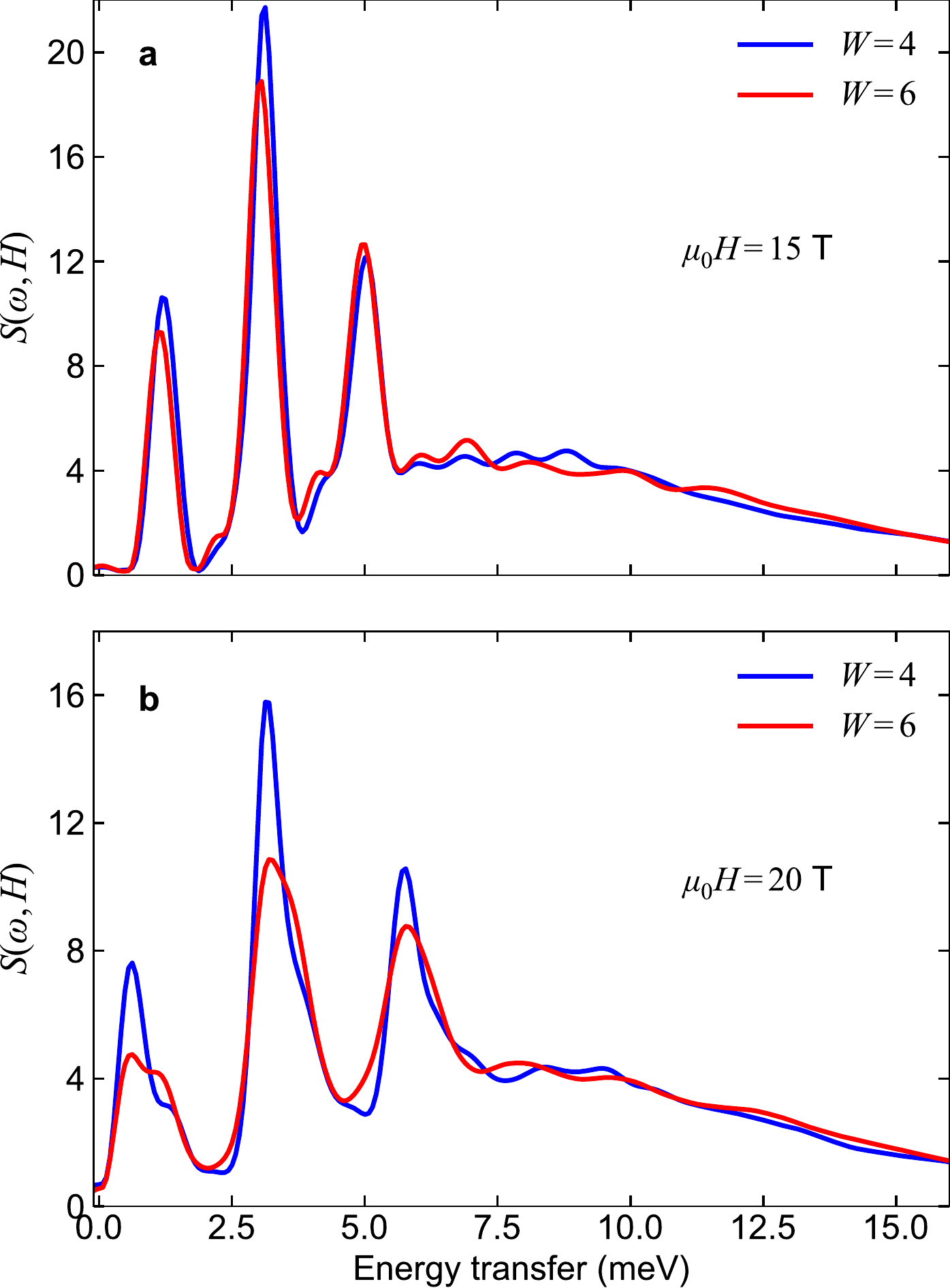}
\caption{Comparison of fully ${\bf Q}$-integrated dynamical structure factors, 
shown as a function of energy transfer, computed using cylinders of widths 
$W = 4$ and 6 at 15 T ({\bf a}) and 20 T ({\bf b}).}
\label{etwv}
\end{figure}

TDVP calculations become extremely time-intensive on wider cylinders for two 
reasons. First the bond dimension of the MPS should increase exponentially 
with $W$ in order to describe the state with equivalent accuracy. Second, 
wider cylinders introduce further-neighbour interactions in the effective 
1D model that increase the virtual dimension of the Matrix Product Operator 
in proportion with $W$. The resulting $O (W^2 \exp (3W))$ scaling of the 
calculation time therefore prevented us from obtaining the DSF with $W = 6$ 
at all values of the magnetic field required for our study, but we did compute 
it at 15 and 20 ~T, and in Fig.~\ref{etwv} we compare the ${\bf Q}$-integrated 
DSFs obtained for $W = 4$ and 6 at both fields. Because the allowed $k_y$ 
values for a $W = 4$ cylinder are 0 and $\pi$, while for a $W = 6$ cylinder 
they are 0, $2\pi/3$ and $4\pi/3$, increasing $W$ results in more spectral 
weight for each mode and thus we normalize the spectra to the same integrated 
weight. At first sight there is remarkably little change from $W = 4$ to 6, 
and on close inspection one may ascertain that the loss of one-triplon peak 
height in the $W = 6$ spectrum reappears at energies between and above the 
three primary branches. This could be expected on the grounds that the wider 
cylinder allows more space for the development of multi-triplon dynamics. 
Overall, however, the effects of $W$ are small and the effects of $L$ are 
negligible beyond $L = 20$; while we do not show the latter null result, it 
is readily understood from the extreme localization of all one- and 
higher-triplon excitations in the ideally frustrated SSM geometry. 

\subsection{Role of Dzyaloshinskii-Moriya interactions}
\label{DM_vary}

As discussed in the main text, a primary consequence of two-triplon bound-state 
condensation is the presence of a one-triplon gap at all fields. However, in 
\SCBO~there are weak DM interactions on both the intra- and inter-dimer bonds, 
as represented in Fig.~1b of the main text, and in particular the intra-dimer 
$D$ term leads to a direct mixing of $S^z$ sectors in the SSM. It is known from 
studies of a spin ladder with intra-dimer DM terms that an avoided crossing can 
occur, such that the one-triplon gap reaches a finite minimum at the expected 
condensation field, and then increases again at higher fields 
\cite{miyahara2007}. To eliminate the possibility that the behaviour we 
observe is a DM-induced avoided crossing, we have performed MPS calculations 
at different values of $D$ to identify the evolution of the lowest-lying 
excitations. We note again that a small but finite $D$ is required to obtain 
a uniformly magnetized ground state in our calculations, and for all the MPS 
results shown in the main text this was set to the experimentally deduced 
values $D = 0.034 J$ \cite{kodama2005}, with $D' = - 0.02 J$ \cite{cepas2001}. 
MPS results for the $D, D' = 0$ limit in the presence of a field may be 
obtained by varying both parameters and extrapolating the results. 

\begin{figure}
\centering
\includegraphics[width=0.92\columnwidth]{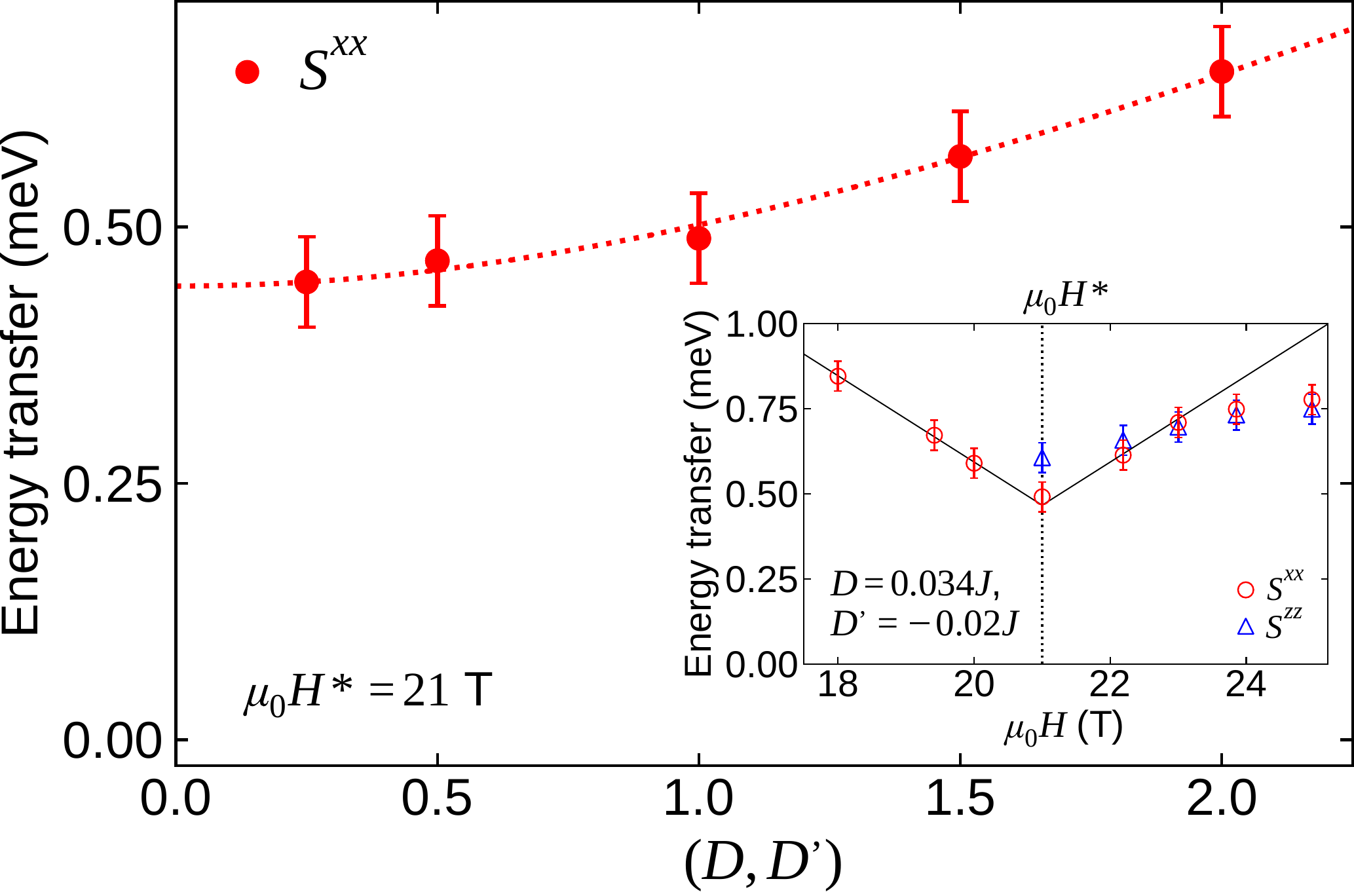}
\caption{Excitation energies obtained by MPS calculations on cylinders of 
width $W = 4$ to illustrate the effect of the DM interactions at a field 
equivalent to $\mu_0 H = 21$~T. On the $x$-axis, $D$ and $D'$ are scaled to 
their experimental values. On the $y$-axis, the position of the $t_+$ branch 
was obtained from the lowest peak in the $S^{xx}$ component of the dynamical 
structure factor and the vertical bar indicates the width of this peak. 
(Inset) Variation of the $t_+$ gap with magnetic field at fixed $D = 0.034 
J$ and $D' = - 0.02 J$. For comparison we show the peak energy of the 
low-energy spectral weight in the $S^{zz}$ component, which arises from 
internal excitations within the $|t_+ t_+ \rangle$ bands (Fig.~4a of the 
main text).}
\label{tgdmv}
\end{figure}

We computed the transverse ($S^{xx}$) and longitudinal ($S^{zz}$) components 
of the dynamical structure factor at $\mu_0 H = 21$ T for five different $D$ 
values, with $D'$ fixed to $0.59 D$, and extracted the lowest-lying 
excitations in each sector. The one-triplon gap, read from $S^{xx}$ 
by fitting a Gaussian to the low-energy spectral weight, is shown in 
Fig.~\ref{tgdmv}. For comparison we show the generic functional form 
$\Delta (D) = \sqrt{\Delta_0^2 + c^2 D^2}$, which interpolates between a 
constant at $D = 0$ and a linear increase at large $D$. This simple form 
offers a convincing fit, demonstrating that the lower bound for the 
one-triplon gap as $D \rightarrow 0$ is $\Delta_0 = 0.45$~meV. In the inset 
of Fig.~\ref{tgdmv}, we show that this gap is also not a consequence of the 
field choice, as varying $H$ confirms the linear increase away from $\mu_0 
H^* = 21$~T with decreasing field, and initially with increasing field, as 
represented qualitatively in Fig.~1e of the main text. For completeness we 
show also the peak energy determined in the same way from $S^{zz}$, which 
appears as a result of internal excitations of the two-triplon bound state 
(Figs.~4a and 4f of the main text), and moves below the $S^{xx}$ peak at 
the highest fields we study. If one accepts the result of Sec.~S2A, that 
this constant minimum gap of order 0.5~meV at $H^*$ is not a finite-size 
effect, it must therefore be an intrinsic property of the pure SSM (i.e.~of 
the model with only Heisenberg interactions). We repeat the analogy to 
superconductivity drawn in the main text, where a single-particle excitation 
formed by pair-breaking in the condensate has an energy gap equal to half 
the binding energy. This implies that the two-triplon binding energy in 
\SCBO~is around 1~meV, a value fully consistent with the results of 
third-order perturbation theory \cite{momoi2000}.  

\begin{figure*}
\centering
\includegraphics[width = 0.8\textwidth]{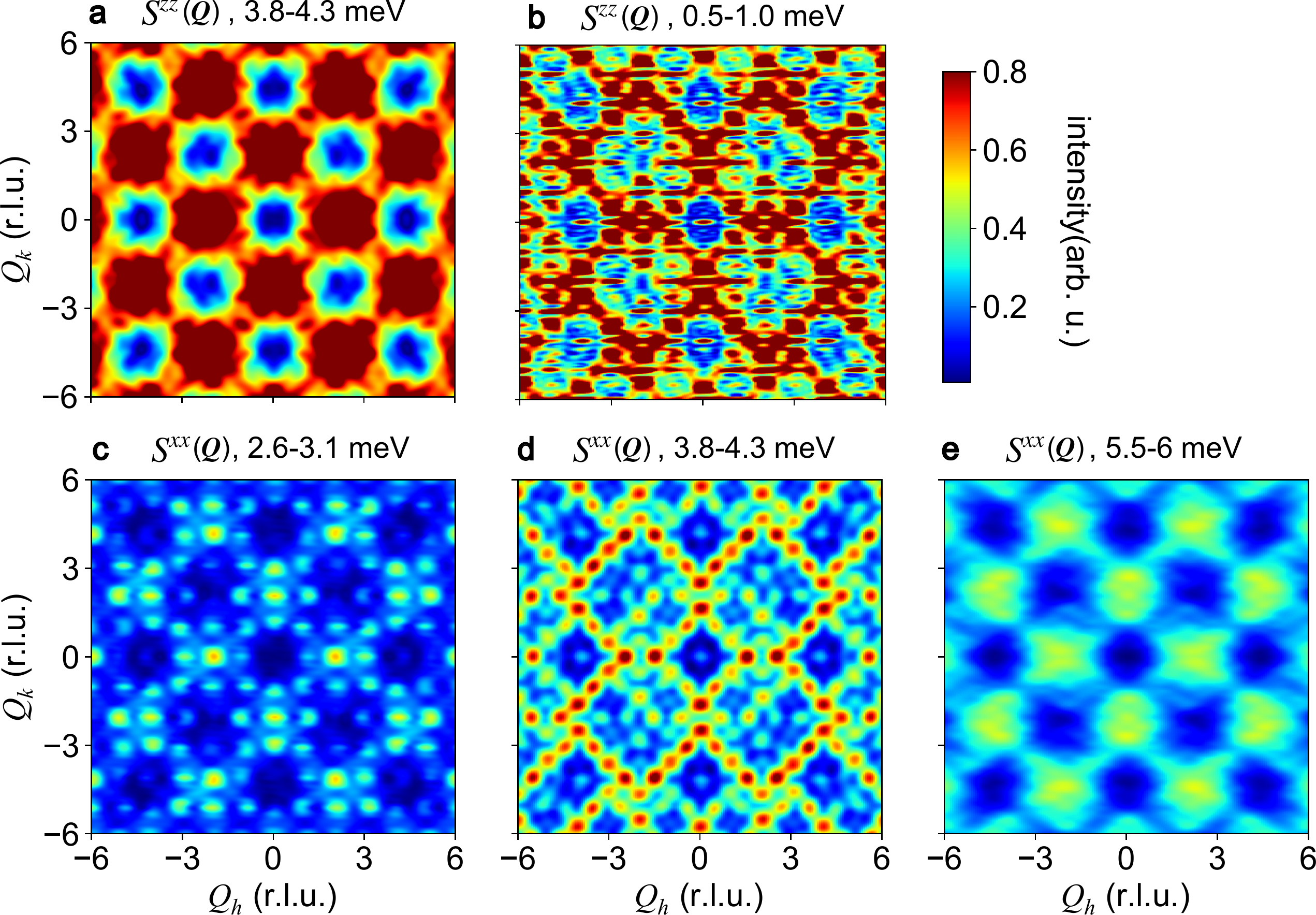}
\caption{Constant-energy slices at zero field displayed in $(Q_h,Q_k)$. 
{\bf a} Slice including the one-triplon $t^0$ branch extracted from $S^{zz}$. 
{\bf a} Slice including the internal excitations of the two-triplon bound 
state, extracted from $S^{zz}$. 
{\bf c}-{\bf e} Slices including different multi-triplon bound states extracted 
from $S^{xx}$ for energy intervals 2.6-3.1 {\bf c}, 3.8-4.3 {\bf d} and 5.5-6.0 
meV {\bf e}. We comment that the weak breaking of fourfold symmetry visible in 
these panels is a consequence of the cylinder geometry (20$\times$4, 
Fig.~\ref{cgvw}a).} 
\label{Qmte}
\end{figure*}

\begin{figure*}
\centering
\includegraphics[width=0.96\textwidth]{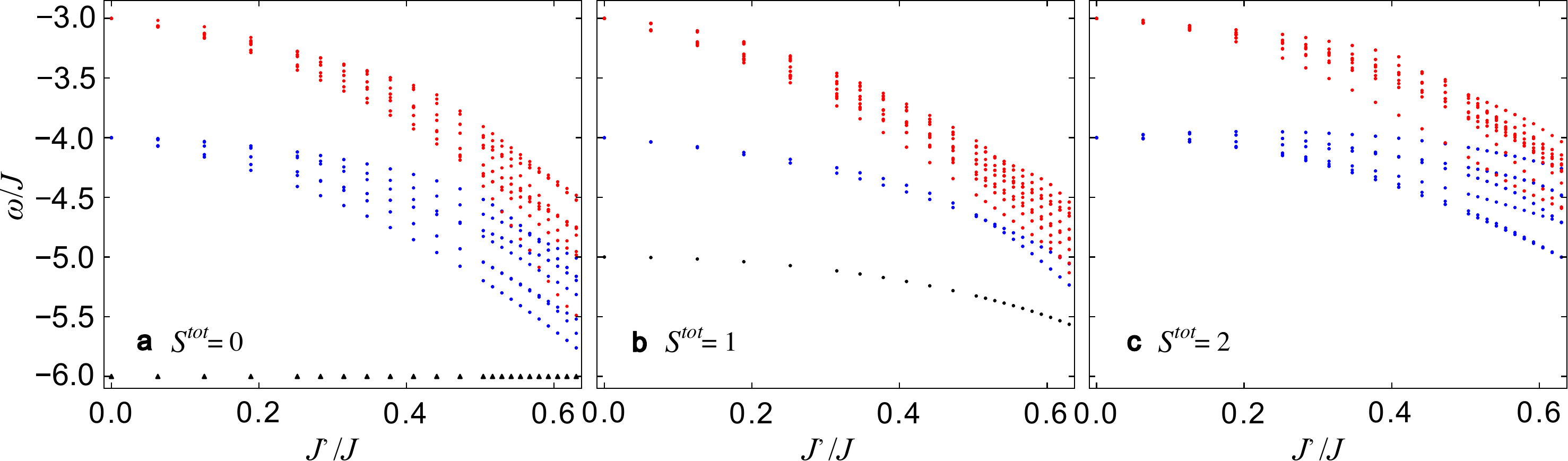}
\caption{Energy levels of the SSM obtained by ED of a 4$\times$4 spin cluster. 
Results are shown as a function of $J'/J$ at zero field for the separate
total-spin ($S$) sectors at momentum $(k_x,k_y) = (0,0)$. Black triangles 
mark the singlet ground state, black circles the three one-triplon states, 
blue circles the two-triplon states and red circles the three-triplon states.}
\label{Imte}
\end{figure*}

\subsection{{\bf Q}-dependence of MPS results}

A full comparison of our cylinder MPS results with the DSF measured in 
experiment requires the inclusion of the structure factor of \SCBO, as noted 
in the Methods section. When the DSF is integrated over a large volume in 
reciprocal space, as in Figs.~1e, 1f, 2, 4a and 4b of the main text, the 
effect of the structure factor is largely invisible, and for this reason we 
do not perform extensive comparisons of the ${\bf {Q}}$-dependence in our INS 
and MPS results. To illustrate the role of the structure factor, for example in 
preparing constant-energy slices of the type shown in Fig.~\ref{fig:supp5}, 
we integrate the DSF obtained from MPS over energy windows selected to 
highlight a particular excitation feature and display the results in the 
space of $(Q_h,Q_k)$. Our MPS calculations allow the further advantage of 
separating the excitations overlapping in a specific energy window by 
considering the transverse and longitudinal channels separately. In 
Fig.~\ref{Qmte} we show a number of constant-energy slices prepared from 
our MPS data at a field equivalent to 25~T. The $t_0$ branch of the 
one-triplon excitation is found in the longitudinal ($S^{zz}$) channel and 
its {\bf Q}-dependence is shown in Fig.~\ref{Qmte}a by selecting an energy 
window of 0.5~meV around the field-induced centre position of 4.05~meV. 
Similarly, the internal excitations of the $|t_+ t_+ \rangle$ pinwheel, 
shown in Fig.~4a of the main text, are found at low energies in $S^{zz}$ 
(Fig.~\ref{Qmte}b). Otherwise, stronger spectral contributions from the 
novel multi-triplon excitations appear in the transverse ($S^{xx}$) channel, 
and in Figs.~\ref{Qmte}c-e we compare the intensity distributions for energy 
windows below the $t_0$ branch (Fig.~\ref{Qmte}c), around the $t_0$ branch 
but in the opposite channel (Fig.~\ref{Qmte}d) and below the $t_-$ branch 
(Fig.~\ref{Qmte}e). In each case we observe distinctive patterns that could 
be analysed in future high-field INS experiments. 

\begin{figure*}
\includegraphics[width=0.9\textwidth]{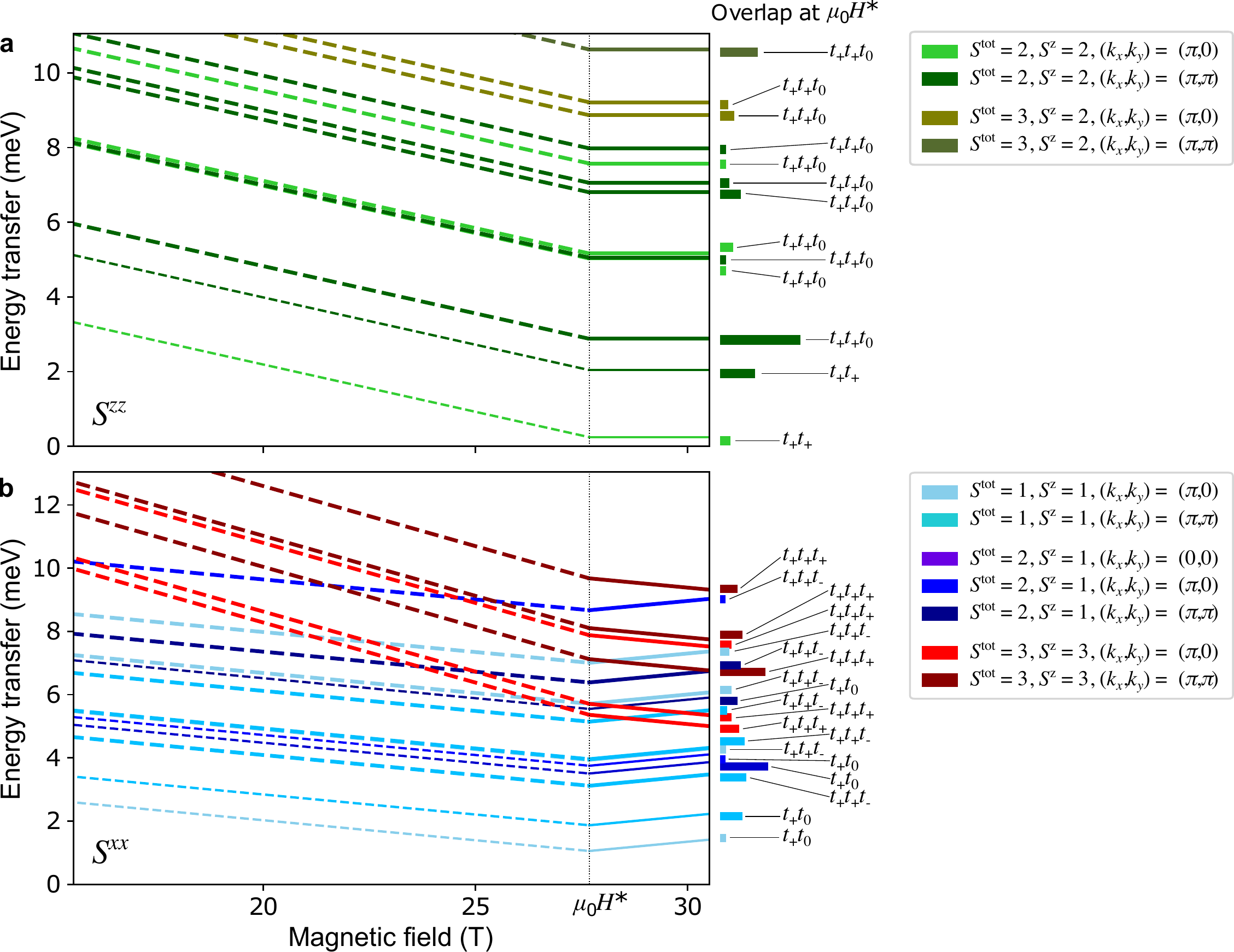}
\caption{Identification of composite multi-triplon states in the spectrum of 
the 4$\times$4 SSM spin cluster in a magnetic field strong enough to condense 
the lowest $|t_+ t_+ \rangle$ state. {\bf a} Excitations accessible in the 
longitudinal component ($S^{zz}(k,\omega,H)$) of the DSF. {\bf b} Excitations 
accessible in the transverse component ($S^{xx}(k,\omega,H)$) of the DSF. On 
this cluster, the lowest-energy component of the two-triplon bound state 
reaches the singlet energy at a field equivalent to $\mu_0 H^* = 27.7$~T when 
using the $g$-factor of \SCBO. We show all of the higher-weight states 
appearing at low energies in the momentum sectors $(0,0)$, $(\pi,0)$ and 
$(\pi,\pi)$. Because the spin Hamiltonian is of pure Heisenberg type, the 
field term causes only a linear evolution whose gradient is determined by 
$S^z$, with a kink and a $\Delta S^z = 2$ change of slope at $H^*$.}
\label{fig:lif}
\end{figure*}

\subsection{Identification of multi-triplet excitations}

To identify the nature of the excited states found in our cylinder MPS 
calculations (Figs.~1f and 4a-b of the main text), we exploit the fact that 
the DM interactions of the system are too weak to introduce confusion (by a 
mixing of spin sectors) and neglect them completely. For identification 
purposes it is fully sufficient to use a pure SSM and to perform ED on a 
cluster of only 4$\times$4 spins, where (as noted above) the available $k_x$ 
and $k_y$ momentum sectors are 0 and $\pi$. The ED spectrum is computed in 
symmetry sectors labelled by the quantum numbers of total spin 
$(S^{\mathrm{tot}})$, total magnetization $(S^z)$ and the momenta $k_x$ and 
$k_y$, and we make use of the separate $S^{zz}$ and $S^{xx}$ components of the 
DSF to identify states from the action of the $S^z$ and $S^x$ spin operators. 

As discussed in the main text, the two-triplon bound state, $|t_+ t_+ \rangle$, 
is the first to condense as the applied magnetic field is increased. This 
state belongs to the symmetry sector $S = 2$, $S^z = 2$, and its lowest-energy 
component appears at $(k_x, k_y) = (0,0)$ on the 4$\times$4 cluster. In the 
limit $J'/J \rightarrow 0$, the energies of $n$-triplon states approach $nJ$, 
and thus all the states in the spectrum are readily identified by following 
the evolution of their energies with $J'/J$, as we show in Fig.~\ref{Imte} 
for the lowest three $S^{\mathrm{tot}}$ sectors. It is clear that the different 
$n$-triplon sectors remain well separated in energy for smaller values of 
$J'/J = 0$, but that their energies show significant overlap as the 
interaction ratio approaches the \SCBO~value of $J'/J = 0.63$. 

For more detailed insight into the character of each $n$-triplon sector, we 
consider the action of the operators $S^z$ and $S^x$ on the two-triplon bound 
state. From the Lehmann representation of the DSF, the excitations of a given 
initial state lie in symmetry sectors allowed by the selection rules
\begin{enumerate} 
\item  $S^{zz}(k_x, k_y)$: $\Delta S = 0, 1$; $\Delta S^z = 0$; $k_x = k_{x,f}
 - k_{x,i}$; $k_y = k_{y,f} - k_{y,i}$;
\item  $S^{xx}(k_x, k_y)$: $\Delta S = \pm 1$; $\Delta S^z = \pm 1$; $k_x = 
k_{x,f} - k_{x,i}$; $k_y = k_{y,f} - k_{y,i}$. 
\end{enumerate}
The excitations allowed by these rules do not contribute equally to the 
spectral weight, because all have different matrix elements. To deduce the 
contribution of each state in the spectrum, we compute the overlap 
\begin{eqnarray}
{\cal N} = \left |\left\langle \psi_{\rm e} | S^{\alpha}_{k_x, k_y} |\psi_0 
\right\rangle \right|^2,
\label{overlap}
\end{eqnarray}
where $\alpha = x,z$, $|\psi_0 \rangle$ is the ground (initial) state and 
$|\psi_{\rm e} \rangle$ is the excited (final) state. We take ${\cal N} \ge 
0.01$ as the criterion to regard $|\psi_{\rm e} \rangle$ as a state making a 
significant contribution to the DSF. We note that, while the initial state 
is always the lowest-energy state of the $(S, S^z, k_x, k_y) = (2, 2, 0, 0)$ 
sector, this is not in fact the ground state at $H < H^*$. 

The final step is to deduce the energies of these high-weight levels in the 
presence of a magnetic field. Because of the purely Heisenberg nature of the 
Hamiltonian, every state undergoes a simple Zeeman shift determined by its 
$S^z$ quantum number, giving 
\begin{equation}
\omega_{\rm{e}} (H) = E_{\rm{e}} - E_0 + (S^z_0 - S^z_{\rm{e}}) H,
\end{equation}
which allows us to track the origin of all the observed states at low fields, 
as we show in Fig.~\ref{fig:lif}. The small cluster size we have used for this 
exercise results in non-trivial finite-size effects on the energy levels: in a 
field, the $t_+$ branch crosses the singlet before the lowest $|t_+ t_+ \rangle$
branch, and the crossing field defined from the latter is $\mu_0 H = 27.7$~T. 
However, these quantitative details are not relevant for the qualitative 
purpose of identifying the high-weight excitations appearing in an INS 
experiment on a system with a two-triplon bound-state condensate (i.e.~a spin 
nematic) as the field-induced ground state. Figure \ref{fig:lif} provides the 
full justification for the statements made in the main text that the many 
excitations visible in the INS and MPS spectral functions are composite two- 
and three-triplon excitations of several different types.  


%